\documentclass[fleqn,usenatbib]{mnras}

\usepackage[T1]{fontenc}

\DeclareRobustCommand{\VAN}[3]{#2}
\let\VANthebibliography\thebibliography
\def\thebibliography{\DeclareRobustCommand{\VAN}[3]{##3}\VANthebibliography}

\usepackage{graphicx}	
\usepackage{amsmath}	
\usepackage{amssymb}	

\usepackage{ulem} 
\usepackage{gensymb}
\usepackage{hyperref}

\usepackage{algorithm, algorithmic}
\usepackage{color}
\newcommand{\tcr}{\textcolor{red}}
\newcommand{\tcb}{\textcolor{blue}}

\usepackage{multirow}
\usepackage{array}
\usepackage{siunitx}
\usepackage{booktabs}
\usepackage{makecell}

\usepackage{caption}
\usepackage{subcaption}
\usepackage{capt-of}

\usepackage{newtxtext,newtxmath}

\title[Deep Learning for Improved Seeing]{Uncertainty-Aware Learning for Improvements in Image Quality of the Canada-France-Hawaii Telescope}

\author[S. Gilda]{Sankalp Gilda$^{1}$,\thanks{This work was conducted while the first author was a Ph.D. candidate in the Department of Astronomy, University of Florida, Gainseville, FL 32611, United States.\newline E-mail: sankalp.gilda@gmail.com}
Stark C. Draper$^{2}$,
S\'ebastien Fabbro$^{3}$,
William Mahoney$^{4}$,
\newauthor
Simon Prunet$^{4,9}$,
Kanoa Withington$^{4}$,
Matthew Wilson$^{4}$,
Yuan-Sen Ting$^{5,6,7,8}$,
\newauthor
and Andrew Sheinis$^{4}$
\\
$^{1}$ML Collective\\
$^{2}$Department of Electrical and Computer Engineering, University of Toronto, Toronto, ON M5S 3G4, Canada\\
$^{3}$National Research Council Herzberg, 5071 West Saanich Road, Victoria, BC, Canada\\
$^{4}$Canada-France-Hawaii-Telescope, Kamuela, HI, United States\\
$^{5}$Institute for Advanced Study, Princeton, NJ 08540, United States\\
$^{6}$Department of Astrophysical Sciences, Princeton University, Princeton, NJ 08540, USA\\
$^{7}$Observatories of the Carnegie Institution of Washington, 813 Santa Barbara Street, Pasadena, CA 91101, USA\\
$^{8}$Research School of Astronomy \& Astrophysics, Australian National University, Cotter Rd., Weston, ACT 2611, Australia \\
$^{9}$Université Côte d'Azur, Observatoire de la Côte d'Azur, CNRS, Laboratoire Lagrange, France \\
}

\date{Accepted 2021 November 04. Received 2021 November 03; in original form 2021 August 30}

\pubyear{2015}

\usepackage{xcolor, soul}

\usepackage{dsfont}

\definecolor{orange}{rgb}{1,0.5,0}
\definecolor{mydarkcyan}{rgb}{0,0.5,0.5}
\newcommand{\yst}[1]{\textcolor{mydarkcyan}{YST: #1}}
\newcommand{\seb}[1]{\textcolor{olive}{SF: #1}}
\newcommand{\bm}[1]{\textcolor{red}{#1}}
\newcommand{\sg}[1]{\textcolor{green}{#1}}

\newcommand{\iqMea}{\textrm{IQ}_{\textrm{Measured}}}
\newcommand{\iqOpt}{\textrm{IQ}_{\textrm{Optics}}}
\newcommand{\iqDome}{\textrm{IQ}_{\textrm{Dome}}}
\newcommand{\iqAtmo}{\textrm{IQ}_{\textrm{Atmospheric}}}
\newcommand{\iqAtmoPrime}{\textrm{IQ}_{\textrm{Atmospheric}}^{'}}

\newcommand{\iqCorr}{\textrm{IQ}_{\textrm{Corrected}}}

\begin{document}
\label{firstpage}
\pagerange{\pageref{firstpage}--\pageref{lastpage}}
\maketitle

\begin{abstract}
We leverage state-of-the-art machine learning methods and a decade's worth of archival data from CFHT to predict observatory image quality (IQ) from environmental conditions and observatory operating parameters. Specifically, we develop accurate and interpretable models of the complex dependence between data features and observed IQ for CFHT's wide-field camera, MegaCam. Our contributions are several-fold. First, we collect, collate and reprocess several disparate data sets gathered by CFHT scientists. Second, we predict probability distribution functions (PDFs) of IQ and achieve a mean absolute error of $\sim0.07''$ for the predicted medians. Third, we explore the data-driven actuation of the 12 dome ``vents'' installed in 2013-14 to accelerate the flushing of hot air from the dome. We leverage epistemic and aleatoric uncertainties in conjunction with probabilistic generative modeling to identify candidate vent adjustments that are in-distribution (ID); for the optimal configuration for each ID sample, we predict the reduction in required observing time to achieve a fixed SNR.  On average, the reduction is $\sim12\%$. Finally, we rank input features by their Shapley values to identify the most predictive variables for each observation. Our long-term goal is to construct reliable and real-time models that can forecast optimal observatory operating parameters to optimize IQ. We can then feed such forecasts into scheduling protocols and predictive maintenance routines. We anticipate that such approaches will become standard in automating observatory operations and maintenance by the time CFHT's successor, the Maunakea Spectroscopic Explorer, is installed in the next decade.


\end{abstract}

\section{Introduction}\label{sec:introduction}

Situated at the summit of the 4,200m volcano of Maunakea on the island of Hawaii, the Canada-France-Hawaii Telescope is one of the world's most productive ground-based observatories~\citep{crabtree:2019}. The productivity of CFHT is due, in part, to the exquisite natural image quality (IQ) delivered at the observatory's location on Maunakea. Image quality is key metric of observatory operations and relates directly  to  realized signal-to-noise ratio (SNR) as well as to achievable spatial resolution. SNR and spatial resolution, in turn, dictate the information content of an image.  They thereby provide a direct measure of the efficacy of scientific observation.

\begin{figure*}
   \centering
   \includegraphics[width=0.95\linewidth]{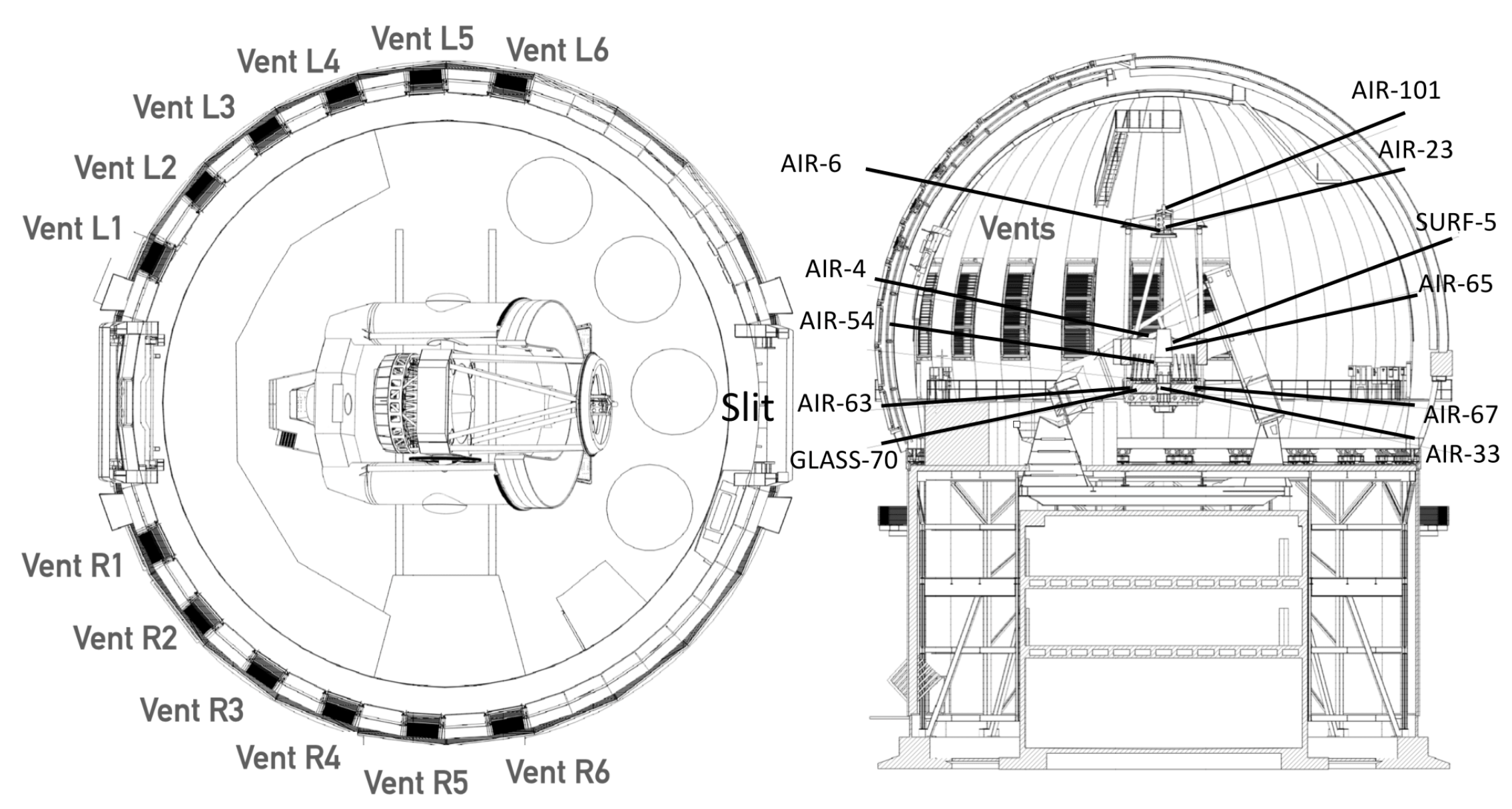}
   \caption{A schematic of the CFHT; top-view and profile. The twelve actionable dome vents are marked.  Important thermal sensors identified in past works (see Section~\ref{sec:relatedWork}) are highlighted. These sensors are detailed in Table~\ref{tab:description_of_tempsensors}.}
   \label{fig:dome}
\end{figure*}

The difference between the theoretically achievable and measured IQ can be attributed to air turbulence in the optical path.  There are two sources of turbulence.  The first is atmospheric.  At the summit of Maunakea atmospheric turbulence is minimal due to the smooth laminar flow of the prevailing trade winds and the height of the summit; this is the reason CFHT and other world-class observatories are located on Maunakea.  The second is turbulence induced by local thermal gradients between the observatory dome itself (and the structures within) and the surrounding air.  There have been continual improvements in the CFHT facility since 1979, many aimed at reducing this source of turbulence. We particularly make note of the December 2012 installation of dome vents.  After a protracted mechanical commissioning period that lasted about 18 months, the vents  came online in July of 2014. By allowing the (generally) hotter air within the observatory to flush faster, the vents accelerate thermal equalization. A schematic of the dome and the vents is provided in Figure~\ref{fig:dome}. A listing of the temperature sensors marked in Figure~\ref{fig:dome} is provided in Table~\ref{tab:description_of_tempsensors}.
Even given these improvements, and as is the case with all major ground-based observatories, the IQ attained at CFHT rarely reaches what the site can theoretically deliver.\footnote{Direct (prime focus) wide field imaging systems that we consider in this paper are not compatible with adaptive optics~\citep{adaptiveoptics0,adaptiveoptics1}, which require a relay or an adaptive secondary mirror. Although such AO systems can be designed to specifically correct for ground layer, enabling imaging of wide fields at improved seeing resolutions~\citep{imaka}, they are not well suited to correct dome induced turbulence, which may not be homogeneously distributed over the pupil or may be at too high a spatial frequencies to be corrected by a deformable mirror.}



Our project is motivated by our strong belief that the ability to model and predict IQ accurately in terms of the exogenous factors that affect IQ would prove enormously useful to observatory operations.  Observing time at world-class facilities like CFHT is oversubscribed many-fold by science proposals. Specifically at CFHT, good seeing time, defined as time when IQ is smaller than the mode seeing of $0.70''$ in the $r$-band, is oversubscribed roughly three-fold. Further, observations frequently either fail to meet, or exceed, the IQ requirements of their respective science proposals~\citep{milli2019nowcasting_paramal}. Through accurate predictions we can better match delivered IQ to scientific requirements.  We thereby aim to unlock the full science potential of the observatory. 
If we can predict the impact on IQ of the parameters the observatory can control (pointing, vent and wind-screen settings, cooling systems), then by adjusting these parameters and (perhaps) the order of imaging, we create an opportunity to accelerate scientific productivity.  In this work we lay the groundwork for  these types of improvements.

In this paper, we leverage almost a decade-worth of sensor telemetry data, post-processing IQ measurements, and exposure information that is collected in tandem with each CFHT observation.  Based on this data we build a predictive model of IQ. Through the implementation of a feed-forward mixture density network (MDN, \citealt{bishop_mdn}), we demonstrate that ancillary environmental and operating parameter data are sufficient to predict IQ accurately. Further, we illustrate that, keeping all other settings constant, there exists an optimal configuration of the dome vents that can substantially improve IQ. Our successes here lay the foundation for the development of automated control and scheduling software.

The IQ prediction system we detail in this paper is developed for MegaPrime~\footnote{\url{https://www.cfht.hawaii.edu/Instruments/Imaging/MegaPrime/}}, a wide-field optical system with its mosaic camera MegaCam \citep{megacam}. MegaPrime is one of CFHT's most scientifically productive instruments. Built by CEA in Saclay, France, and deployed in 2003, MegaCam is a wide-field imaging camera. It is used extensively for large surveys covering thousands of square degrees in the sky and ranging in depth from 24 to 28.5 magnitude. MegaCam is placed at the prime focus of CFHT.  It includes an image stabilization unit and an auto-focus unit with two independent guide charge-coupled device (CCD) detectors. MegaCam consists of $40$ CCDs, each $2048 \times 4612$ pixels in size, for a total of $378$ megapixels. The image plane covers a $1 \degree \times 1 \degree$ square field of view at a resolution of $0.187''$ (arc-seconds) per pixel. The CFHT archive at the Canadian Astronomy Data Centre (CADC) contains close to 300k Megacam science exposures with 24 filter pass bands. These images have a median IQ of $\sim0.7''$ in the $r-$band. 
One of our main results is that, based purely on environmental and observatory operating conditions, we can predict the \textit{effective} MegaPrime IQ (MPIQ for the rest of the paper) to a mean accuracy of about $0.07''$.

\begin{table}
    \centering
    \caption{Brief description of the temperature sensors marked in Figure~\ref{fig:dome}.}
    \label{tab:description_of_tempsensors}
    \begin{tabular}{ll}
        \toprule
        Probe label & Description\\ \midrule
        AIR-4 & Air temperature -- Caisson, west \\
        SURF-5 & Steel temperature -- Caisson, east \\
        AIR-6 & Air temperature -- Upper end, west \\
        AIR-23 & Air temperature -- Under end, east \\
        AIR-33 & Air temperature -- Under primary, west \\
        AIR-54 & Air temperature -- Mirror cell, west \\
        AIR-63 & Air temperature -- Under primary, south \\
        AIR-65 & Air temperature -- Inside spigot, north \\
        AIR-67 & Air temperature -- Under primary, north \\
        GLASS-70 & Glass temperature -- Under primary, south \\
        AIR-86 & Air temperature -- Weather tower \\
        AIR-101 & Air temperature -- MegaPrime exterior \\ \bottomrule
    \end{tabular}
\end{table}

\begin{table*}
\caption{Data fields in the MegaCam dataset.}
\label{table:megaCamData}
\centering
\begin{tabular}{p{0.15\linewidth} p{0.095\linewidth}p{0.05\linewidth} p{0.125\linewidth}p{0.45\linewidth}}\toprule\toprule
Parameter & Units & \#Features & Range & Description \\\midrule
\multicolumn{5}{l}{Environmental} \\ \midrule
Temperature & \degree C & 57 & [-8,20] / [-200,850] & Temperature values from sensors in and around the dome. Three sensors are placed within the  dome.  The rest are external.\\
Wind speed & knots & 1 & [0,35] & Wind speed at the weather tower.\\
Wind azimuth & NONE & 2 & [-1,1] & Sine and cosine of wind azimuth with respect to true north. \\
Humidity & \% & 2 & [1.4,100] & Measured both at the top of the observatory dome, and at the weather tower.\\
Dew point & \% & 2 & [1.4,100] & Measured both in the basement of the observatory building, and at the telescope mirror cell (near GLASS 70 in Figure \ref{fig:dome})\\
Barometric pressure & mm of Hg & 1 & [607,626] & Atmospheric pressure measured on the fourth floor of the observatory building.\\
MPIQ & $''$ & 1 & [0.35,2.36] & Measured seeing from MegaCam/MegaPrime.\\
\midrule
\multicolumn{5}{l}{Observatory} \\ \midrule
Vents & NONE; NONE; NONE & 36 & \{0,1\}; [0,1]; [0,1] & For each sample, we have three types of vent values: vent configuration (`OPEN' or `CLOSE'), and Sine and cosine of vent$_{\rm{AZ}}$ \\
Dome azimuth & NONE & 2 & & Sine and cosine of the angle of the slit-center from true North. \\
Pointing altitude & NONE & 1 & [0.15,1] & Sine of the angle of the telescope focus from the horizontal.\\
Pointing azimuth & NONE & 2 & [-1,1] & Sine and cosine of angle of the telescope focus from true north. \\
Wind screen position & NONE & 2 & [0,1] & Fraction that the wind screen is open. (The wind screen is located at the `Slit' position in the left of Figure \ref{fig:dome}.)\\
Central Wavelength & nm & 1 & [354,1170] & Central wavelengths of each of the 22 filters.\\
Dome Az $-$ Pointing Az & NONE & 2 & [-1,1] & Sine and cosine of difference between dome and pointing azimuths.\\
Dome Az $-$ Wind Az & NONE & 2 & [-1,1] & Sine and cosine of difference between dome and wind azimuths.\\
Pointing Az $-$ Wind Az & NONE & 2 & [-1,1] & Sine and cosine of difference between wind and pointing azimuths.\\
\midrule
\multicolumn{5}{l}{Other} \\ \midrule
Exposure time & seconds & 1 & [30,1800] & Observation time per sample.\\
Observation Time & NONE & 4 & [-1,1] & Sine and cosine of $\frac{\rm{hour\_of\_day}}{23}$ and $\frac{\rm{week\_of\_year}}{51}$.\\
\bottomrule
\end{tabular}
\end{table*}



We train our models to predict MegaPrime IQ (MPIQ) using CFHT observations dating back to July 23, 2014.  While the CFHT data catalogue dates back to 1979, we use  data only for the period in which the dome vents have been present.
 The collected measurements include temperature, wind speed, barometric pressure, telescope altitude and azimuth, and configurations of the dome vents and windscreen. \footnote{We have made our data set publicly available at \url{https://www.cfht.hawaii.edu/en/science/ImageQuality2020/}.}  In Table~\ref{table:megaCamData} we summarize the environmental sensors, observatory parameters, and miscellaneous features used in this work.  

Our goal is to toggle the twelve CFHT vents based on our predictions of MPIQ. We must thus err on the side of caution -- CFHT is already oversubscribed by a factor of $\sim 3$, and any mis-prediction of vent configurations would waste valuable time in re-observing targets. We therefore eschew point predictions in favor of making a prediction of the MPIQ distribution (the conditional PDF) for each data sample. We followed this procedure when presenting some preliminary results in \cite{gilda_cfht_neurips}. Here we extend that work significantly and make the following contributions:
\begin{enumerate}
    \item We compile and collate several sets of measurements from various environmental sensors, metadata about observatory operating conditions, and measured IQ from MegaCam on CFHT. We curate and combine these sources of data into a single dataset.  We publish the curated dataset.
   \item We use supervised learning algorithms to predict IQ at $0.07''$ accuracy. We present results for a gradient boosting tree algorithm and for a mixture density network (MDN). For the latter we provide a detailed analysis of {\it feature attributions}, assigning the relative contribution of each input variable to predicting MPIQ.
  

\item The IQ predictions we produce are robust. We perform an uncertainty quantification analysis.  Guided by a robust variational autoencoder (RVAE) that models the density of the data set, we identify non-representative configurations of our sensors.

 \item We use our MDN to find the optimal vent configurations that would have resulted in the lowest IQ. We use these predictions to estimate the annual increase in science return and scientific observations.  We find the improvement to be $\sim 12\%$. This improvement results from increased observational efficiency at CFHT, in particular minimizing the observation times for hypothetical \textit{r-}band targets of the $25^{\rm th}$ magnitude to achieve an SNR of 10; these figures are representative of deep observations of faint targets of large imaging programs at CFHT like the Canada France Imaging Survey \footnote{\url{https://www.cfht.hawaii.edu/Science/CFIS/}}.
\end{enumerate}

We structure the rest of this paper as follows. In Section~\ref{sec:relatedWork} we discuss relevant previous work. In Section~\ref{sec:data} we explore in depth the various sources of input data and the processing pipeline we implement to collate and convert the data sources into the final usable dataset. In Section~\ref{sec:method} we describe in detail our methodology, including attributes of our gradient boosting tree and neural network methods, feature importance method, and our predictions for best vent configurations. In Section~\ref{sec:results} we present our results. We conclude in Section~\ref{sec:conclusion}.  To help keep our focus on astronomy, some supporting figures that help detail our machine-learning implementations are deferred to Appendix~\ref{sec.workflowFigs}.

\section{Related Work}\label{sec:relatedWork}

The summit of Maunakea was selected as the site for CFHT due to its excellent astronomical observing properties: low turbulence, low humidity, low infrared absorption and scattering, excellent weather,  clear nights.  Image quality, or ``seeing'', is quantified using the full-width half-max (FWHM) measure. FWHM, expressed in arcseconds ($''$), is calculated as the ratio of the width of the support of a distribution, measured at half the peak amplitude of the distribution, to  half the peak-amplitude value; smaller FWHM is better.  For example, the FWHM  of the Gaussian distribution is $2 \sqrt{2 \log 2} \sigma$, roughly $2.4$ times the standard deviation $\sigma$.  In our application, FWHM operationally quantifies the degree of blurring of uncrowded and unsaturated images of point sources (such as a star or a quasar) on the central CCDs of a MegaCam frame. 
The FWHM measured this way, referred to as image quality or IQ, is an aggregate of multiple sources.\footnote{Note that larger FWHM $\implies$ higher ``seeing'' $\implies$ poorer image quality (IQ) $\implies$ more arc-sec.  So, lower FWHM which equates to a better IQ (fewer arc-sec) is desired.} The main contributors to FWHM / IQ are: imperfections in the optics ($\iqOpt$), turbulence induced by the dome ($\iqDome$), and  atmospheric turbulence ($\iqAtmo$). These contributions are well-modeled as being independent and as combining to form the measured IQ ($\iqMea$) according to
\begin{align}
\iqMea^{5/3} =  \iqOpt^{5/3} + \iqDome^{5/3} + \iqAtmo^{5/3}.
\label{eqn:iq_measured}
\end{align}
If the contributions were modeled by a Gaussian distribution, the exponents in Equation (\ref{eqn:iq_measured}) would be $2 = 6/3$ (because variances of independent Gausian random variables add).  The $5/3^{\rm rd}$ power is due to the spectrum of turbulence which was characterized by Kolmogorov in 1941~\citep{tartarskii}.  We note that while the contribution of the optics is not due to turbulence, we still use the power of $5/3$ in our model for consistency. Finally, we note that of the three contributors we can only influence $\iqDome$ through actuation of various observatory controls.
    
\begin{figure*}
\begin{subfigure}{0.99\textwidth}
    \centering
    \includegraphics[width=.98\linewidth]{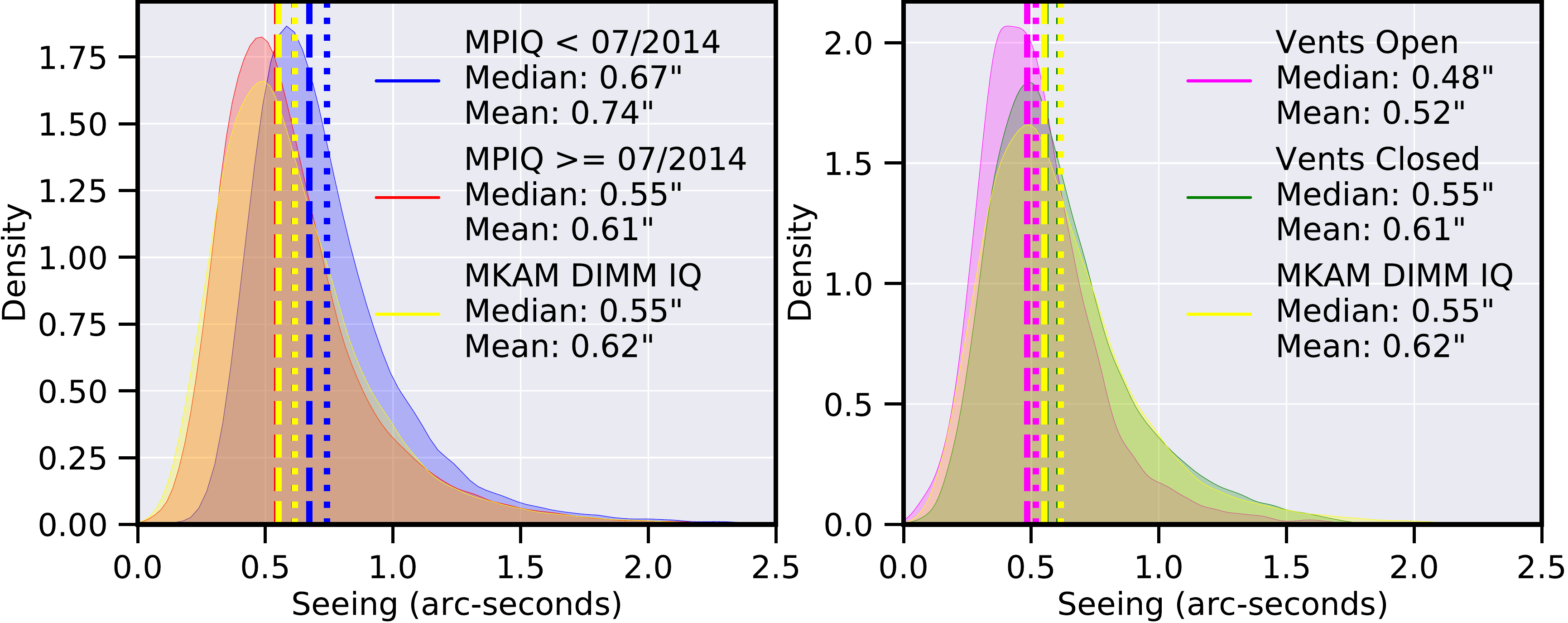}
    \label{fig:mpiq_mkamiq}
\end{subfigure}
\newline
\begin{subfigure}{0.99\textwidth}
    \centering
    \includegraphics[width=.98\linewidth]{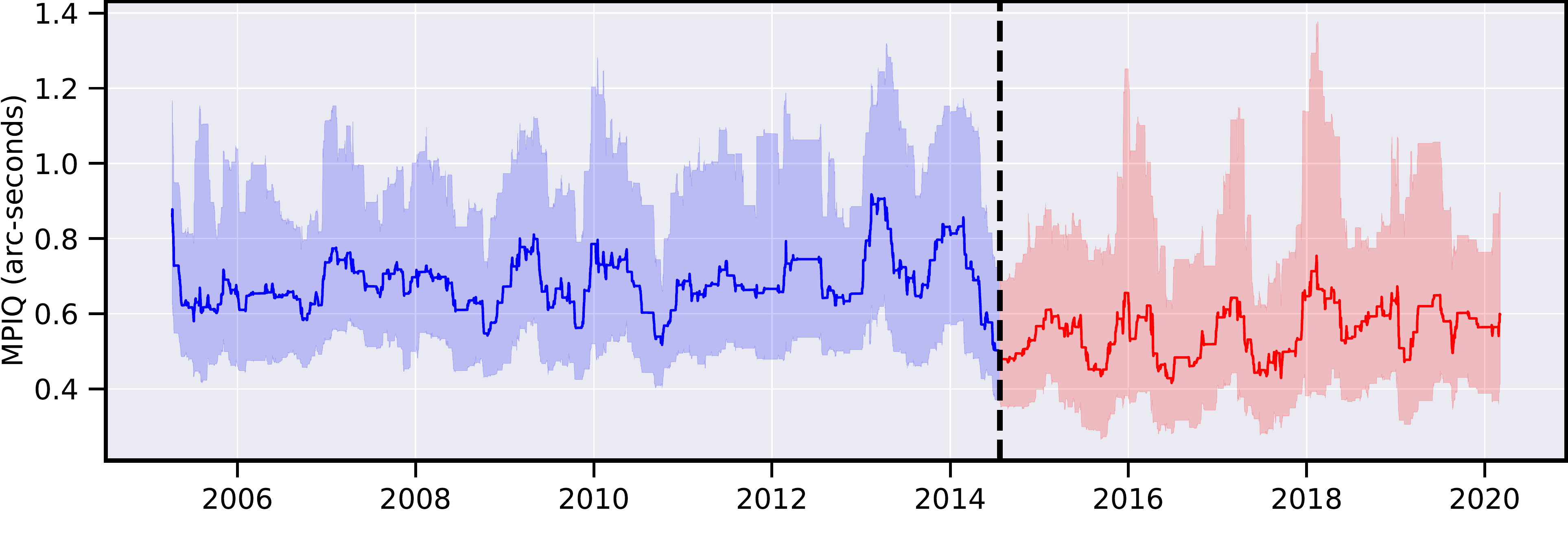}
    \label{fig:fwhm_history}
\end{subfigure}
\caption{Seeing evolution and distribution. \textbf{Upper left:}  
Distribution of seeing measured by 
the MKAM Differential Image Motion Monitor (DIMM) and effective IQ measured at MegaPrime since 2010. Effective MPIQ is the measured IQ less the contributions from optics -- see Equation~(\ref{eqn:iq_measured}) and Section \ref{sec:data_cleaning}. Both the MKAM and the MPIQ curves peak at $\sim 0.55''$; the former contains relatively higher seeing contribution from the ground layer, while the latter includes contribution from the dome itself (see Section \ref{sec:relatedWork} for detailed discussion). Both peaks are higher than the best possible seeing at the site of $\sim0.43''$. \textbf{Upper right:} Distribution of seeing {\it after} the installation of the vents as a function of vent configuation: \textit{all-open} or \textit{all-closed}.  Observe the statistics of the former are much better than those of the latter.  We also plot the MKAM histogram, which is basically unchanged from prior to the installation of the vents.  \textbf{Lower:} Quarterly averaged MegaPrime IQ of the CFHT. The wiggly curve of decaying mean and oscillation amplitude is a model that peaks in midwinter, when the outside air temperature tends to be colder than the dome air. The drop in July of 2014 corresponds to when the vents first started to be used; this is why in the data input to our models we do not use samples from before this month.}
\label{fig:hist_preliminaries}
\end{figure*}

While the mean free atmosphere ($\textrm{IQ}_{\textrm{Atmosphere}}$) seeing on Maunakea is estimated to be about $0.43''$~\citep{salmon2009cfht}, in practice, the IQ realized at CFHT is usually worse (i.e. the seeing is higher). Through 40 years of effort by the CFHT staff and  consortium scientists,  $\iqMea$ has steadily decreased, from early values of $2''$ or greater to its current median value of around $0.71''$. Removal of  $\iqOpt$ further reduces this figure to $0.55''$ (see~Figure~\ref{fig:hist_preliminaries}, Equation \ref{eqn:iq_optics_correction}, and Section \ref{sec:feature_engineering}).\footnote{\url{https://www.cfht.hawaii.edu/Science/CFHLS/T0007/T0007-docsu11.html}} In the remainder of this section we discuss prior efforts to quantify IQ and to reduce IQ. Later, in Section~\ref{sec:data} when we discuss our data sources, we return to~(\ref{eqn:iq_measured}) and step through a number of sources of variation in observing conditions (e.g., wavelength of observation, elevation of observation) that we correct to produce a normalized dataset in which IQ measurements from distinct observations can be directly compared.

Early published efforts to quantify and reduce the IQ at CFHT (e.g. \cite{racine1984}) detail campaigns to minimize turbulence inside and around the dome, including analysis and measurements of the opto-mechanical imperfections of the telescope. The team led by Ren\'e Racine estimated that if the in-dome turbulence was corrected and the telescope imperfections were removed, the natural Maunakea seeing would  offer images with IQ below $0.4''$ FWHM on one-quarter of the nights. Later efforts ~\citep{racine1991mirror} used data from the (then) new HRCam, a high-resolution imaging camera at the prime focus of CFHT, to develop a large and homogeneous IQ data set. They correlated their IQ data with thermal sensor data, through which they were able to identify and quantify ``local seeing'' effects.  Their main findings, relevant to our work, are listed below.
\begin{enumerate}
\item The contribution of mirror optics $\iqOpt$ amounts to about $0.4'' (\Delta T)^{6/5}$ where $\Delta T$ is the temperature difference between the primary mirror and the surrounding air.
\item The dome contribution $\iqDome$ amounts to about $0.1'' (\Delta T)^{6/5}$ where $\Delta T$ is the temperature difference between the air inside and outside of the dome.
\item The median natural atmospheric seeing at the CFHT site $\iqAtmo$ is $0.43'' \pm 0.05''$.  The 10th and 90th percentiles are roughly $0.25''$ and $0.7''$.
\end{enumerate}

More recent follow-up work is presented in~\cite{salmon2009cfht}.  The authors correlate measured IQ using the (then) new MegaCam with temperature measurements. They analyze $36,520$ MegaCam exposures made in the $u$, $g$, $r$, $i$, and $z$-bands in the three year period between August 2005 and  August 2008.  They find strong dependencies of the measured IQ on temperature gradients. Furthermore, in  Table 4 of~\cite{salmon2009cfht} the authors categorize important factors that contribute to the seeing -- atmosphere, dome, optics -- and provide estimates of their respective contribution.
As the authors discuss, these estimates update the findings of~\cite{racine1991mirror}.  The most significant findings of~\cite{salmon2009cfht} can be summarized as follows.
\begin{enumerate}
\item The orientation of the dome slit with respect to the wind direction has important effects on IQ.
\item The median dome induced seeing $\iqDome$ before the installation of the vents in 2013 was $0.43''$.
\item The seeing contribution from optics and opto-mechanical imperfections $\iqOpt$ varied from $0.46''$  in the u-band  to $0.28''$  in the i-band. 
\item Atmospheric seeing $\iqAtmo$ at the CFHT site at a wavelength of $500$ nm and an elevation of $17$m above ground was measured using a separate imager.  The median $\iqAtmo$ measured was $0.55''$.  This estimate of atmospheric seeing is independent of effects related to the dome and the optics.
\end{enumerate}

The culminating result of these studies that analyzed the delivered IQ was the December 2012 installation, and July 2014 initial use, of the 12 dome vents depicted Figure~\ref{fig:dome}. Since their installation, CFHT operators have kept all 12 vents completely open as often as possible, barring conditions of mechanical failure and strong winds. As already mentioned, this allows  faster venting of internal air and  equalization of internal and external temperatures.\footnote{\url{http://www.cfht.hawaii.edu/AnnualReports/AR2012.pdf}}  The vent-related improvement in seeing has been dramatic, 
with median $\iqMea$ improving from about $0.67''$ to $0.55''$~\footnote{\url{http://www.cfht.hawaii.edu/AnnualReports/AR2014.pdf}}. 

In order to have an external, regularly sampled seeing reference, we use the Maunakea Atmospheric Monitor \citep[MKAM,][]{Skidmore+09,mkam2}. This telescope, dedicated to seeing monitoring, is mounted on top of a weather tower just outside of CFHT. It has a composite instrument, including a Multi Aperture Scintillation Sensor (MASS) and a Differential Image Motion Monitor (DIMM). We only use data from the latter as the former is insensitive to the lower layers of turbulence. DIMMs measure seeing by computing the variance of the relative motion of the images formed by two separate sub-apertures, therefore probing the curvature of the wave-front. This variance can be directly related to the the full-width at half-max (FWHM) of the point spread function (PSF) in long exposures given the wavelength of observation and the sub-apertures diameter \citep[see e.g. ][]{Sarazin+90}. While MKAM measurements are free of the CFHT dome contribution to seeing, they are also sensitive to part of the ground layer contributions to seeing that are not seen by the CFHT instruments, partly due to the lower altitude of the weather tower as compared to the CFHT aperture, and to localized differences in the summit turbulence in the first few meters above ground. MKAM thus serves as a slightly noisy seeing reference for Megacam, free of CFHT dome seeing contributions.

In the left sub-plot of Figure \ref{fig:hist_preliminaries}, we plot histograms of the \textit{corrected} seeing values from MegaCam both before (starting February $2^{\rm nd}$ 2002) and after July $7^{\rm th}$ 2014, when the vents started being used; the latter is the the start date of our data set used in the remainder of this work. We compare these with the seeing distribution from MKAM and MegaCam for observations since 2002. Corrected seeing removes the contribution of the telescope optics from the measured seeing, and is defined in Equation \ref{eqn:iq_optics_correction}. In the top-right plot in Figure \ref{fig:hist_preliminaries}, ``Vents Open'' refers to samples where the 12 vents are either all open, or at most one of them is closed, whereas ``Vents Closed'' incorporates samples where all 12 vents are closed. As can be seen, the introduction of the vents has reduced the median MPIQ by $0.21'' = \left((0.55^{5/3}-0.48^{5/3})^{3/5}\right)$, and the mean MPIQ by $0.27'' = \left((0.61^{5/3}-0.51^{5/3})^{3/5}\right)$. However, there is still money on the table -- the estimated free-air, observatory-free IQ at the CFHT site is estimated to be $\sim0.43-0.44''$ \citep{salmon2009cfht}. This means that there is still a possible improvment in median IQ of $0.15'' = \left((0.48^{5/3}-0.435^{5/3})^{3/5}\right)$, and a mean IQ of $0.23'' = \left((0.52^{5/3}-0.435^{5/3})^{3/5}\right)$. This range of improvement was independently verified by another CFHT team in 2018 \citep{racine2018}. They found that, when open, the dome vents on average reduce IQ by $0.37''$. While this number is significantly larger their either of our estimates of $0.21''$ or $0.27''$, their estimate of degradation of IQ of about $0.20''$, caused by residual eddies induced by thermal differences in the dome, closely matches our own. It is precisely this residual part of $\iqDome$ that we aim to capture in our work. We later show in Figures \ref{fig:common_deltaoptimalreference_vs_nominal_iqs} and \ref{fig:common_deltaoptimalreference_vs_nominal_iqs_robust}, and in Section \ref{sec:resultsRelContribIQ} that our models are indeed able to capture these improvements.

We remind the reader, as mentioned above, that CFHT operators have kept all 12 vents completely open as much as possible.  They have chosen this manner of operation as they had no basis upon which to choose a more varied configuration of the dome vents.  Although, we note that fluid flow modeling conducted during the vent design process predicted that intermediate settings (i.e., neither all-optimal nor all-closed) would optimally reduce internal turbulence \citep{wind_tunnel_test}. By tuning the dome vent configurations, based on current environmental conditions, to a setting between all-open and all-closed, we aim to reduce this residual.


From a programmatic perspective, our work is a natural extension of~\cite{racine1984, racine1991mirror,salmon2009cfht, racine2018}. While these prior investigations correlated IQ with measurements of temperature gradients, our work tries to relate all of the metrics (not solely the temperature metrics) with the $\iqMea$ through the application of advanced machine learning techniques. Further, rather than only establishing correlation, we also seek to understand whether by actuating the dome parameters under our control we can improve the delivered $\iqMea$. Recent work at the Paranal Observatory by~\cite{milli2019nowcasting_paramal} similarly collected 4 years of sensor data, and trained random forest and neural networks to model and forecast over the short term ($<2$ hrs) the DIMM seeing and the MASS-DIMM atmospheric coherence time and ground layer fraction. Their early results demonstrate good promise, especially for scheduling adaptive optics instruments.

Finally, we mention recent work~\citep{lyman2020} by the Maunakea Weather Center (MKWC) which takes a macro approach to predict $\iqAtmo$.  The authors tap into large meteorological modeling models.  They start from the NCEP/Global Forecasting System (GFS) which outputs a 3D-grid analyses for standard operational meteorological fields: pressure, wind, temperature, precipitable water, and relative humidity.  Coupling these predictions with advanced analytics and decades of  MKAM DIMM seeing data, \cite{lyman2020} predict the free air contribution ($\iqAtmo$) to seeing on the mountain.  Their work is complementary to ours in that we take in our local sensor measurements to predict (and reduce) the effect of $\iqDome$ on $\iqMea$, while \cite{lyman2020} directly predict $\iqAtmo$. In the long term these two models can be combined to yield improved seeing estimates, forecasts, and decisions.

\section{Data} \label{sec:data}

\begin{figure*}
    \centering
    \includegraphics[width=\textwidth]{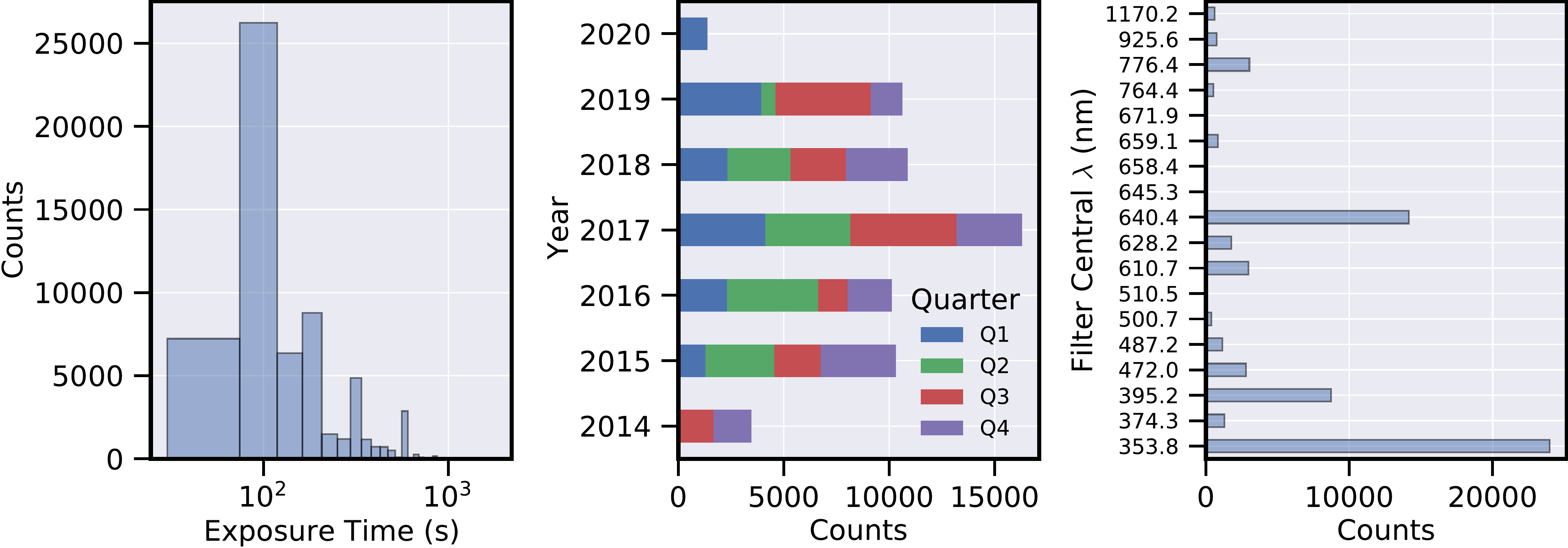}
    \caption{\textbf{Left:} Histogram of exposure times of the $\sim$60,000 samples/exposures used in this paper. \textbf{Middle:} Collected observations per year, broken down by quarter. \textbf{Right:} Same observations group by filter as represented by the filter central wavelength.}
    \label{fig:dataexploration}
\end{figure*}

In this section we discuss how we curated and prepared the data for use in our models. As mentioned already, our efforts began with  almost a decade's worth of sensor measurements archived at CFHT, together with IQ measured on the MegaCam exposures retrieved from the Canadian Astronomy Data Center (CADC) web services. 
At the start pertinent variables were spread across multiple data sets, sensor measurements were missing due to sensor failures,  data records contained errant values due to mis-calibrated data reduction pipelines. We therefore spent substantial effort cleaning the data.
In Section~\ref{sec:rawData} we discuss the various data sources that we collate to form our final data set. We then discuss our data cleaning and feature engineering procedures in Section~\ref{sec:data_cleaning} and Section~\ref{sec:feature_engineering}.


\subsection{Data sources}\label{sec:rawData}
Our first step in data collection was to build a data archive that contains one record per MegaCam exposure. In the remainder of this paper, we refer to each exposure and its associated sensor measurements interchangeably as a ``sample'', a ``record'', or an ``observation''. Each record contains three distinct types of predictive variables: 
\begin{enumerate}
    \item \textit{Observatory parameters}. These can be divided into operating (controllable) and non-operating (fixed) parameters. The former include the configurations of the twelve dome vents (open or closed), and the windscreen setting (degrees open).  These are examples of the variables that we can adjust the settings of in real time. The non-operating features include measurements of the telescope altitude and azimuth (which correspond to pointing of the astronomical object being observed) and the central wavelength of the observing filter.
    \item \textit{Environmental parameters}. These include exposure-averaged wind speed, wind direction, barometric pressure, and temperature values at various points both inside and outside the observatory. 
    \item \textit{Ancillary parameters}. Each exposure come with metadata. Relevant to our work are the date and time of the observation and the length of exposure. All predictive variables have been summarized in Table \ref{table:megaCamData}.
The median time of each exposure is $\sim$150 seconds, while the median time between two consecutive exposures made on the same night is $\sim$240 seconds\footnote{We emphasize that in this work we forego temporal dependencies and treat all exposures as independent. We provide the time between exposures for the sake of context.}. 
\end{enumerate}

In total, there are 160,341 observations, and 86 variables (\textit{including} image quality) are provided with each exposure. The records span February 2005 to March 2020. An overview of the data is provided in Table~\ref{tab:datasets_overview}, where we note the expanded number of features created using {\it feature engineering}, which we expound upon below.

\subsection{Data Cleaning}
\label{sec:data_cleaning}
We now list the data cleaning we performed.  In short these included the removal of data records corresponding to (i) non-sidereal targets, (ii) data records associated with too-short or too-long exposures, (iii) data records associated with IQ estimates deemed unrealistic, and (iv) data records containing missing or errant data values. 


\begin{enumerate}
    \item \textit{ Non-sidereal:} We remove moving, non-sidereal, targets.   The IQ measurements for these data records are not valid as the data pipeline that calculates IQ assumes sidereal observation.  Therefore these records are not appropriate for training.  We note that as part of the configuration data recorded along with each observation the astronomer specifies whether the observation is sidereal or not. Hence, these data records are easy to remove.
    \item \textit{Non-trustworthy IQ estimates:} We remove MegaCam exposures associated with IQ estimates of less than $0.15''$ or greater than $2''$.  Such  IQ numbers are deemed unrealistic.  It is believed that an IQ of $\sim 0.2''$ is the best possible at Maunakea.  Anything below this is deemed to result from an erroneous calculation when converting from the raw exposure data. On the other hand, IQ $>2''$ is too large for useful  science.
    \item \textit{Missing and errant measurements:} Not all sensor measurements are available at all times of the exposure.  We refer to these as "missing data". As is tabulated in the first row of Table~\ref{tab:datasets_overview}, prior to considering missing data, our cleaned data set (cleaned of non-sidereal and non-trustworthy) contains $120$ features ($86$ original + rest engineered features + 1 MPIQ, see Section~\ref{sec:feature_engineering}) and $160,341$ samples. Of these, just under $100,000$ samples \textit{do not} contain all measurements; we specify the fraction of missing measurements in the last column of Table~\ref{tab:datasets_overview}. We refer to the dataset with {\it all} samples as $\mathcal{D_{F_S,S_L}}$, i.e., $\mathcal{D}$ata set with a $\mathcal{S}$mall number of $\mathcal{F}$eatures, and a $\mathcal{L}$arge number of $\mathcal{S}$amples. By removing those samples that contain at least one missing feature, we obtain $\mathcal{D_{F_S,S_S}}$: $\mathcal{D}$ata set with a $\mathcal{S}$mall number of $\mathcal{F}$eatures, and a $\mathcal{S}$mall number of $\mathcal{S}$ samples. This latter dataset consists of 63,082 samples (second and third rows in Table~\ref{tab:datasets_overview}. 
    In this paper, we train our models on $\mathcal{D_{F_S,S_S}}$, since feed-forward neural networks cannot handle missing values without non-trivial modifications. In future work, we will use a variational autoencoder capable of imputing missing values \citep{vae_missingvalues} to enable us to leverage the larger dataset, $\mathcal{D_{F_L,S_L}}$: $\mathcal{D}$ata set with a $\mathcal{L}$arge number of $\mathcal{F}$eatures, and a $\mathcal{L}$arge number of $\mathcal{S}$amples.

\begin{table}
    \caption{Summary statistics of data sets described in Section \ref{sec:data_cleaning}. `\#Original Features' includes the MegaPrime Image Quality (MPIQ), while `\#Engineered Features' are additional hand-crafted ones added to enhance the predictive capability of our models (see Section \ref{sec:feature_engineering} for details). However for the remainder of this paper, we use `features' to refer to the union of original and engineered features less the MPIQ column: these are predictive, independent variables. Similarly, going forward MPIQ -- the dependent variable -- is referred to as the `target'.}
    \label{tab:datasets_overview}
    \centering
    \begin{tabular}{ccccc}
        \toprule
        \toprule
        \thead{Dataset \\ Identifier} & \#Samples & \thead{\#Original \\ Features} & \thead{\#Engineering \\ Features} & \thead{Percentage \\ Missing} \\ \midrule
        $\mathcal{D_{F_S,S_L}}$ & 160,341 & 86 & 34 & 62\% \\
        $\mathcal{D_{F_S,S_S}}$ & 63,082 & 86 & 34 & 0\% \\
        $\mathcal{D_{F_L,S_S}}$ & 63,082 & 86 & 1115 & 0\% \\
        $\mathcal{D_{F_L,S_L}}$ & 160,341 & 86 & 1115 & 62\% \\        
        \bottomrule
    \end{tabular}
\end{table}

\end{enumerate}

\begin{figure*}
   \centering
   \includegraphics[width=1.0\linewidth]{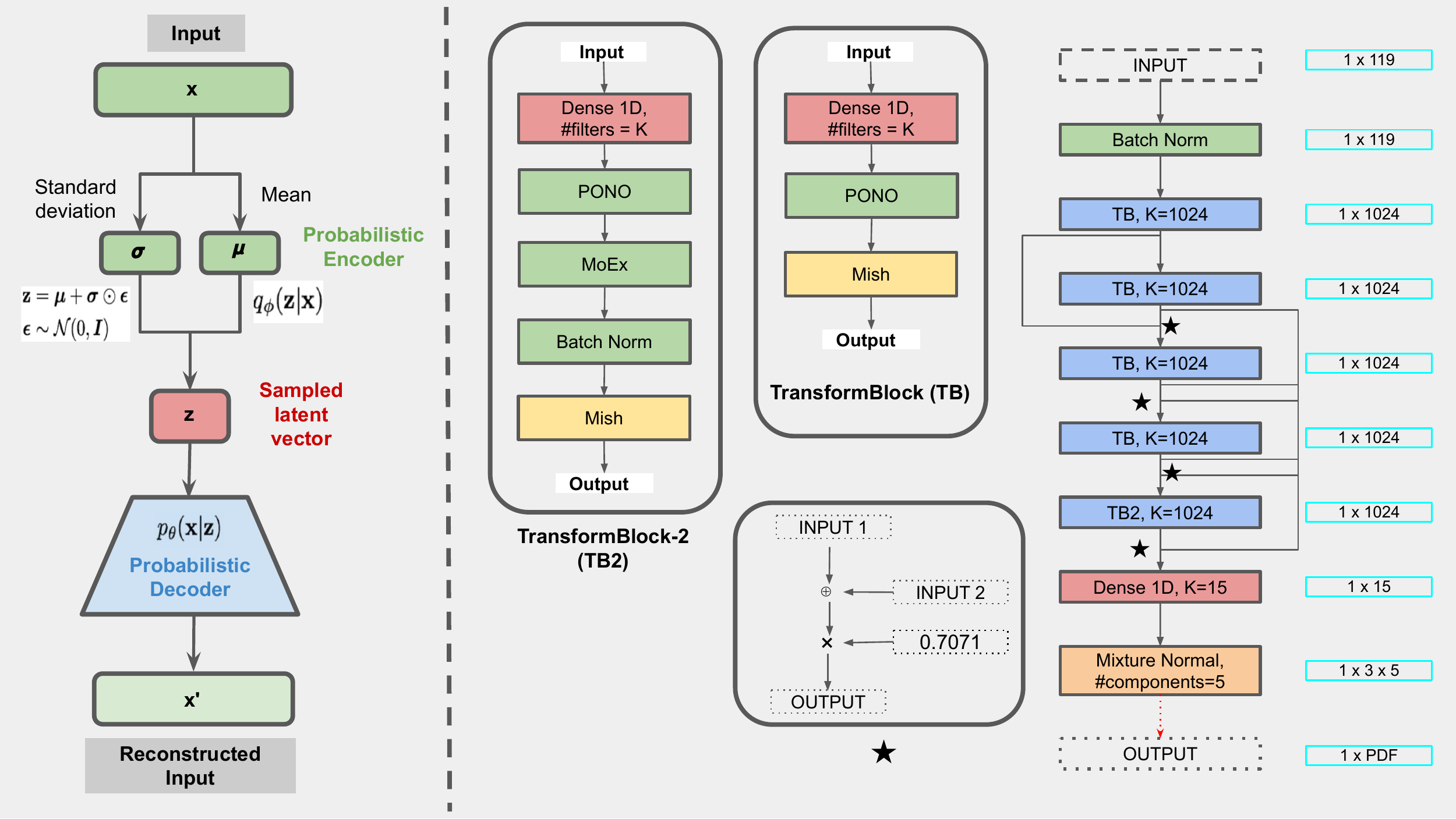}
   \caption{The architectures of the two networks used in this study. To the \textbf{left} of the dashed vertical line, we show the overview of a variational autoencoder, but note that the \textit{robust} VAE (i.e. RVAE) used in this work uses a special reconstruction loss (comparing \textbf{x} and \textbf{x'}) which is not depicted in this cartoon. On the \textbf{right} we show a dense feed-forward network with skip connections, mish activations \citep{mish_activation}, positional normalization \citep{positional_normalization_pono}, batch normalization \citep{batchnorm},  
   and momentum exchange \citep{moex} augmentation layers. This mixture density network (MDN) has 5 components. Near the right edge of the figure, we indicate in cyan colored rectangles the output shapes after an input sample has passed through each layer.}
   \label{fig:rvae_plus_mdn}
\end{figure*}

\subsection{Feature Engineering}
\label{sec:feature_engineering}
Feature engineering is the process of modifying existing features, using either domain expertise, statistical analysis, or intuition derived from scientific expertise.  The goal is to create predictive variables that are more easily understood by an ML algorithm. We now describe the feature engineering we performed. 

\begin{enumerate}
\item \textit{Optics IQ correction:} We remove the fixed, but wavelength dependent, contributions of the telescope optics to IQ, $\iqOpt$. These corrections are based on work by \cite{salmon2009cfht}, and range from $0.31''$ in the $i$-band to $0.53''$ in the $u$-band \citep{racine2018}; cf. second column of Table \ref{tab:per_filter_contrib}. 
After removing the contribution of optics, we are  left with a convolution of dome seeing and atmospheric seeing. This is because dome seeing, referred here to as ${\rm IQ}_{\rm Dome}$, is enmeshed with ${\rm IQ}_{\rm Measured}$ in a complicated way that does not lend itself to easy separation; the relationship between these is governed by Equation~(\ref{eqn:iq_optics_correction}), a rearranged version of Equation~(\ref{eqn:iq_measured}):

\begin{align}
&\iqAtmo^{5/3} + \iqDome^{5/3} = \iqMea^{5/3} - \iqOpt^{5/3}, \nonumber \\
&\rm{IQ}_{\rm Atmospheric}^{'} = \left(\iqMea^{5/3} - \iqOpt^{5/3}\right)^{3/5}
\label{eqn:iq_optics_correction}
\end{align}


\begin{table}
    \centering
    \caption{IQ$_{\rm Optics}$ for different bands, calculated according to the prescription of \citet{salmon2009cfht}. The average seeing across all bands is about $0.33''$, as noted in  Table 4 of \citet{salmon2009cfht}.}
    \label{tab:per_filter_contrib}
    \begin{tabular}{ccc}
        \toprule
        \toprule
        Filter & Central $\lambda$ (nm) & IQ$_{\rm Optics} ('')$ \\
        \midrule
        Ha.MP7605 &	645.3 &	0.284 \\
        HaOFF.MP7604 &	658.4 &	0.280 \\
        Ha.MP9603 &	659.1 &	0.280 \\
        HaOFF.MP9604 &	671.9 &	0.276 \\
        TiO.MP7701 &	777.7 &	0.260 \\
        CN.MP7803 &	812 &	0.260 \\
        u.MP9301 &	374.3 &	0.441 \\
        u.MP9302 &	353.8 &	0.459 \\
        CaHK.MP9303 &	395.2 &	0.424 \\
        g.MP9401 &	487.2 &	0.358 \\
        g.MP9402 &	472 &	0.368 \\
        OIII.MP7504 &	487.2 &	0.358 \\
        OIII.MP9501 &	487.2 &	0.358 \\
        OIIIOFF.MP9502 &	500.7 &	0.350 \\
        r.MP9601 &	628.2 &	0.290 \\
        r.MP9602 &	640.4 &	0.285 \\
        gri.MP9605 &	610.68 & 0.296 \\
        i.MP9701 &	777.6 &	0.261 \\
        i.MP9702 &	764.4 &	0.261 \\
        i.MP9703 &	776.4 &	0.261 \\
        z.MP9801 &	1170.2 & 0.397 \\
        z.MP9901 &	925.6 &	0.276 \\
        \bottomrule
    \end{tabular}
\end{table}

At the risk of being redundant with information presented towards the tail-end of Section \ref{sec:relatedWork}, we remind the readers that \cite{racine2018} and \cite{salmon2009cfht} estimate ${\rm IQ}_{\rm Atmospheric}$ to be in the range of $0.43''$ to $0.45''$, and ${\rm IQ}_{\rm Atmospheric}^{'}$ to be about $0.55''$. They demonstrate that opening all 12 vents completely allows one to reduce ${\rm IQ}_{\rm Atmospheric}^{'}$ to about $0.51''$, which leaves a residual median $0.20''$ $\left((0.51^{5/3} - 0.43^{5/3})^{3/5}\right)$ on the table, which is what we aim to capture in this paper. These numbers also agree with our own calculations, as described in Section \ref{sec:relatedWork} and visualized in the top two sub-figures of Figure \ref{fig:hist_preliminaries}. Our argument, introduced in Section \ref{sec:introduction} and expounded upon in Section \ref{sec:introduction}, is that for any given observation, there is an optimal set of vent configuration, somewhere between all-open and all-closed, that allows us to bite into this $0.20''$ residual ${\rm IQ}_{\rm Dome}$. 

\item \textit{Wavelength IQ correction:}
Each MegaCam exposure is taken using one of 22 band-pass filters.  The right-hand subfigure in Figure~\ref{fig:dataexploration} plots a histogram of  observations across  bands. The use of the filters results in a wavelength-dependent IQ variation.   To make IQ measurements consistent we scale IQ to a common wavelength of $500$nm. The formula for the scaling is provided in Equation~(\ref{eqn:iq_filter_plus_zenith_correction}), which we present in conjunction with a zenith angle correction, discussed next.

\item \textit{Zenith angle correction:} IQ is also affected by the amount of atmosphere through which the observation is made. The contribution of air mass is, to first degree, predictable, and can be removed together with the wavelength correction via~(\ref{eqn:iq_filter_plus_zenith_correction}), where $z$ is the zenith angle in degrees and $\lambda$ is the central wavelength of a given filter in nm.

\begin{align}
\iqCorr = \iqAtmoPrime \times \left(\frac{\lambda}{500}\right)^{1/5} \times \left(\cos{z}\right)^{3/5} \label{eqn:iq_filter_plus_zenith_correction}
\end{align}
    
\item \textit{Cyclic encoding of time-of-day and day-of-year:}
Every observation has an associated time stamp, indicating the beginning of image acquisition. Using this `timestamp' feature, we derive two time-features, the hour-of-day and the day-of-year.  These features better capture latent cyclical relationship between weather events and IQ. We represent each of these two features into a pair cyclical `sinusoidal' and `co-sinusoidal' component. For example, for the day-of-week feature values -- which can range from 0 to 6 -- we encode it as day-of-week-sine and day-of-week-cosine.  These can each respectively take on values from $\sin{\left(0\times180\degree/6\right)}$, to $\sin{\left(6\times180\degree/6\right)}$, and $\cos{\left(0\times180\degree/6\right)}$, to $\cos{\left(6\times180\degree/6\right)}$. In this way we replace the timestamp feature with four new, and more easily digestible features.

\item \textit{Cyclic encoding of azimuth:} Similar to the temporal information, we cyclically encode the telescope azimuth, splitting it into two features. We note that since the altitude of observation ranges from $0$ to $90$ degrees, and is not cyclical in nature, we leave that feature unmodified.
    
\item \textit{Temperature differences features:} 
As argued in our discussion, and evidenced by the prior work, temperature differences are the prime source of turbulence. In recognition of this key generative process, we engineer new temperature features that consist of the pairwise differences of existing temperature measurements.

We note that, given sufficient data, a deep enough neural network should be able to discover that temperature differences are important features.  We engineer in such features as, from our knowledge of physics, we understand temperature difference are important and providing them explicitly to the network eases the inference task faced by the network. In addition, unlike a neural network, the boosted-tree model that we use for comparative analysis is, by design, unable to create new features. The boosted-tree therefore benefits quite significantly from increased feature representation.

We implement two different flavors of engineering here. First, for every temperature feature in our two data sets of 160,341 and 63,082 samples, we subtract it from every other temperature feature.  We calculate the Spearman correlation of these newly generated features with the MPIQ values.  We then rank them by magnitude in descending order and pick the top three features. This increases our original 85 input features to 119, and this is how we get $\mathcal{D_{F_S,S_L}}$ and $\mathcal{D_{F_S,S_S}}$. For the second variation, we do not pick the top 3, but retain all the newly generated temperature-difference features.  This increases the number of features from 86 to 1115. This is how we get $\mathcal{D_{F_L,S_L}}$ and $\mathcal{D_{F_L,S_S}}$. This is summarized in Table~\ref{tab:datasets_overview}. As a reminder, in this work, we only use $\mathcal{D_{F_S,S_S}}$; empirical results showed that our neural networks' performance did not significantly improve by using $\mathcal{D_{F_L,S_S}}$.
\end{enumerate}

\section{Methodology}\label{sec:method}

The raw sensor data is a collection of time series and ultimately it would best to model the multiple sensors in their native data structure. In the analysis we perform in this paper, we compiled the sensor data into a large table to ease exploration, consisting of heterogeneous and categorical data. The heterogeneity is caused by the wide variety of sensors (wind speed, temperature, telescope pointing) each recorded in specific units. Categorical features emerged because certain measurements values were binned. For instance, due to the unreliability of wind speed measurement, we have binned these values -- wind speed below $5$ knots, $5$-$10$ knots, etc. Similarly, for simplicity, each of the twelve vents have been encoded into either completely open or completely closed. These characteristics induce a discontinuous feature space. Our training data set is thus tabular in nature. 
At hand with our curated data set, we are equipped to work on our two objectives: making accurate predictions of  MegaPrime Image Quality, and use our predictor to, on a per-sample basis, explore the importance of each feature on IQ. 



Decision tree-based models ~\citep{decision_trees} and their popular derivatives such as random forests~\citep{randomforests}, and gradient boosted trees \citep{gradient_boosted_decision_trees_gbdt} are well-matched to tabular data. and often are the best performers. Tree-based models select and combine features greedily to whittle down the list of pertinent features to include only the most predictive ones. Feature sparsity and missing data is naturally accommodated by tree models, they simply do not include feature cells containing such values in their splits. We show below our implementation of a variant of gradient boosted tree with uncertainty quantification, and feature exploration.

However tree-based models require the human process of feature engineering and are known (e.g. \cite{bengio2010decision}) to poorly generalize.
In contrast to tree-based models, deep neural networks (DNNs) are powerful feature pre-processors. Using back-propagation, they learn a fine-tuned hierarchical representation of data by mapping input features to the output label(s). This allows us to shift our focus from feature engineering to fine-tuning the architecture, designing better loss functions, and generally experimenting with the mechanics of our neural network. In reported comparison cases, DNNs yield improved performance with larger sized datasets \citep{airbnb}. As we will show, our neural network implementation, with the feature engineering steps described above, performs better than the alternative tree-based boosted model. We therefore deepen our analysis of the deep neural networks further: we quantify its probabilistic predictions, and we attempt to model the feature space.

\subsection{Probabilistic Predictions with a Mixture Density Network}
\label{sec:mdn}

Mixture density networks (MDNs) are composed of a neural network, the output of which are the parameters of a mixture model \citep{bishop_mdn}. They are of interest here because the relationship between the feature vectors $\mathbf{x}$ and target labels $\mathbf{y}$ can be thought of stochastic nature. Therefore, MDNs express the probability distribution parameters of MPIQ as a function of the input sensor features. In a one-dimensional mixture model, the overall probability density function (PDF) is a weighted sum of $M$ individual PDF $p_\theta^m(\mathbf{y}|\mathbf{x})$ parameterized by a neural network of parameters $\theta$ \footnote{In their initial form~\cite{bishop_mdn}, MDNs used a Gaussian mixture model (GMM). They can easily be generalized to other distributions.}:


\begin{equation*}
p_\theta(\mathbf{y}|\mathbf{x})=\sum_{m=1}^{M} \alpha_\theta^m(\mathbf{x}) p_\theta^m(\mathbf{y}|\mathbf{x}) \quad \textrm{with} \quad
\sum_{m=1}^{M} \alpha_\theta^m(\mathbf{x})=1.
\end{equation*}

Under the assumption of $N$ independent samples $\mathbf{x}$ from the features distribution, and the corresponding conditional samples $\mathbf{y}$ of MPIQ, we minimize over the negative log-likelihood of the density mixture to obtain the neural network weights:
\begin{equation}
\theta^* = \mathop{\mathrm{argmin}}_{\theta} -\frac{1}{N}\sum_{n=1}^{N} \log p_\theta(\mathbf{y}_n|\mathbf{x}_n)
\label{eqn:mdn_loss}
\end{equation}

To train the neural network we take as the network input the data record of sensor readings, observatory operating conditions, etc.  The network outputs are the (per-vector) mixture model parameters modeling the MPIQ conditional distribution, implicitly parameterized by the neural network. 
In our experiments, we use $\beta$ distributions, and set $M=5$ as it gave sensible results.

\subsection{Complementary Predictions and Interpretation with Gradient Boosted Decision Trees}
\label{sec:gbdt}

We complement the MDN IQ predictions by another algorithm to secure our results: a gradient boosted decision tree (GBDT) to predict IQ from the sensor data. This is in fact one of the main reason so much of feature engineering was performed on the sensor data.
A set of consecutive decision trees is fit where each successive model is fit to obtain less overall residuals than the previous ones by weighting more the larger sample residuals. Once converged, we can obtain a final predictions from the trained boosted tree as the weighted mean of all models. The optimization can be performed with gradient descent. Several implementations of this popular algorithm exist and we selected the \textsc{catboost}\footnote{\url{https://catboost.ai/}} one for our modelling, with a loss optimized for both the mean and the variance of the predictions. We first perform nested cross-validation as for the MDN, obtain the best hyper-parameters. We then train ten  GBDT models with the same hyper-parameters with a stochastic optimization, each with a different initialization of the model parameters. Aleatoric and epistemic uncertainties are estimated with a simple ensemble averaging method \citep{malinin2021uncertainty} of each model predictions. We show our results in Figure \ref{fig:MDN_moneyplot_catboost}, and discuss them in detail in Section \ref{sec:results}.


\subsection{Density Estimation with a Robust Variational Autoencoder}
\label{sec:rvae}



An autoencoder~\citep{hinton2006reducing_autoencoderintroduction} is a neural network that takes high dimensional input data, encode it into a common efficient representation (usually of lower dimension), and then recreates a full-dimensional approximation at the other end. Through its mapping of the input to a smaller or sparser, and more manageable, but information-dense, latent vector, the autoencoder finds application in many areas including  compression, filtering, and accelerated search. A variational encoder \citep{sgvb, vae_intro_2013, vae2} is a probabilistic version of the autoencoder.  Rather than mapping the input data to a specific (fixed) approximating vector, it maps the input data to the parameters of a probability distribution, e.g., the mean and variance of a Gaussian. VAEs produce a latent variable $\mathbf{z}$, useful for data generation. We refer to the \textit{prior} for this latent variable by $p(\mathbf{z})$, the observed variable (\textit{input}) by $\mathbf{x}$, and its conditional distribution ({\it likelihood}) as $p_{\theta}(\mathbf{x} | \mathbf{z})$. In a Bayesian framework, the relationship between the \textit{input} and the \textit{latent variable} can be fully defined by the \textit{prior}, the \textit{likelihood}, and the \textit{marginal} $p_{\theta}(\mathbf{x})$ as:
\begin{align}
p_{\theta}(\mathbf{x})=\int_{\mathcal{Z}} p_{\theta}(\mathbf{x} | \mathbf{z}) p(\mathbf{z}) d \mathbf{z}. \label{eq:fullDist}
\end{align}


It is not not easy to solve Equation~(\ref{eq:fullDist}) as the integration across $\mathbf{z}$ is most often computationally intractable, especially in high dimensions.  To tackle this, variational inference \citep[e.g.][]{Jordan1999} is used to introduce an approximation $q_{\phi}(\mathbf{z}|\mathbf{x})$ to the posterior $p_{\theta}(\mathbf{z}|\mathbf{x})$. In addition to maximizing the probability of generating real data, the goal now is also to minimize the difference between the real $p_{\theta}(\mathbf{z}|\mathbf{x})$ and estimated $q_{\phi}(\mathbf{z}|\mathbf{x})$ {\it posteriors}. We state without proof (see \cite{kingma2019introduction} for detailed derivation):
\begin{align}
    &-\log p_{\theta}(\mathbf{x}) + D_{\rm KL}\left(q_{\phi}(\mathbf{z}|\mathbf{x}\right) \| p_{\theta}(\mathbf{z}|\mathbf{x})) \nonumber \\ 
    &= -\mathbb{E}_{\mathbf{z} \sim q_{\phi}(\mathbf{z}|\mathbf{x})} [ \log p_\theta(\mathbf{x} | \mathbf{z})] + D_{\rm KL}\left(q_\phi(\mathbf{z}|\mathbf{x}) \| p_\theta(\mathbf{z})\right) \label{eq:VAEeq}
\end{align}

The approximating distribution $q$ is chosen to make the right-hand-side of Equation~(\ref{eq:VAEeq}) tractable and differentiable. Taking the right-hand-side as the objective to simultaneously minimize both the divergence term on the left-hand-side (making $q$ a good approximation to $p$) and $-\log p(x)$. This is exactly the loss function that we want to minimize via backpropagation:
\begin{align}
L_{\mathrm{VAE}}(\theta, \phi; \mathbf{x}) & = -\mathbb{E}_{\mathbf{z} \sim q_{\phi}(\mathbf{z} | \mathbf{x})} \left[\log(p_{\theta}(\mathbf{x} | \mathbf{z}))\right] \nonumber \\
&+D_{\mathrm{KL}}\left(q_{\phi}(\mathbf{z} | \mathbf{x}) \| p_{\theta}(\mathbf{z})\right) \nonumber \\
-L_\mathrm{ELBO} &= L_{\mathrm{REC}}(\theta, \phi; \mathbf{x}) + D_{\mathrm{KL}}(\theta, \phi; \mathbf{x}) \label{eq:VAEeq2}
\end{align}
where $\theta^{*}, \phi^{*} = \mathop{\mathrm{argmin}}_{\theta, \phi} L_{\mathrm{VAE}}$. Since KL-divergence is non-negative, Equation (\ref{eq:VAEeq2}) above can be thought of as the lower bound of $p_{\theta}(\mathbf{x})$, and is the loss function to minimize. It is commonly called the ELBO, short for {\it {evidence based lower bound}}. $L_{\rm{REC}}$ minimizes the difference between input and encoded samples, while $D_{\rm{KL}}$ acts as a regularizer \citep{KLregularizer}.

The typical choice for $q$ (that we also make) is an isotropic conditionally Gaussian distribution whose mean and (diagonal) covariance depend on $\mathbf{x}$.  The result is that the divergence term has a closed-form expression where the mean and variance are learned, for example by using a neural network.  To be able to backpropagate through the first term (the expectation) in the loss function, a reparameterization  is introduced.  For each sample from $\mathbf{x}$ take one sample of $\mathbf{z}$ from the conditionally Gaussian distribution $q_\phi(\mathbf{z}|\mathbf{x})$.  Without loss of generality we can generate an isotropic Gaussian $\mathbf{z}$ by taking a Gaussian source $\boldsymbol{\epsilon} \sim \mathcal{N}(0, \mathrm{\mathbf{I}})$ shifting it by the $\mathbf{x}$-dependent mean $\mathbf{\mu}$ and scaling by the standard deviation $\boldsymbol{\sigma}$ to get $\mathbf{z} = \boldsymbol{\mu}+\boldsymbol{\sigma}\odot \boldsymbol{\epsilon}$, where $\odot$ is the element-wise product.  Approximating the first (expectation) term in the objective with a single term using this value for $\mathbf{z}$ allows one to backpropagate gradients through this objective.  Note that, in the terminology of autoencoders, the $q$ and $p$ functions play the respective roles of encoder and decoder; $q_{\phi}(\mathbf{z} | \mathbf{x})$ generate the latent representation from a data point and  $p_{\theta}(\mathbf{x} | \mathbf{z})$ defines a generative model.





The VAE described so far, which we refer to as the `vanilla' VAE, is not the optimal model for our purposes.
This is because our data set $\mathcal{D_{F_S,S_S}}$ 
can contain outliers caused mostly by sensor failures, and sometimes by faulty data processing pipelines. The ELBO for `vanilla' VAE contains a log-likelihood term (first term in RHS of Equation \ref{eq:VAEeq}) that will give high values for low-probability samples \citep{akrami2019robustvae}. 

We state without proof (see \citet{akrami2019robustvae, akrami2020robustvaetabular} for details) that $L_{\rm {REC}}$ for a single sample can be re-written as:
\begin{align}
L_{\rm {REC}}^{(i)} = \mathbb{E}_{\mathbf{z} \sim q_{\phi}(\mathbf{z} | \mathbf{x}^{(i)})}\left[D_{\rm {KL}} \left(\hat{p}(\mathbf{X}) | p_\theta(\mathbf{X}|\mathbf{Z})\right)\right], \label{eq:lrec_mod}
\end{align}
where $\hat{p}(\mathbf{X}) = \frac{1}{N} \sum_{i=1}^{N} \delta (\mathbf{X}- \mathbf{x}^{(i)})$ is the empirical distribution of the input matrix $\mathbf{X}$, and $N$ is the number of samples in a mini-batch. We then substitute the KL-divergence with the $\beta$-cross entropy \citep{ghosh2016robust_variationalbayesianinference} which is considerably more immune to outliers:
\begin{align}
    L_{\rm{REC}, \beta}^{(i)} = \mathbb{E}_{\mathbf{Z} \sim q_{\phi}(\mathbf{Z} | \mathbf{x}^{(i)})}\left[H_{\beta}(\hat{p}(\mathbf{X}),p_{\theta}(\mathbf{X} | \mathbf{Z}))\right], \label{eq:lrec_beta}
\end{align}
where the $\beta$ cross-entropy is given by \cite{formula_betacrossentropy,adaptiveoptics0,rvae_orig}:
\begin{align}
    &H_{\beta}(\hat{p}(\mathbf{X}),p_{\theta}(\mathbf{X} | \mathbf{Z})) = \nonumber \\
    &-\frac{\beta+1}{\beta}\!\int\!\hat{p}(\mathbf{X})\left(p_{\theta}(\mathbf{X} | \mathbf{Z})^{\beta}-1\right) d \mathbf{X}+\int\!p_{\theta}(\mathbf{X} | \mathbf{Z})^{\beta+1} d \mathbf{X} \label{eq:ce_beta_def}
\end{align}
Here $\beta$ is a constant close to 0. This makes the total loss function for a given sample:
\begin{align}
L_{\beta}\left(\theta, \phi ; \mathbf{x}^{(i)}\right) =& \mathbb{E}_{\mathbf{Z} \sim q_{\phi}(\mathbf{Z} | \mathbf{x}^{(i)})}\left[H_{\beta}(\hat{p}(\mathbf{X}),p_{\theta}(\mathbf{X} | \mathbf{Z}))\right] \nonumber \\
&+ D_{K L}\left(q_{\phi}\left(\mathbf{Z} | \mathbf{x}^{(i)}\right) \| p_{\theta}(\mathbf{Z})\right) \label{eq:loss_beta_prelim}
\end{align}

To draw from the continuous $\mathbf{Z}$, we use an empirical estimate of the expectation, and convert the above into a form of the Stochastic Gradient Variational Bayes (SGVB) cost \citep{sgvb} with a single sample $\mathbf{z}^{(j=1)}$ from $\mathbf{Z}$.
Next, for each sample we calculate $H_{\beta}(\hat{p}(\mathbf{X}),p_{\theta}(\mathbf{X} | \mathbf{z}^{(1)}))$ when $\mathbf{x}^{(i)} \in [0,1]$. We substitute $\hat{p}(\mathbf{X})=\delta\left(\mathbf{X}-\mathbf{x}^{(i)}\right)$ and model $p_{\theta}(\mathbf{X} | \mathbf{z}^{(1)})$ with a mixture of Beta distributions with weight vector $\mathbf{\omega}$. That is,
\begin{align}
    p_{\theta}(\mathbf{X} | \mathbf{z}^{(1)}) = \sum_{k=1}^{k} \omega_k (\mathbf{X}^{p_k-1})(1-\mathbf{X})^{q_k-1} \times \frac{\Gamma(p_k+q_k)}{\Gamma(p_k)\Gamma(q_k)} \label{eq:pxgivenz_mixturedensity}
\end{align}
Using Equations (\ref{eq:ce_beta_def}) and (\ref{eq:pxgivenz_mixturedensity}), we obtain:
\begin{align}
    &H_{\beta}(\delta(\mathbf{X}-\mathbf{x}^{(i)}),p_{\theta}(\mathbf{X} | \mathbf{Z})) = \nonumber \\
    &-\frac{\beta+1}{\beta} \left(\sum_{d=1}^{D}\left(\sum_{k=1}^{K} \omega_k (\mathbf{x}_{d}^{(i) \cdot p_k-1})(1-\mathbf{x}_{d}^{(i)})^{q_k-1} \Lambda_{d,k}\right)\right) \nonumber \\
    &+\sum_{d=1}^{D} \sum_{k=1}^{K}\frac{\left((p_{d,k}-1)^{1+\beta}+1\right) \left((q_{d,k}-1)^{1+\beta}+1\right)}{(p_{d,k}-1)^{1+\beta}+(q_{d,k}-1)^{1+\beta}+2} \label{eq:ce_final}
\end{align}
where $\Lambda_{d,k} = \frac{\Gamma(p_{d,k}) + \Gamma(q_{d,k})}{\Gamma(p_{d,k})\Gamma(q_{d,k})}$, $D$ is the number of dimensions in a single sample, and $K$ is the number of components in the mixture. Equations (\ref{eq:loss_beta_prelim}) and (\ref{eq:ce_final}) together give us the total loss across all $N$ samples in a given mini-batch:
\begin{align}
L_{\beta}\left(\theta, \phi ; \mathbf{X}\right) = &\frac{1}{N}\sum_{i=1}^{N} \left[H_{\beta}^{(i)}(\hat{p}(\mathbf{X}),p_{\theta}(\mathbf{X} | \mathbf{Z})) \right. \nonumber \\
& \left. + D_{K L}\left(q_{\phi}\left(\mathbf{z}^{(1)} | \mathbf{x}^{(i)}\right) \| p_{\theta}(\mathbf{z}^{(1)})\right)\right], \label{eq:loss_beta_final}
\end{align}
where the superscript $^{(1)}$ implies a single draw from \textbf{z} from \textbf{Z}.

The final, robust variational autoencoder architecture is denoted in the left of Figure \tcr{\ref{fig:rvae_plus_mdn}}.

\subsection{Uncertainty Quantification}
\label{sec:uncertainty_quantification}

Our predictions will be safer for decision making if for each input vector, in addition to the prediction of IQ, we also predict the degree of (un)certainty. This is especially true since we aim to toggle the twelve vents based on our predictions, which is an expensive manoeuvre -- a configuration of vents that ends up increasing observed IQ as opposed to decreasing it would require re-observation of the target, when CFHT is already oversubscribed by a factor of $\sim 3$. For this reason, we predict a probability density function of MPIQ for every input sample, as described in Section \ref{sec:mdn}.

Higher error (corresponding to lower model belief or confidence in the estimate) can result from absence of predictive features, error or failure in important sensors, or an input vector that value has drifted from the training distribution. We decompose the sources of predictive uncertainties into two distinct categories: {\it aleatoric} and {\it epistemic}.
Aleatoric uncertainty captures the uncertainty inherent to the data generating process. To analogize using an everyday object, this is the entropy associated with an independent toss of a fair coin. Epistemic uncertainty, on the other hand captures the uncertainty associated with improper model-fitting. In contrast to its aleatoric counterpart, given a sufficiently large data set epistemic uncertainty can theoretically be reduced to zero\footnote{The word ``aleatoric'' derives from the Latin ``aleator'' which means ``dice player''. The word ``epistemic'' derives from the Greek ``episteme'' meaning ``knowledge''~\citep{gal_thesis}.}. Aleatoric uncertainty is thus sometimes referred to as {\it irreducible} uncertainty, while epistemic as the {\it reducible} uncertainty. High aleatoric uncertainty can be indicative of noisy measurements or missing informative features, while high epistemic uncertainty for a prediction could be a pointer to the outlier status of the associated input vector.

The architecture of the MDN (Section \ref{sec:mdn}) allows us to predict a PDF of MPIQ for each sample. For each sample and mixture model component, let $\mu_m$, $(\sigma_m)^2$, and $\alpha_m$ respectively denote the mean, variance, and normalized weight (weights for all mixture model components must sum to 1) in the mixture model. We obtain the predicted IQ value as the weighted mean of the individual means: 
\begin{equation}
\mu = \sum_{m=1}^M \alpha_m \mu_m \label{eq.calcIQpredict}
\end{equation}
Aleatoric uncertainty is the weighted average of the mixture model variances, calculated as \citep{choi2018uncertainty}:
\begin{equation}
\sigma_{\mathrm{al}}^2 =  \sum_{m=1}^{M} \alpha_m\sigma_m^2, \label{eq.calcAleaUncert}
\end{equation}
while epistemic uncertainty is the weighed variance of the mixture model means: 
\begin{equation}
\sigma_{\mathrm{epis}}^2 = \sum_{m=1}^{M} \alpha_m \mu_m^{2} - \mu^{2} \label{eq.calcEpisUncert}
\end{equation}
The total uncertainty is computed by adding Equations ~(\ref{eq.calcAleaUncert}) and~(\ref{eq.calcEpisUncert}) in quadrature.
\subsection{Probability Calibration}\label{sec:DL_probability_calibration}

In Section~\ref{sec:uncertainty_quantification} we describe how to derive both aleatoric and epistemic errors. While these variance estimates yield a second-order statistical characterization of the distribution of output errors, they can at times mislead the practitioner into a false sense of overconfidence
~\citep{lakshminarayanan_probability_calibration0, probability_calibration1, crude_probability_calibration}.  

It therefore becomes imperative to {\it calibrate} our uncertainty estimates to more closely match the true distribution of errors. In other words, we ensure that 68\% confidence intervals for MPIQ predictions (derived from the epistemic uncertainty) contain the true MPIQ values $\sim 68\%$ of times. The confidence interval is the range about the point prediction of IQ in which we expect, to some degree of confidence, the true IQ value will lie.  For example, if our error were conditionally Gaussian, centered on our point prediction, then we would expect that with about $68.2$\% probability the true IQ value would lie within $\pm 1\sigma$ of our IQ prediction where $\sigma$ is the standard deviation of the Gaussian. To accomplish this we reserve some of our data which we use to estimate the distribution of errors -- this is the validation set. Using the inverse cumulative distribution function of this estimated distribution, scaled by the predicted standard deviation and shifted by the predicted mean, allows us to obtain a calibrated estimate of the output realization corresponding to any particular percentile of the distribution. The specific approach we use is the CRUDE method~\cite{crude_probability_calibration}.

However, calibrating the error estimates is not the only thing we care about if, through the calibration process we loose substantial accuracy. For instance, one can increase predicted uncertainties to arbitrarily high values to obtain a perfectly calibrated model; however, this would make these predictions useless for practically any downstream task. Therefore, CRUDE not only calibrates our post-processed predictions, but also ensures that they are  {\it sharp}~\citep{measuring_calibration_in_deep_learning}. Sharpness refers to the concentration of the predictions, akin to the inverse of the posterior error variance.  The more peaked (the sharper) the predictions are, the better, provided the sharpness does not come at the expense of calibration.

\subsection{Performance Metrics}\label{sec:DL_metrics}
For each input sample \textbf{x} we derive the predicted IQ, the aleatoric uncertainty, and the epistemic uncertainty, respectively, $\mu$, $\sigma_a$, and $\sigma_e$, cf.~Equations (\ref{eq.calcIQpredict}), (\ref{eq.calcAleaUncert}), (\ref{eq.calcEpisUncert}). In Section \ref{sec:metrics_det} we compare the median of predicted IQ values against their ground truth values. In Section \ref{sec:metrics_prob} we evaluate the quality of the predicted PDF.

\subsubsection{Metrics for Deterministic Predictions}\label{sec:metrics_det}
We present three measures to quantify the quality of the IQ prediction, Root-mean-square error (RMSE), mean absolute error (MAE), and bias error (BE).  Respectively, these three measures are defined as
\begin{align*}
        \mathrm{RMSE} & = \sqrt{\frac{1}{N} \sum_{i=1}^{N}\left(\mu_i-y_i\right)^{2}},\\  \mathrm{MAE} & = \frac{1}{N} \sum_{i=1}^{N}\left|\mu_i-y_i\right|,\\
        \mathrm{BE} & = \frac{1}{N} \sum_{i=1}^{N} \left(\mu_i-y_i\right).
    \end{align*}
In the above definitions, $y_i$ and $\mu_i$ are the true and predicted IQ values corresponding to an input sample and $N$ is the number of samples.

\subsubsection{Metrics for Probabilistic Predictions}\label{sec:metrics_prob}


As discussed, for each sample our model yields a prediction tuple $\{\mu, \sigma_a, \sigma_e\}$.  We further use $\sigma$ to denote total uncertainty where $\sigma^2 = \sigma_{a}^2 + \sigma_{e}^2$.  Considering $68^{\rm th}$ percentile (``one-sigma'') confidence intervals, the lower and upper bounds of the interval are $L_\alpha = \mu - \sigma_e$ and $U_\alpha = \mu+ \sigma_e$ where, for this example of a one-sigma confidence interval, $\rm{CI} = 0.682$ and $\alpha = 1-\rm{CI} = 1 - 0.682 = 0.318$. The parameter $\alpha$ is the fraction of time the model predicts the true IQ will fall outside the confidence interval. We denote by $l_\alpha$ and $u_\alpha$ the cumulative distribution function of the (assumed Gaussian) PDF respectively evaluated at $L_\alpha$ and $U_\alpha$, i.e., $l_\alpha = 0.5 - \rm{CI}/2 = 0.159$, $u_\alpha = 0.5 + \rm{CI}/2 = 0.841$. 


We are now ready to introduce our two measures of the quality of our probabilistic predictor: average coverage area (ACE) and interval sharpness (IS).
Given $N$ predictions, let the true IQ for one sample be denoted $y$.  We also define an indicator function $\mathds{1}_\alpha$ that evaluates to 1 if the true IQ of a sample falls within the corresponding predicted confidence intervals, and zero elsewhere:
\begin{align*}
        \mathds{1}_\alpha=\left\{\begin{array}{lll}
1 & {\rm if} &  y \in\left[L_{\alpha}, U_{\alpha}\right] \\
0 & {\rm else}
\end{array}\right. .
    \end{align*}
The average coverage estimator is defined as for all samples:
\begin{align*}
        \mathrm{ACE}_{\alpha}=\frac{1}{N} \sum_{i=1}^{N} \mathds{1}_\alpha^{i} - (1-\alpha)
\end{align*}
and is a measure of the how well the confidence interval captures the realized distribution of predictions.  A value of zero tells us that exactly a fraction $1-\alpha$ of the predicted confidence intervals encapsulate the respective true IQs.  Generally if $\mathrm{ACE}_\alpha$ is small in magnitude then the prediction interval is well matched to the realized distribution of predictions.

    
While the average coverage area gives us a sense of the match between the predicted and realized distributions, it doesn't give us a sense of the concentration of the error.  By letting $\alpha \rightarrow 0$ all data points will fall in the bounds and so $\mathrm{ACE}_\alpha \rightarrow 0$ too.  Therefore we need a second measure of probabilistic prediction.  We use interval sharpness/interval score (IS) as this second measure \citep{gneiting_metrics, bracher_metrics}. Interval sharpness for a single sample is defined as:
\begin{align*}
    \mathrm{IS}_{\alpha}=\left\{\begin{array}{lll}
 \alpha( U_{\alpha}-L_{\alpha}) + 2\left[L_{\alpha}-y\right] \ \ {\rm if} \ \ y<L_{\alpha}, \\
 \alpha( U_{\alpha}-L_{\alpha}) \hspace{4.5em} {\rm if} \ \ L_{\alpha} \leq y \leq U_{\alpha}, \\
 \alpha( U_{\alpha}-L_{\alpha}) +2\left[y-U_{\alpha}\right] \, \ {\rm if} \ \ y>U_{\alpha}, 
\end{array}
\right.
\end{align*}
We normalize this against similar values for all samples, such that the final value lies between 0 and 1:
\begin{align}
    \mathrm{IS}_{\alpha, \rm{norm}} = \frac{\mathrm{IS}_\alpha - \min{(\mathrm{\mathbf{IS}}_\alpha)}}{\max{(\mathrm{\mathbf{IS}}_\alpha)} - \min{(\mathrm{\mathbf{IS}}_\alpha)}} \nonumber
\end{align}
and finally average the normalized values across the samples in the test set:
\begin{equation}
\overline{\mathrm{IS}}_\alpha = \frac{1}{N} \sum_{i=1}^N \mathrm{IS}_{\alpha, \rm{norm}}^i. \label{eq.defIntSharp}
\end{equation}
To understand Equation~(\ref{eq.defIntSharp}) we note first that $0 \leq \overline{\mathrm{IS}}_\alpha \leq1$ and higher sharpness (less positive) corresponds to more concentration and therefore more useful predictions.  The first term, $\alpha( u_{\alpha}-l_{\alpha})$, is a constant, parameterized by $\alpha$. In our experiments we set $\alpha = 0.318$ corresponding to $\pm 1$ standard deviation. Then a smaller variance will lead to a narrower confidence interval and a smaller $\mathrm{IS}_\alpha$ {\it if} the sample falls within the confidence interval.  The sharpness is decreased ($\mathrm{IS}_\alpha$ increases) if the prediction $y$ falls outside of the confidence interval, and the penalty applied is proportional to the distance between the ground truth value and the nearest interval limit. 
Generally a $\mathrm{IS}_\alpha$ small in magnitude means the estimates both fall in the confidence interval {\it and} the confidence interval is narrow.

We calculate ACE and IS for all three uncertainties -- aleatoric, epistemic, and total.

\subsection{Feature Ranking}\label{sec:featureRank}

One of our goals in this work is to understand the physical mechanisms that yield high and low IQ values so that, in the future, we can actuate the observatory to improve the realized IQ.  To accomplish this we need to understand the insights that the ML models decision making processes reveal. To this end, we utilize the methods of integrated Hessians and Shapley values \citep{explaining_explanations_hessians, gilda_mirkwood, gilda_mirkwood_software} for the MDN model. 
We use an implementation provided by the \texttt{pathexplainer} software package which compute feature attributions (or importances). The attributions plot ranks the 119 input features, guiding us on how important each feature is, relative to all other ones, in explaining the predicted MPIQ. These enables us to understand the model's decision making process, and to ascertain that the features deemed important by the model make sense physically.

\subsection{Putting It All Together}\label{sec:putting_it_all_together}

{\bf Training and test sets:} For both the MDN (Section \ref{sec:mdn}) and the RVAE (Section \ref{sec:rvae}) we partition $\mathcal{D_{F_S,S_S}}$ into two unequally-sized subsets -- a {\it training super-set} containing 90\% of the samples and a {\it test set} containing the rest. We are following a nested cross-validation scenario. We partition the data sets carefully, to ensure that the distribution of MPIQ values in both the test and training sets reflect the distribution in the original data set. To accomplish this we sort the samples by MPIQ values and, starting from the lowest MPIQ value allocate each sample in a round-robin fashion to one of ten buckets generated. We then iterate this process for the training super-set -- again producing a 90-10 split -- to respectively produce the final {\it training set} and the {\it validation set}. We train the models on the training set and record its predictions on the validation and test sets. The validation set guards against over-fitting -- we want our models to learn patterns from the training set, but not to the extent where they fail to generalize to unseen samples. Before making predictions on the test set, we revert the weights of both the MDN and RVAE models to their respective epochs where their respective losses on the validation data set were minimal, as shown in Figure \ref{fig:mdn_training_curve} for the MDN. As a quick reminder, a `prediction' for the MDN is a three-tuple consisting of mean $\mu$, aleatoric uncertainty $\sigma_a$, and epistemic uncertainty $\sigma_e$ for the MPIQ, whereas for the RVAE it is the reconstructed input sample.

{\bf Learning rate and optimizer:} We use a cyclical learning rate scheduler to vary the learning rate from an initial high to a final low value, in multiple cycles; this has been shown to result in a considerably better convergence than using step-wise or constant learning rate schedules \citep{cyclical_learning_rate}. To determine these limits for the MDN and the RVAE, we pick arbitrarily high ($10^{-1}$) and low ($10^{-7}$) limits, exponentially increase the learning rate from the latter to the former in a mere 20 epochs, and evaluate the behavior of the respective loss functions. For the MDN, we determine that at $10^{-3}$ and $10^{-6}$, the loss begins to plateau, as can be seen from Figure \ref{fig:mdn_find_lr}. We thus pick these as the higher and lower limits, respectively, and indicate them by dashed vertical lines. Similarly, from Figure \ref{fig:vae_find_lr} we can see that these limits for the RVAE are $10^{-3}$ and $10^{-5}$. We use the Yogi optimizer \citep{yogi_optimizer} for stochastic gradient descent; this optimizer is an improvement over the commonly used Adam \citep{adam_optimizer}, and we find experimentally that it provides faster convergence. We wrap this optimizer in the Stochastic Weight Averaging optimizer \citep{swa_stochastic_weight_averaging} -- accessible via the \texttt{TensorFlow Addons} library\footnote{\url{https://github.com/tensorflow/addons}} -- and average the model weights every 20 epochs, to overlap with the length of a training cycle. The batch size when using both models is 128.

{\bf Feature normalization and data augmentation:} Finally, we apply strong feature normalization and data augmentation to regularize against over-fitting. Specifically, we use Positional Normalization \citep[PONO]{positional_normalization_pono} layers to capture both the first and second moments of latent feature vectors, and use Momentum Exchange \citep[MoEx]{moex} to mix the moments of one input sample with that of another, to encourage our models to draw out training signal from the moments as well as from the normalized features. In each mini-batch of 128 samples, every feature vector for every sample is added with the feature vector for a randomly picked sample; the probability that this happens is set to 0.5 - this is, half the times, there is no mixing. In case of mixing, the weight assigned to the original sample is picked from a $\beta$ distribution with both concentration parameters set to 100, while the weight of the randomly picked sample is the difference of this from 1 (so that both weights sum to unity). The same random ordering of samples and the same weights are carried over to the model outputs as well (MPIQ for the MDN, the reconstructed input for the RVAE). This augmentation scheme has shown to produce state-of-the-art results, and our own experiments confirm excellent performance. This can be seen in Figure \ref{fig:mdn_training_curve}, where we plot the training and validation losses for one of ten folds; the training loss is significantly higher than the validation loss for a large part of the training process. We insert a PONO layer after each Dense layer in the MDN, and after the penultimate encoding layer in the RVAE. The MoEx layers are inserted before the ultimate Dense layer in the MDN, and the ultimate layer in the RVAE. Each PONO layer is followed by a Group Normalization layer \citep[GN]{group_normalization} with a channel size of 16 (see the MDN in Figure \ref{fig:rvae_plus_mdn}), except when a MoEx layer directly follows the PONO layer, where the former is followed by a Batch Normalization layer \citep{batchnorm}.

{\bf Calibration:} For the MDN, we implement additional steps to calibrate the predicted MPIQ PDFs. We treat each of the 10 training sets (these are obtained after splitting the respective training {\it super-sets} into training and validation sets, as explained at the beginning of this section) as a training {\it super-set}, and the associated validation set as the test set. In other words, we sub-divide the training set into 10 training and validation sets , train on the new training data sets and use the new validation sets as guardrails against over-fitting, and predict MPIQ on the new test sets. After repeating this process a total of 10 times, we now have predictions for the mean and both uncertainties for all samples in the original training set. Finally, we {\it calibrate} our model's predictions on the original test set by using the predictions on the original training set, by following the method described in \cite{crude_probability_calibration}. This is the post-processing step discussed in Section~\ref{sec:DL_probability_calibration}. We repeat this entire process a total of 10 times to cover all samples in $\mathcal{D_{F_S,S_S}}$. We illustrate this workflow in Figure~\ref{fig:cfht_mdn_overview} in Appendix~\ref{sec.workflowFigs}, where in the interest of saving space we show only 3 splits instead of 10.

{\bf RVAE tuning:} For the RVAE, there are a couple of additional considerations. For one, we adopt an annealing methodology to handle the problem of vanishing KL-divergence \citep{cyclical_wkl_annealing}. It is known that the KL-divergence loss term in Equation \ref{eq:VAEeq2} very quickly collapses to 0 if both L$_{\rm REC}$ and L$_{\rm KL}$ are equally weighted. We therefore adopt the methodology suggested by \cite{cyclical_wkl_annealing}: we modify Equation \ref{eq:VAEeq2} by multiplying the second term by a weight scalar W$_{\rm KL}$, and vary this from 0 to 1 in a cyclical fashion, as shown in Figure \ref{fig:vae_weightkl_vs_epoch}. Next, there is the requirement to choose an appropriate $\beta$ in Equation \ref{eq:ce_final}. We choose $\beta=0.005$ based as suggested by \cite{rvae_orig}, and leave the task of finding an optimal $\beta$ to future work. 
Finally, since W$_{\rm KL}$ is annealed with epochs, we need to ensure that our lower and upper learning rates help with convergence for all values of this scalar. From Figure \ref{fig:vae_find_lr}, we see that between learning rates of $10^{-5}$ and $10^{-3}$, the total loss decreases for all values of W$_{\rm KL}$. 

{\bf Overall workflow:} Our overall workflow is as follows:
\begin{enumerate}
    \item For a given train-test split (out of a total of 10) of $\mathcal{D_{F_S,S_S}}$, we use the training set with the MDN, record predictions on the test set, and calibrate them using the methodology described above. We save the weights of the MDN at the epoch of minimum validation loss -- this is shown by the dashed vertical line in Figure \ref{fig:mdn_training_curve}, and for the specific split shown, occurs at epoch 38.
    \item Next, we train the RVAE using the same training set. Similar to the process with the MDN, we revert the model weights back to the epoch of minimum loss, and make predictions on the test set. We gather for the training, validation, and test sets the total loss -L$_{\rm ELBO}$, reconstruction loss L$_{\rm REC}$, and the KL-divergence loss L$_{\rm KL}$. These are plotted in Figures \ref{fig:vae_trainingcurve1} and \ref{fig:vae_trainingcurve2}. We save the $95^{\rm th}$ percentile of -L$_{\rm ELBO, Train}$ as the L2; this is our cut-off between ID and OoD samples. 
    \item Next, we create a small data set of only those samples from the test set where all twelve vents are open. While our goal is to hypothesize the gains in seeing/MPIQ we could have gotten had the vents been in their optimal configuration instead of in the all-open configuration, we believe it is important to be conservative in our estimates. Thus we select only those samples for further processing where we are confident that there were no mechanical malfunctions, high wind conditions, or other system errors that could have prevented the telescope operator from opening all vents.
    \item As a first filter, we select only those samples for which L$_{\rm ELBO, Test} <$ L2, with the intention of filtering out samples for which we are not extremely confident about the ID characteristic.
    \item From this newly created test set, we further only select those samples where our MDN from Step (i) predicts that the true MPIQ is covered by $68^{th}$ percent spread about the median in the predicted MPIQ PDF. This is again enacted in the interest of obtaining conservative predictions downstream.
    \item From the filtered test set in Step (v), we create a permutated data set by toggling all twelve vents ON (==1) and OFF (==0). For a total of 12 vents, this results in $4095$ new samples for each input sample, where the remaining 107 features remain unchanged. The $4096^{\rm th}$ sample is the input test sample itself, since its vents are already in the all-open configuration. For each of the these $4095$ samples, we again apply the same filter as in Step (iv) -- filtering out those vent configurations which, given the training set, are OoD.
    \item Finally, we obtain MPIQ predictions using the MDN for all samples in the permutated data set, created by collation ID permutations for all selected test samples.
\end{enumerate}

{\bf Identifying predictable vent variations \& separating in-distribution from out-of-distribution samples:} In Figure \ref{fig:vae_hist_mll_ood}, we demonstrate our methodology for separating ID samples from OoD ones. As should be expected, most test samples are ID, as are $95\%$ training samples (by definition). A striking yet expected result is that only a very small sample of possible permutations are ID. The reason for this becomes clear from Figure \ref{fig:vae_hist_hds}, where we plot histograms of the different vent configurations in the training set -- 0 on the x-axis corresponds to the all-open configuration, while 1 to all-closed. The vast majority of samples, $\sim80\%$, have all vents closed, while $\sim20\%$ have either all vents or most vents open. Thus the vast majority of samples in the permutated dataset, where the twelve vents can take arbitrary configurations -- say half open and half closed, corresponding to a Hamming distance (x-axis in Figure \ref{fig:vae_hist_mll_ood}) of 0.5 -- are those that the RVAE has not seen before, and thus classifies as OoD.

{\bf Process illustration:} Finally, we illustrate the workflow delineated in Steps (ii) through (vii) above in Figure \ref{fig:cfht_rvae_overview}.

\section{Results}
\label{sec:results}
In Section~\ref{sec:resultsPredictingIQ} we present results on using our model to predict the image quality given the current environmental and dome operating conditions. In Section~\ref{sec:resultsImprovingIQ} we discuss how we might better operate the dome to improve IQ.  In particular, we investigate the potential improvement that could result from smart actuation of the configuration of the dome vents. In Section~\ref{sec:resultsRelContribIQ} we present results on the relative contribution of different features to the predicted mean MPIQ.  Through these results we verify observations by earlier groups and we start to understand better what information our models use in its inference process.


\subsection{Predicting image quality}
\label{sec:resultsPredictingIQ}

\begin{figure*}
\begin{subfigure}{0.49\textwidth}
    \centering
    \includegraphics[width=.98\linewidth]{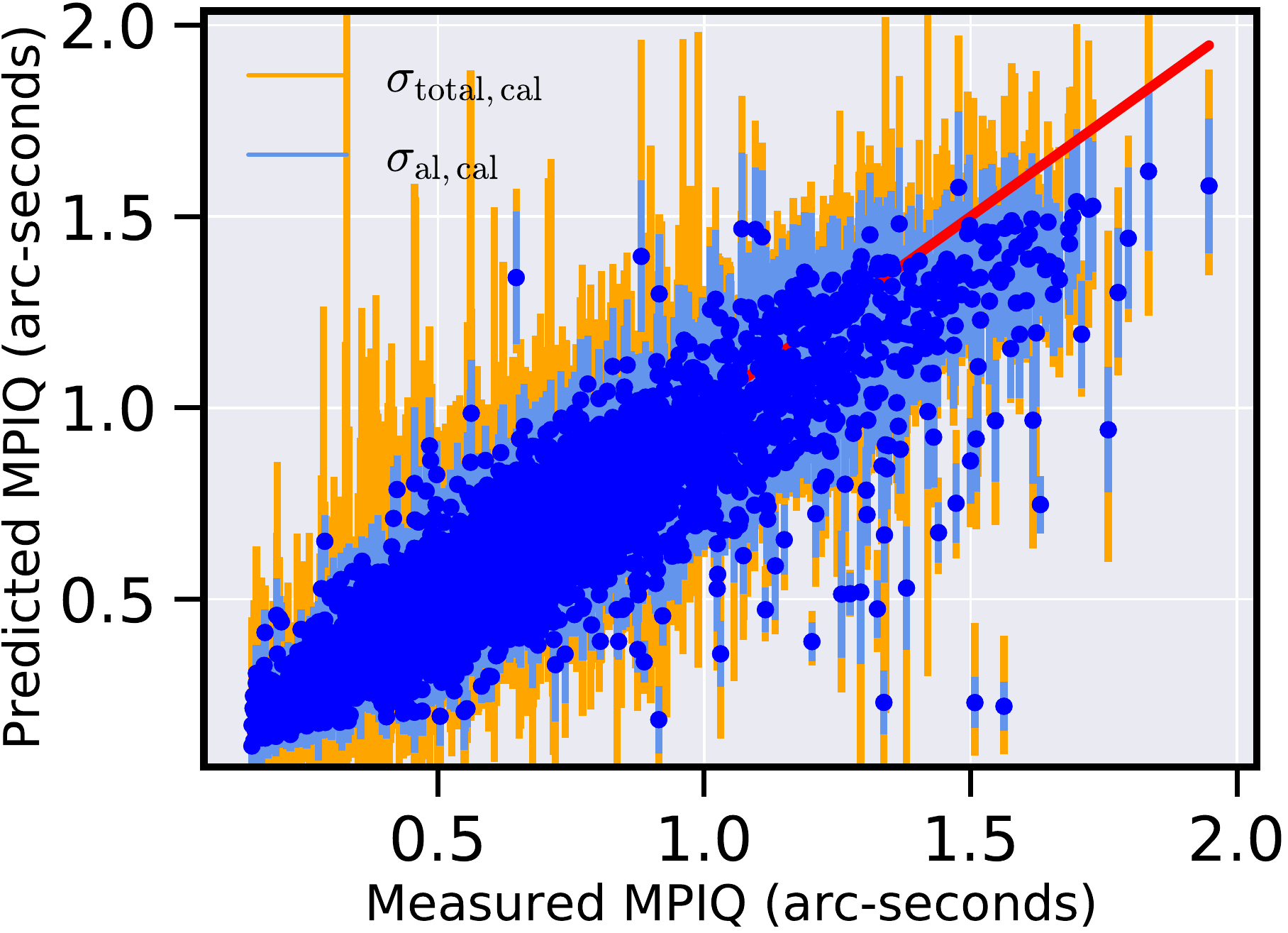}
    \caption{Predicted versus measured MPIQ. For the former we show the calibrated median, and upper and lower quantiles for calibrated uncertainties.}
    \label{fig:mdn_one_to_one}
\end{subfigure}
\hfill
\begin{subfigure}{0.49\textwidth}
    \centering
    \includegraphics[width=.98\linewidth]{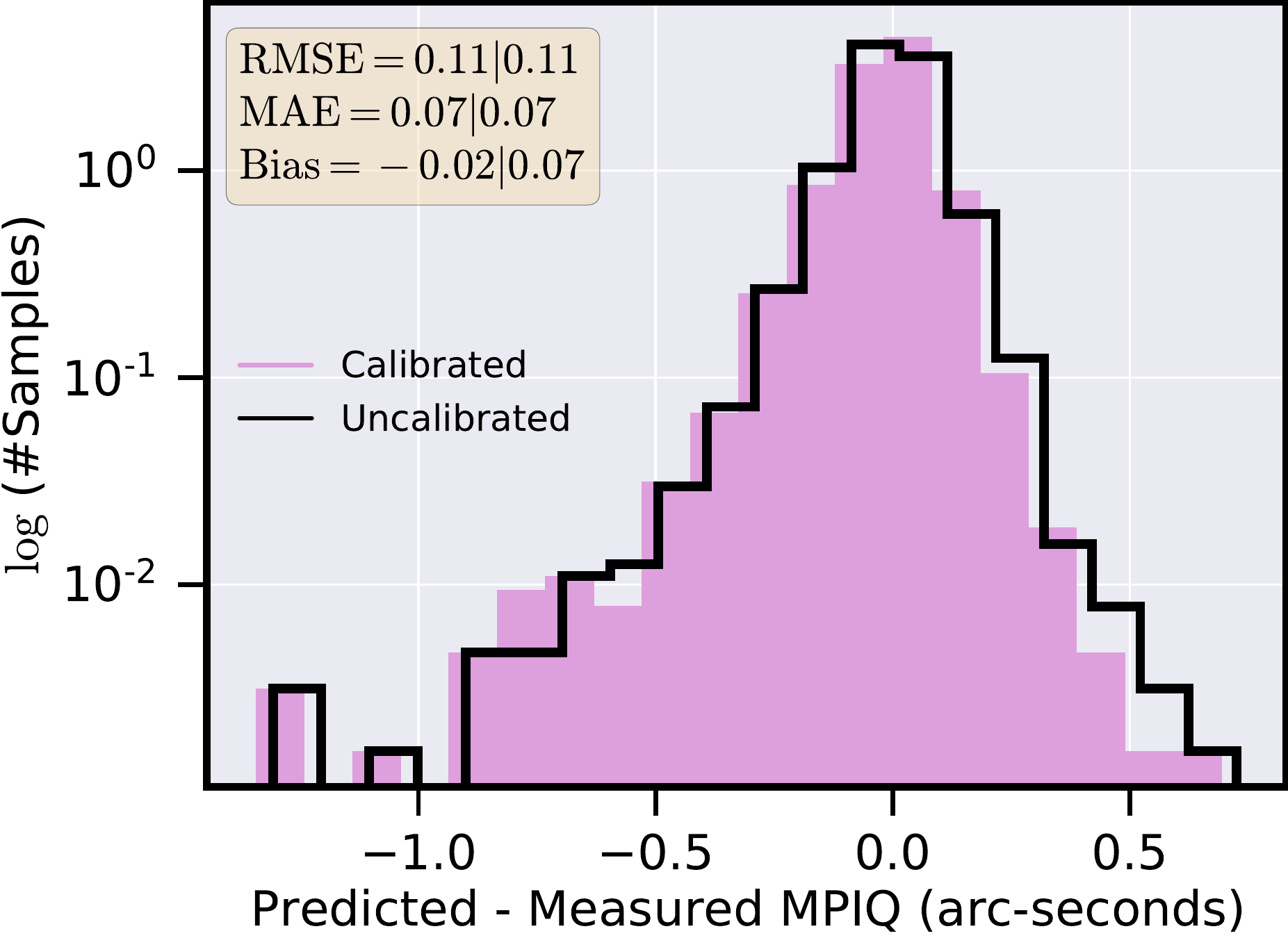}
    \caption{Histogram of prediction errors, along (in inset box) with three metrics that compare performance with and without calibration.}
    \label{fig:mdn_one_to_one_hist}
\end{subfigure}
\newline
\begin{subfigure}{0.49\textwidth}
    \centering
    \includegraphics[width=.98\linewidth]{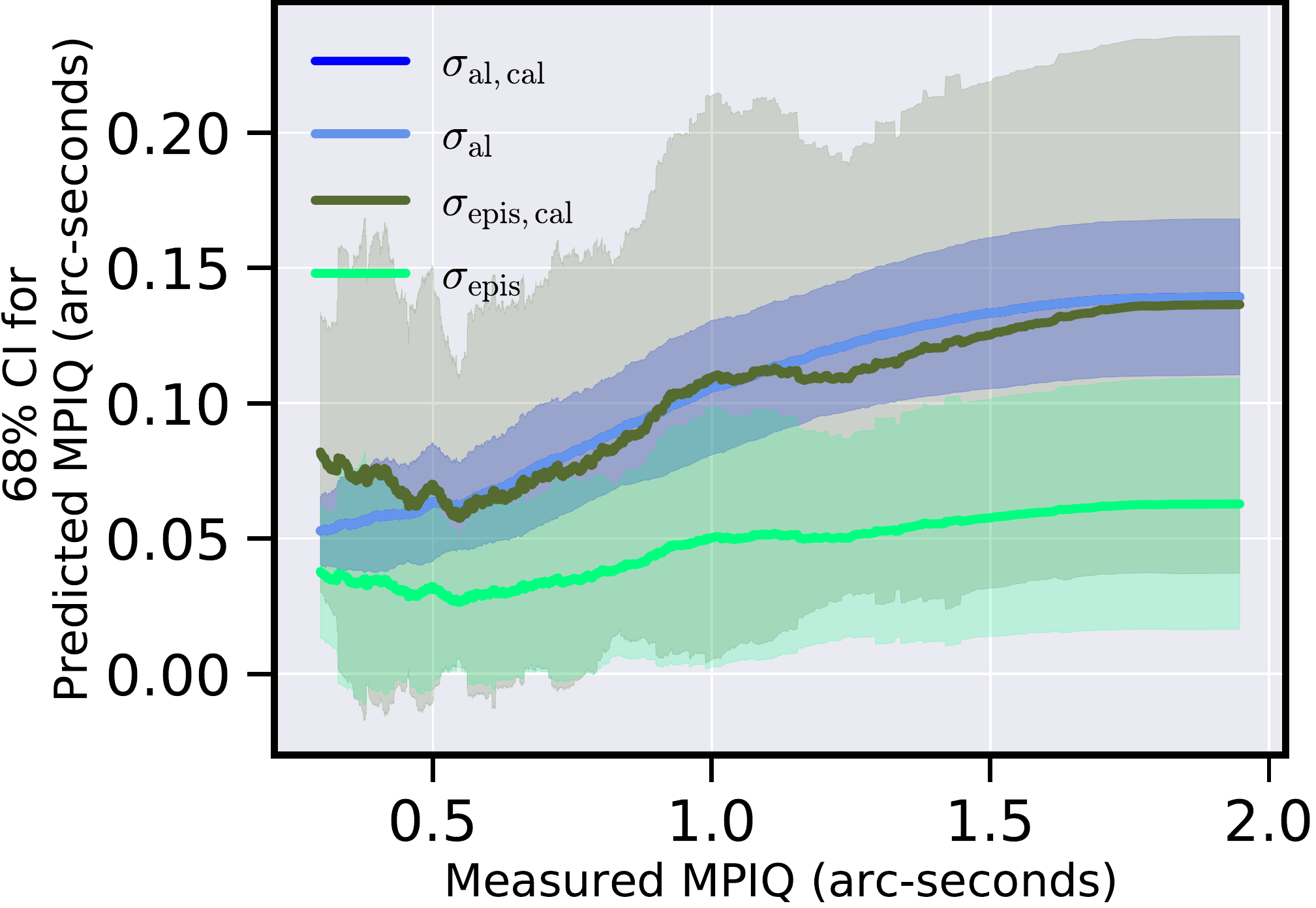}
    \caption{68\% spread in calibrated and uncalibrated aleatoric and epistemic uncertainties in predicted MPIQ, plotted as a function of measured MPIQ.}
    \label{fig:mdn_CI_vs_mpiq1}
\end{subfigure}
\hfill
\begin{subfigure}{0.49\textwidth}
    \centering
    \includegraphics[width=.98\linewidth]{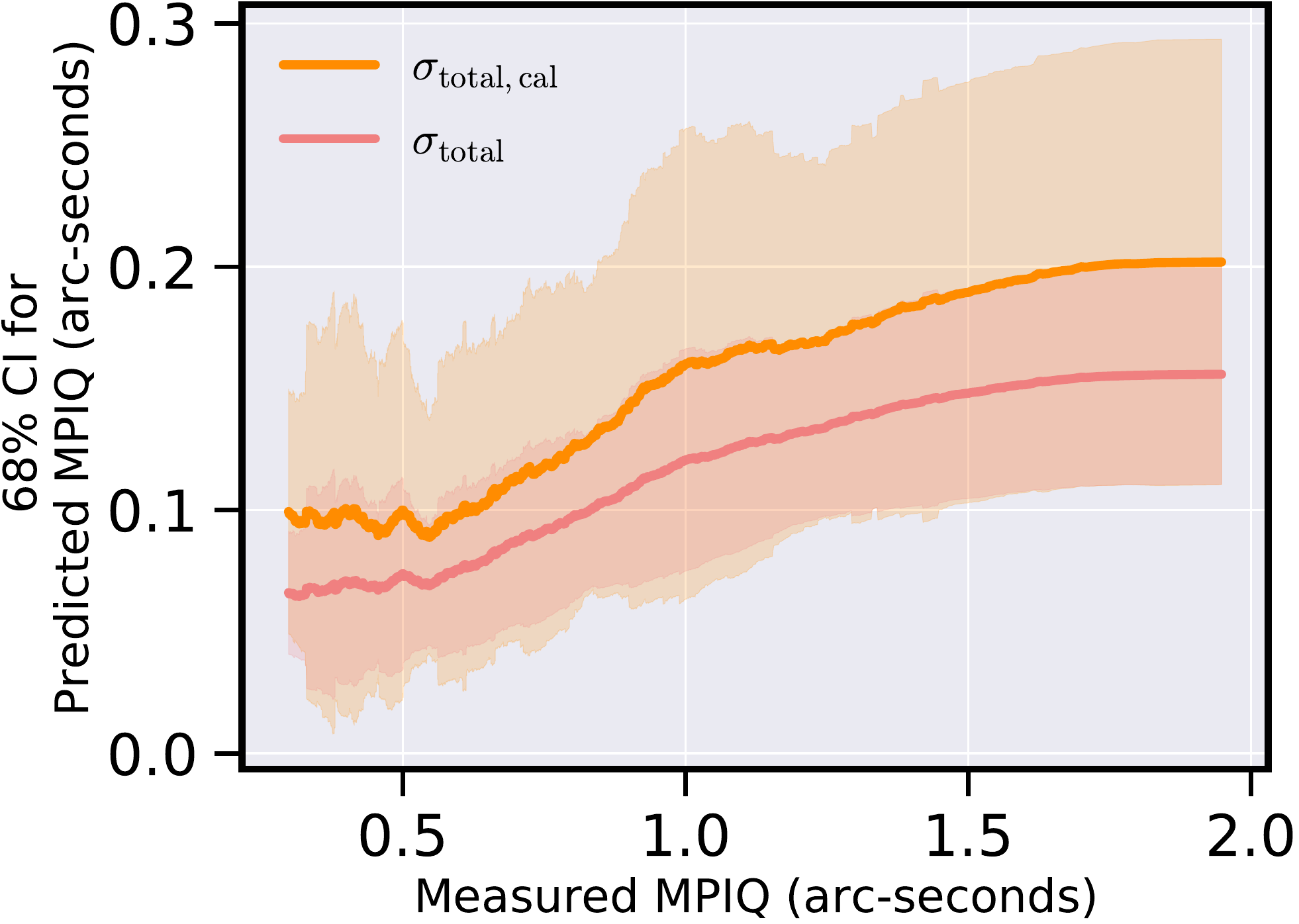}
    \caption{68\% spread in total uncertainties in predicted MPIQ, plotted as a function of measured MPIQ.}
    \label{fig:mdn_CI_vs_mpiq2}
\end{subfigure}
\newline
\begin{subfigure}{0.32\textwidth}
    \centering
    \includegraphics[width=.98\linewidth]{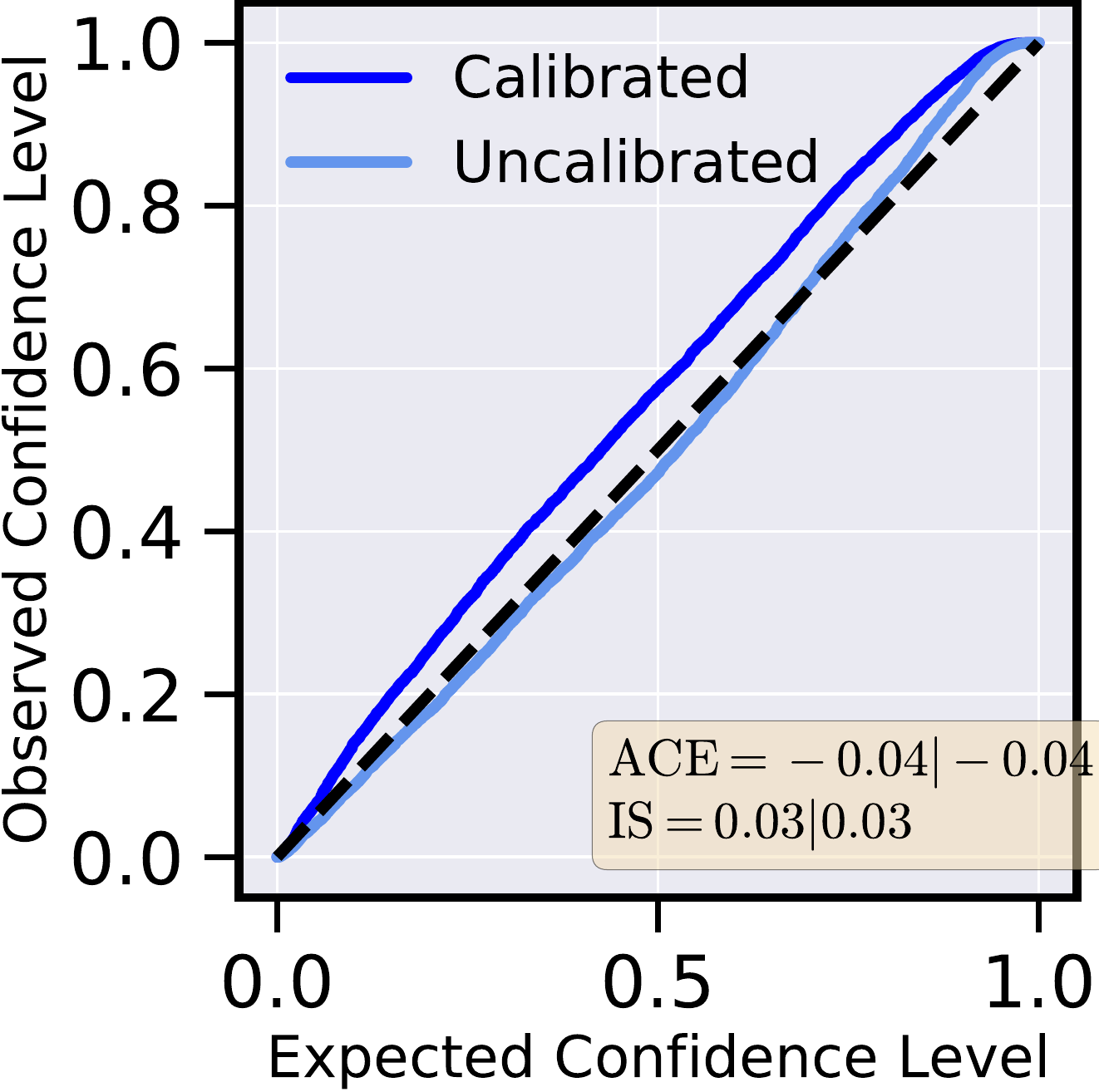}
    \caption{Calibration curves for aleatoric uncertainty.}
    \label{fig:mdn_calibration_curve_al}
\end{subfigure}
\begin{subfigure}{0.32\textwidth}
    \centering
    \includegraphics[width=.98\linewidth]{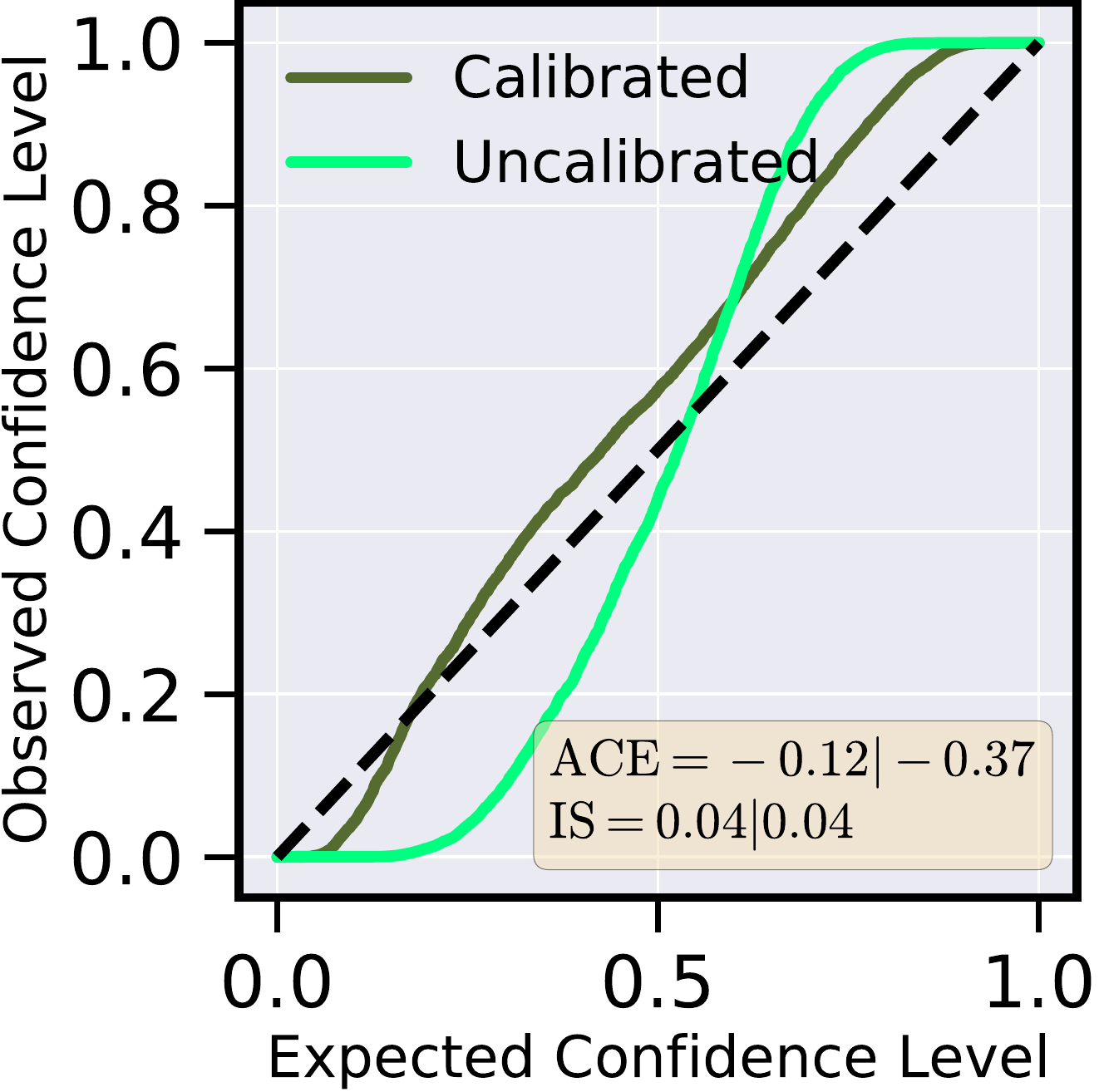}
    \caption{Calibration curves for epistemic uncertainty.}
    \label{fig:mdn_calibration_curve_epis}
\end{subfigure}
\begin{subfigure}{0.32\textwidth}
    \centering
    \includegraphics[width=.98\linewidth]{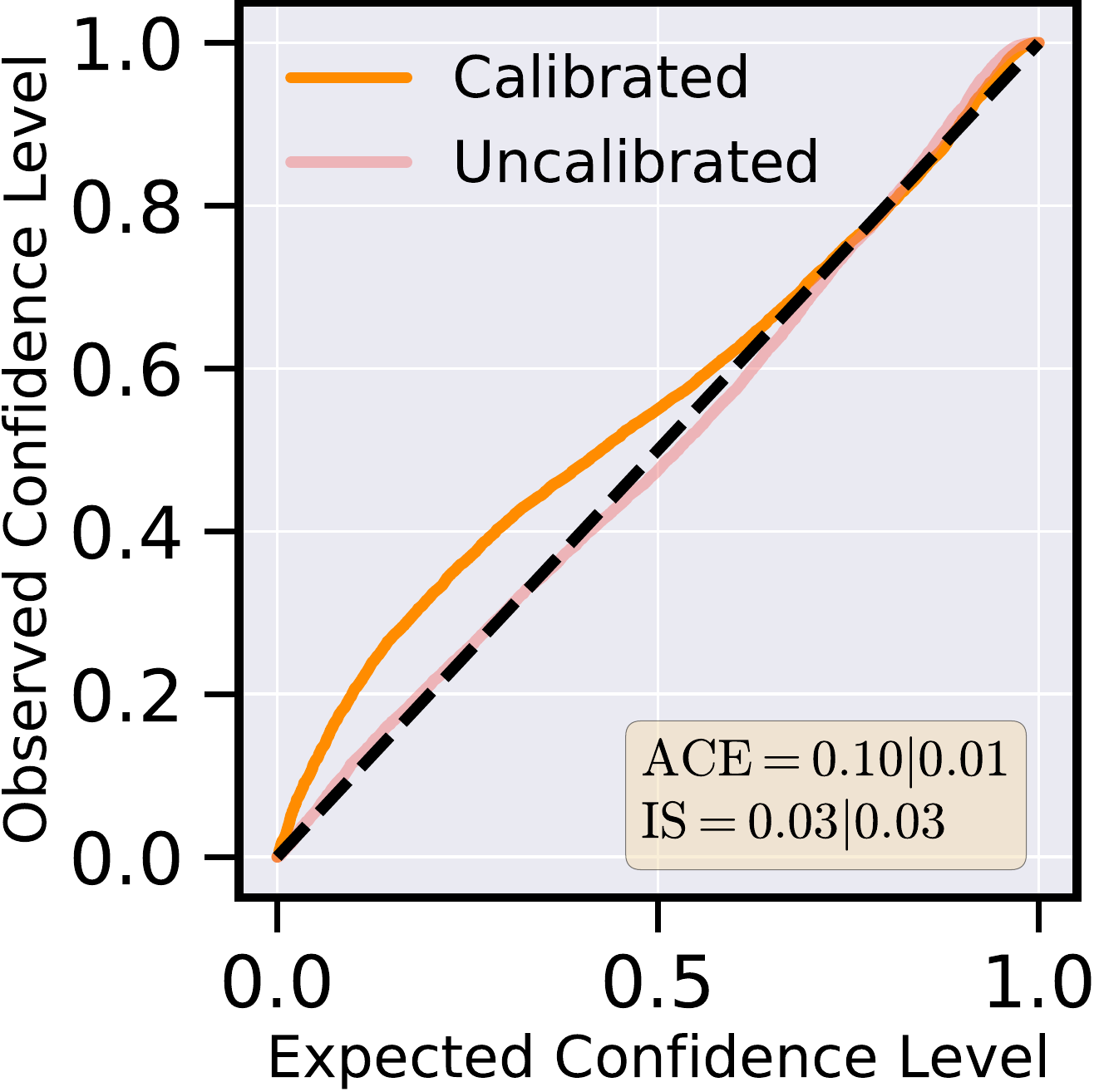}
    \caption{Calibration curves for total uncertainty.}
    \label{fig:mdn_calibration_curve_total}
\end{subfigure}
\caption{Predictions using the Mixture Density Network. $\sigma \implies 0.5\times(84^{\rm th} - 16^{\rm th})$ quantiles. In \textbf{(a)} we plot the predicted MPIQ, characterized by the $16^{\rm th}, 50^{\rm th}$, and $84^{\rm thn}$ percentiles of their respective calibrated PDFs.  These are plotted versus measured (i.e., true) MPIQs. In \textbf{(b)} we subtract the ground-truth MPIQ from the $50^{\rm th}$ percentile predictions, from both the raw uncalibrated, and the calibrated PDFs, and plot their histograms. We also quantify the quality of both calibrated and uncalibrated predictions using root mean squared error (RMSE), mean absolute error (MAE), and  bias error (BE); in the inset box read calibrated on left and uncalibrated on right.  Calibration results in better BE. In \textbf{(c)} and \textbf{(d)}, we plot the smoothed mean and standard deviations of the aleatoric, epistemic, and total uncertainties as a function of the measured MPIQs. All three uncertainties increase when $\sigma_{\rm epis}$ is calibrated.  Uncertainty is highest near the low-end and high-end MPIQ values; in these regimes we have the least number of observations.  Uncertainty is lowest near the mode of the histogram of measured MPIQs where data is plentiful (cf., the histograms in Figure~\ref{fig:hist_preliminaries}). Finally, in \textbf{(e)}, \textbf{(f)} and \textbf{(g)} we visualize the benefits of calibrating $\sigma_{\rm epis}$. The ideal is the 1:1 line; closer is better. In the inset box the interval sharpness (IS) and average calibration error (ACE) metrics, with and without calibration, are provided.}
\label{fig:MDN_moneyplot}
\end{figure*}

\begin{figure*}
\begin{subfigure}{0.49\textwidth}
    \centering
    \includegraphics[width=.98\linewidth]{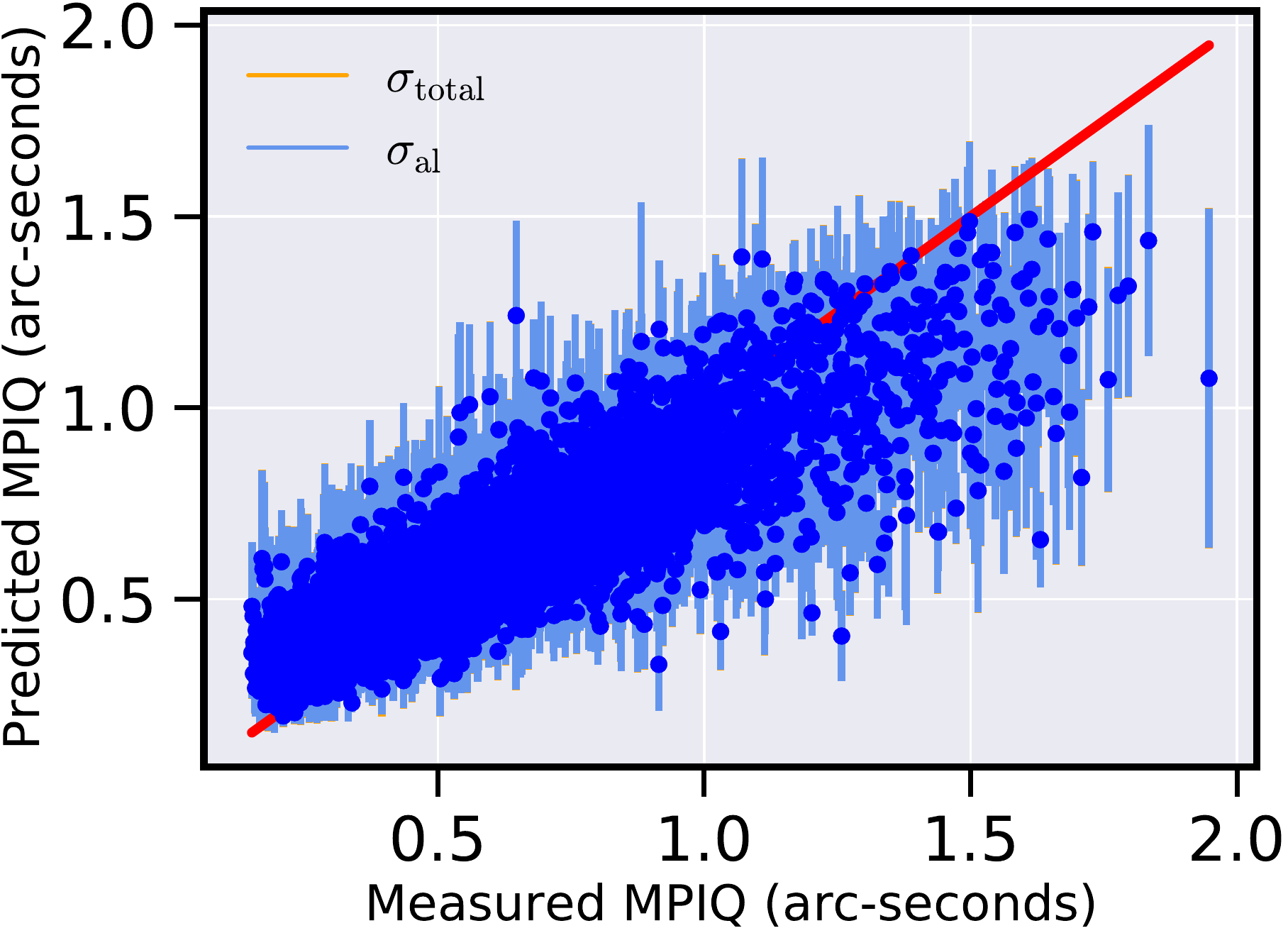}
    \caption{Predicted versus measured MPIQ. We do not calibrate the predictions from the boosted tree model.}
    \label{fig:mdn_one_to_one_cat}
\end{subfigure}
\hfill
\begin{subfigure}{0.49\textwidth}
    \centering
    \includegraphics[width=.98\linewidth]{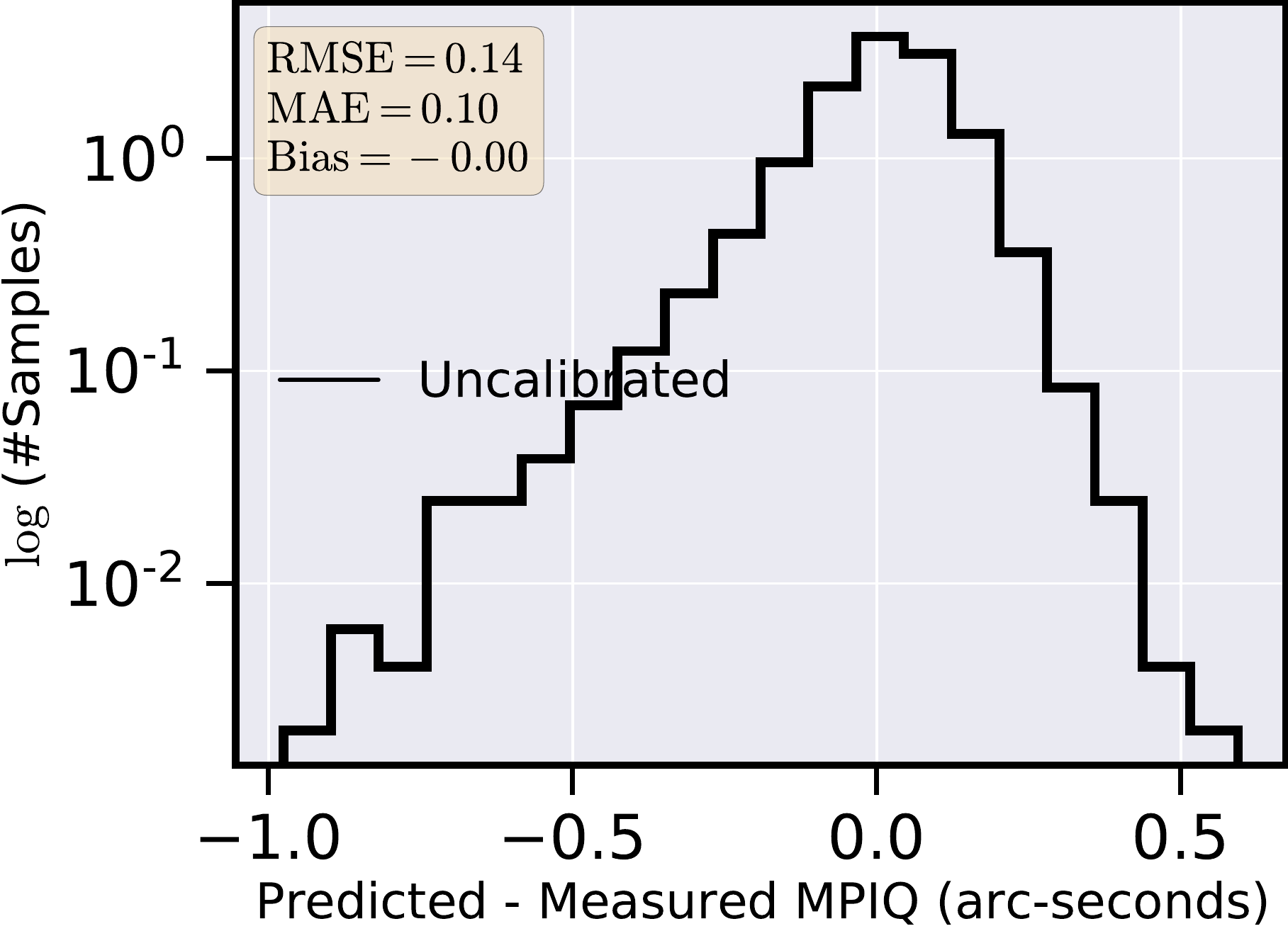}
    \caption{Histogram of prediction errors.  In the inset box we provide three metrics that compare performance with and without calibration.}
    \label{fig:mdn_one_to_one_hist_cat}
\end{subfigure}
\caption{Prediction results using gradient boosted decision trees (GBDTs).}
\label{fig:MDN_moneyplot_catboost}
\end{figure*}

In Figure~\ref{fig:MDN_moneyplot} we present our main results on the accuracy of probabilistic predictions of MPIQ using the MDN.  In Figure~\ref{fig:MDN_moneyplot_catboost} we present comparative results for the graient-boosted tree model.  Table~\ref{tab:ml_vs_dl_results} tabulates summary results.  We describe each set of results in turn.

Figure~\ref{fig:mdn_one_to_one} quantifies the accuracy of our predictions.  The horizontal axis displays measured (a.k.a. nominal) MPIQ, while the y-axis displays predicted MPIQ.  The units of both are arc-seconds ($''$). Perfect prediction is represented by the red $45\degree$  line. True MPIQ varies from a bit below $0.5''$ to just over $2''$.  The blue dots depict the point-predictions (the medians of the output PDFs).  The light blue bars plot the estimated aleatoric uncertainties ($\sigma_a$) of the point predictions.  These are superimposed on the total uncertainty, the differences are the epistemic uncertainties ($\sigma_e$), visible in orange.   
As is tabulated in Table~\ref{tab:ml_vs_dl_results}, the mean absolute error (MAE) between the true MPIQ values and the medians of our calibrated predictions is $\sim 0.07''$. 

Figure \ref{fig:mdn_one_to_one_hist} help us understand the improvement due to calibration. We plot the histograms of the differences between the calibrated predictions and the true MPIQ values, and between the uncalibrated predictions and the true MPIQ values.  These histograms are respectively plotted in pink and black. We use three metrics (cf., Section \ref{sec:metrics_det}) to quantify the improvement resulting from calibration: root mean squared error (RMSE), mean absolute error (MAE), and bias error (BE). The values in the first row are for uncalibrated medians while those in the second row are for calibrated models.  We remind the reader that the calibration using CRUDE \citep{crude_probability_calibration} is enacted only for the epistemic uncertainties, $\sigma_{\rm epis}$, which we observe is significantly decreased for the calibrated model.

\begin{table*}
     \caption{Comparative performance of methods, across all four data sets (cf. Section \ref{sec:data}), and all six metrics (cf. Section \ref{sec:DL_metrics}). The top row shows uncalibrated results for the MDN and the GBDT models.  To ease direct comparison we present results as an ordered tuple (MDN, GBDT).  The bottom row shows calibrated results for the MDN.  We do not provide calibrated results for the GBDT.  For each metric the performance of the best-performing model is highlighted in bold; in all cases the MDN performs at least as well as the GBDT.}
    \begin{tabular}{cccccccccc}
        \toprule
        \toprule
        $\rm{Metric}$ & RMSE & MAE & BE & ACE$_{\rm {al}}$ & ACE$_{\rm {epis}}$ & ACE$_{\rm {total}}$ & IS$_{\rm {al}}$ & IS$_{\rm {epis}}$ & IS$_{\rm {total}}$ \\ \midrule
        Uncalibrated & ({\bf 0.11}, {\bf 0.11}) & ({\bf 0.07}, 0.08) & ({\bf 0.00}, {\bf 0.00}) & (\textbf{-0.01}, -0.09) & (\textbf{-0.31}, -0.59) & (\textbf{0.04}, -0.08) & (\textbf{0.03}, 0.06) & (\textbf{0.04}, 0.08) & (\textbf{0.03}, 0.06) \\
        \midrule
        Calibrated & 0.11 & 0.07 & 0.03 & -0.02 & -0.10 & 0.11 & 0.02 & 0.04 & 0.03 \\
        \bottomrule
    \end{tabular}
    \label{tab:ml_vs_dl_results}
\end{table*}

\begin{figure*}
\begin{subfigure}{0.49\textwidth}
    \centering
    \includegraphics[width=.98\linewidth]{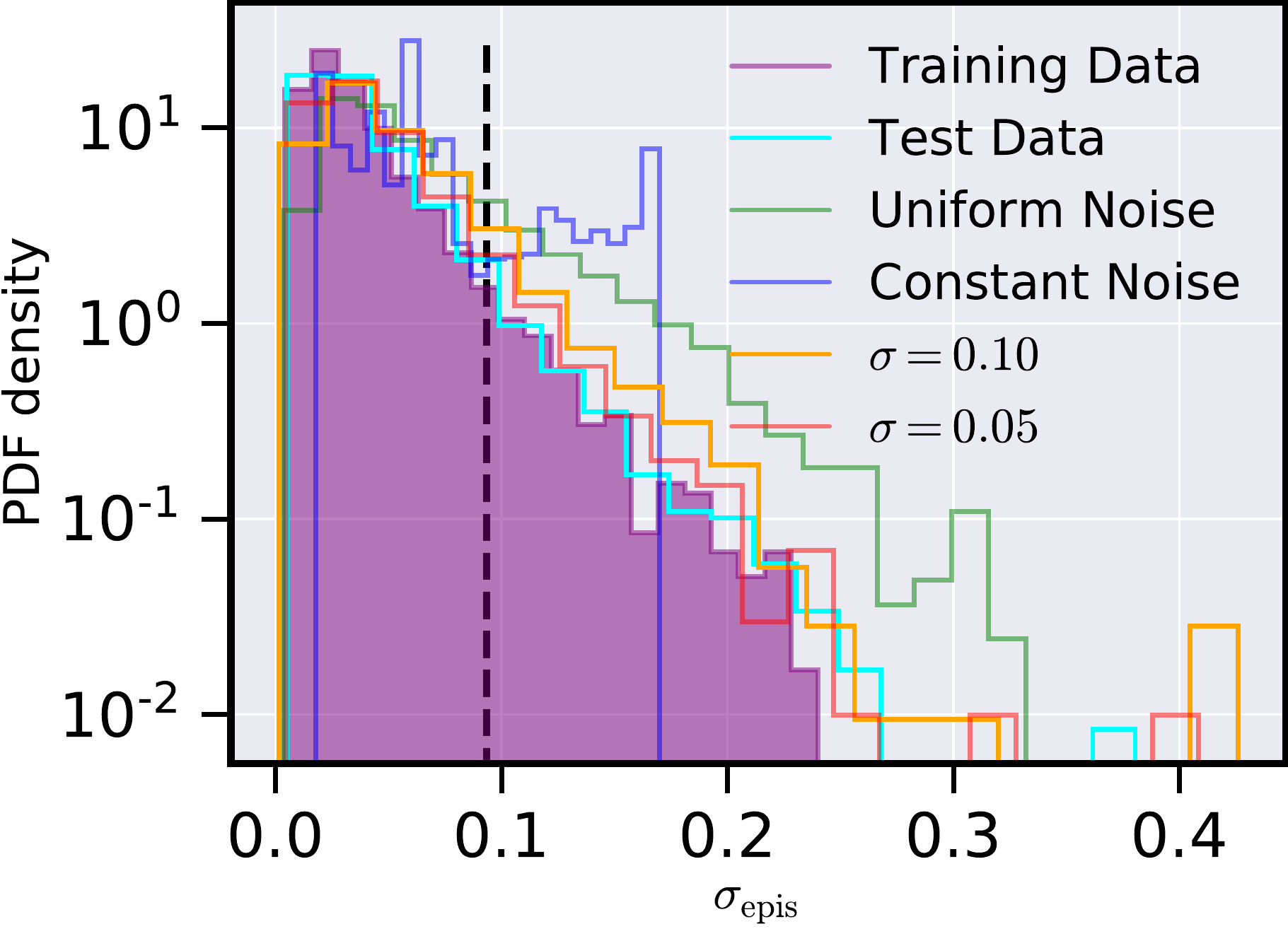}
    \caption{Histograms of {\it uncalibrated epistemic} uncertainty from the MDN, for various data sets. The vertical line is the $95^{\rm th}$ percentile for the training set.}
    \label{fig:vae_hist_epis}
\end{subfigure}
\hfill
\begin{subfigure}{0.49\textwidth}
    \centering
    \includegraphics[width=.98\linewidth]{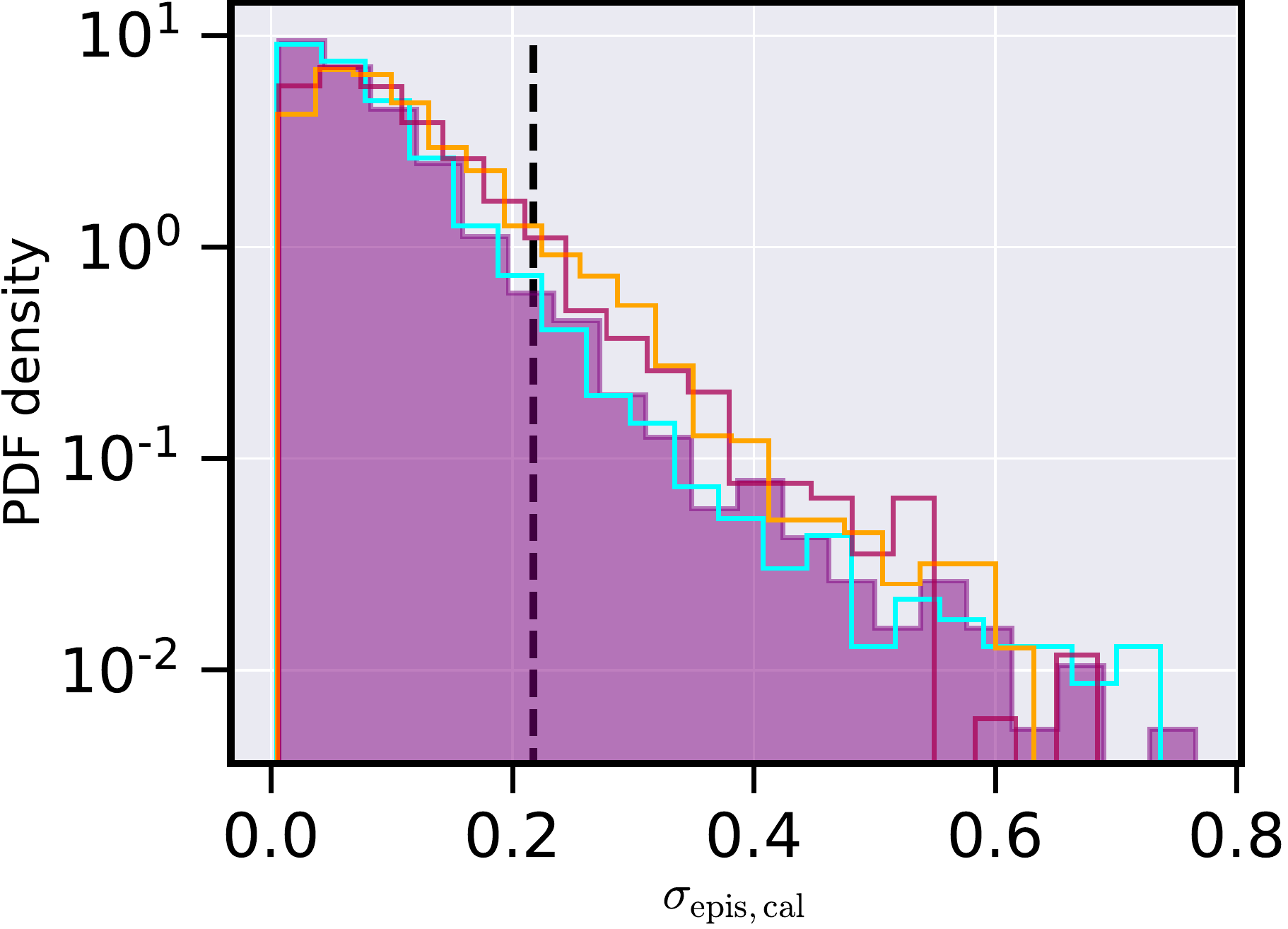}
    \caption{Histograms of {\it calibrated epistemic} uncertainty from the MDN, for various data sets. The vertical line is the $95^{\rm th}$ percentile  for the training set.}
    \label{fig:vae_hist_calepis}
\end{subfigure}
\newline
\begin{subfigure}{0.49\textwidth}
    \centering
    \includegraphics[width=.98\linewidth]{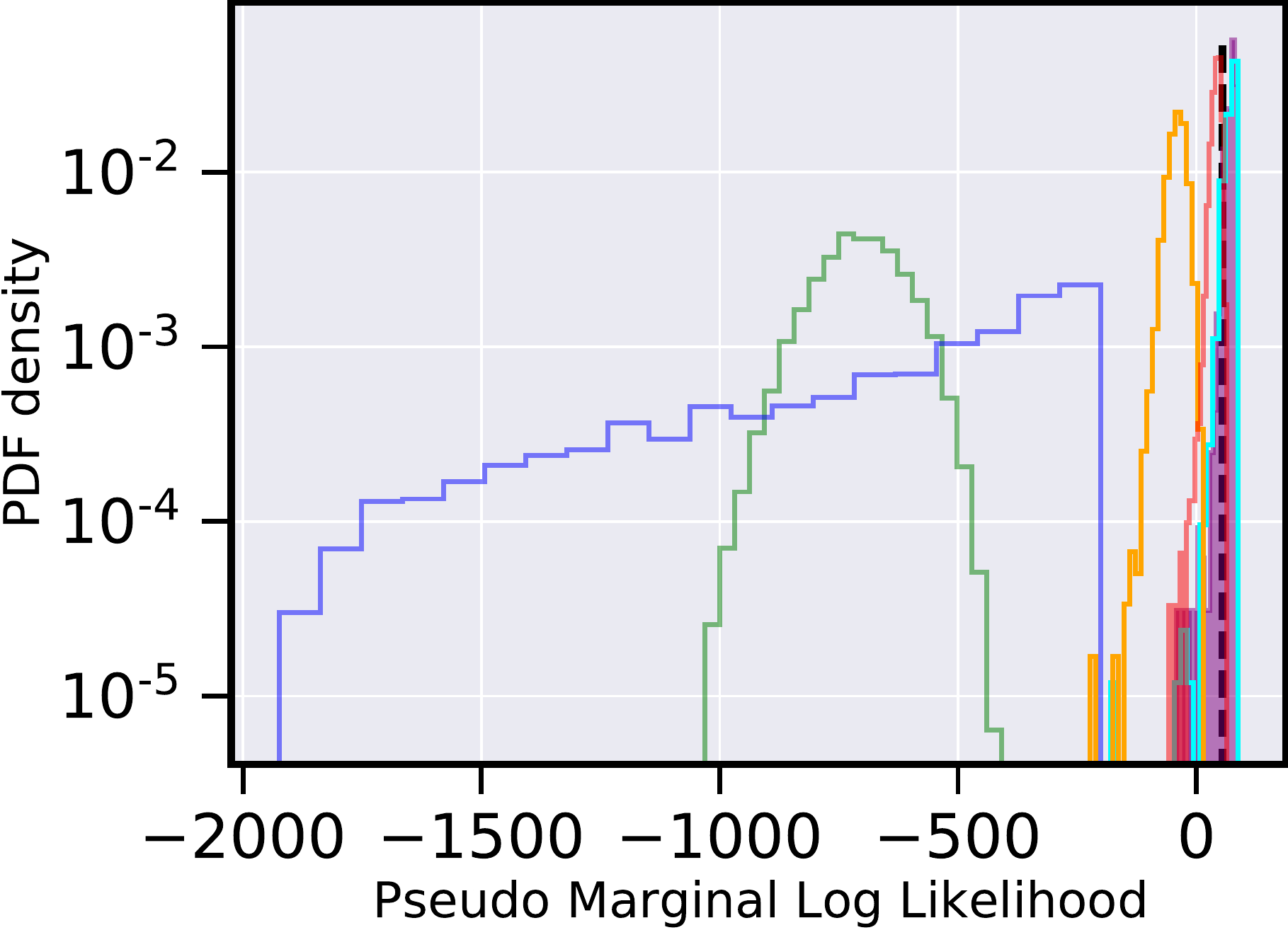}
    \caption{Histograms of the pseudo marginal log likelihood (-L$_{\rm ELBO}$) from the RVAE. The vertical line is the $95^{\rm th}$ percentile  for the training set.}
    \label{fig:vae_hist_mll}
\end{subfigure}
\hfill
\begin{subfigure}{0.49\textwidth}
    \centering
    \includegraphics[width=.98\linewidth]{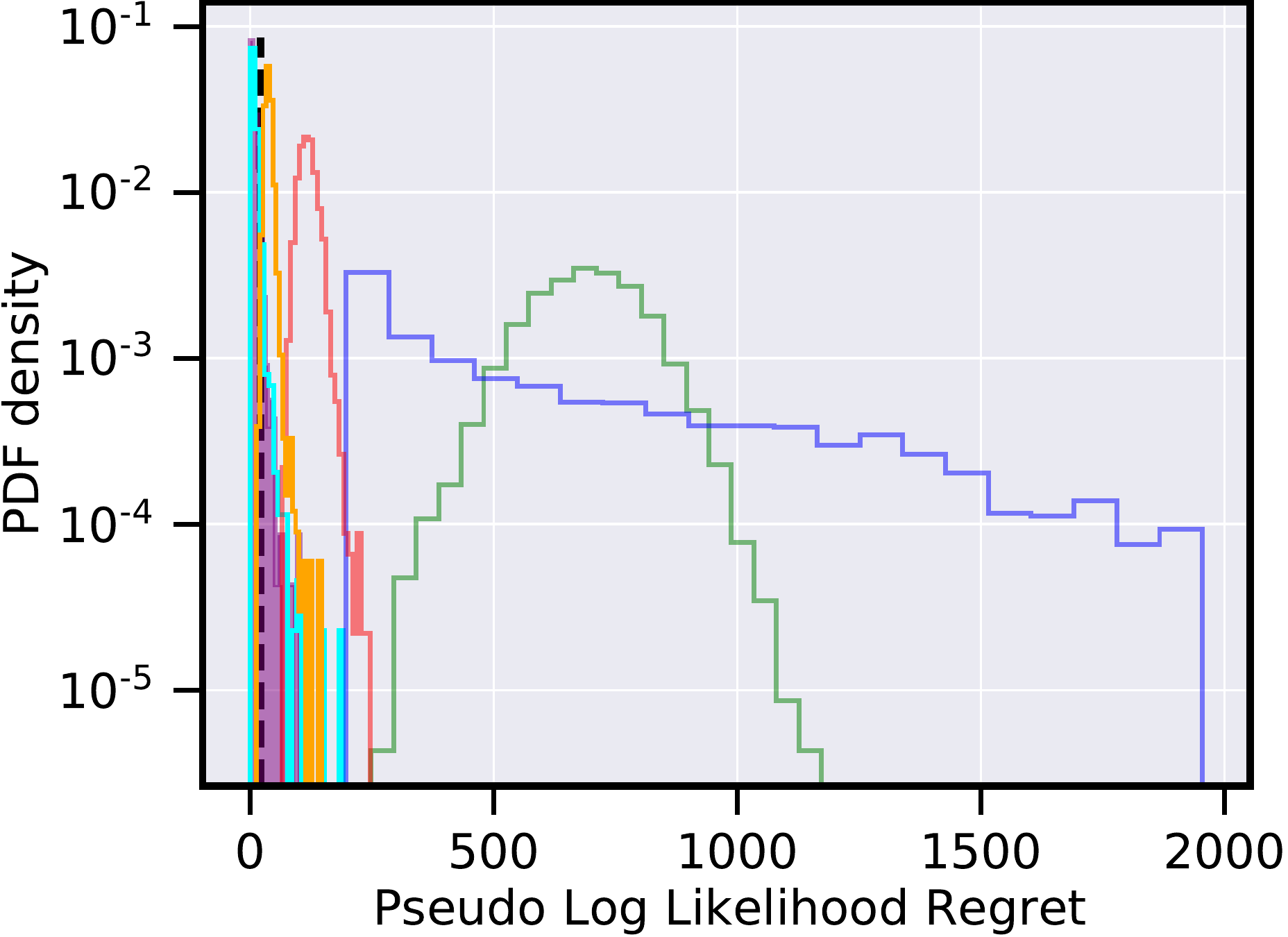}
    \caption{Histograms of log likelihood regret for the same data sets, from the RVAE. The vertical line is the $95^{\rm th}$ percentile  for the training set.}
    \label{fig:vae_hist_regret}
\end{subfigure}
\newline
\begin{subfigure}{0.49\textwidth}
    \centering
    \includegraphics[width=.98\linewidth]{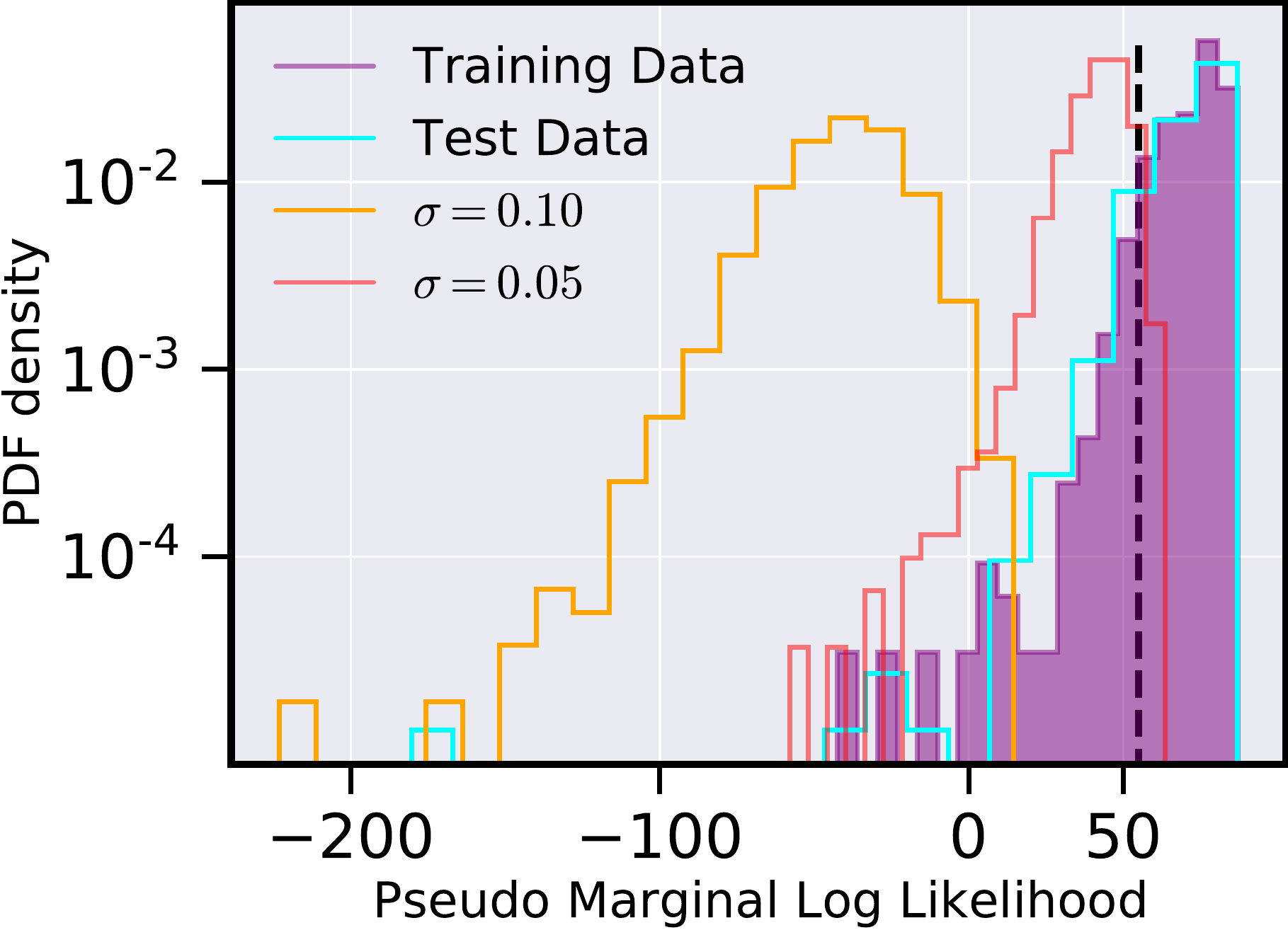}
    \caption{Zoom-in of subfigure~\textbf{(c)}, focusing on the hard OoD data sets.}
    \label{fig:vae_hist_mll_zoomed}
\end{subfigure}
\hfill
\begin{subfigure}{0.49\textwidth}
    \centering
    \includegraphics[width=.98\linewidth]{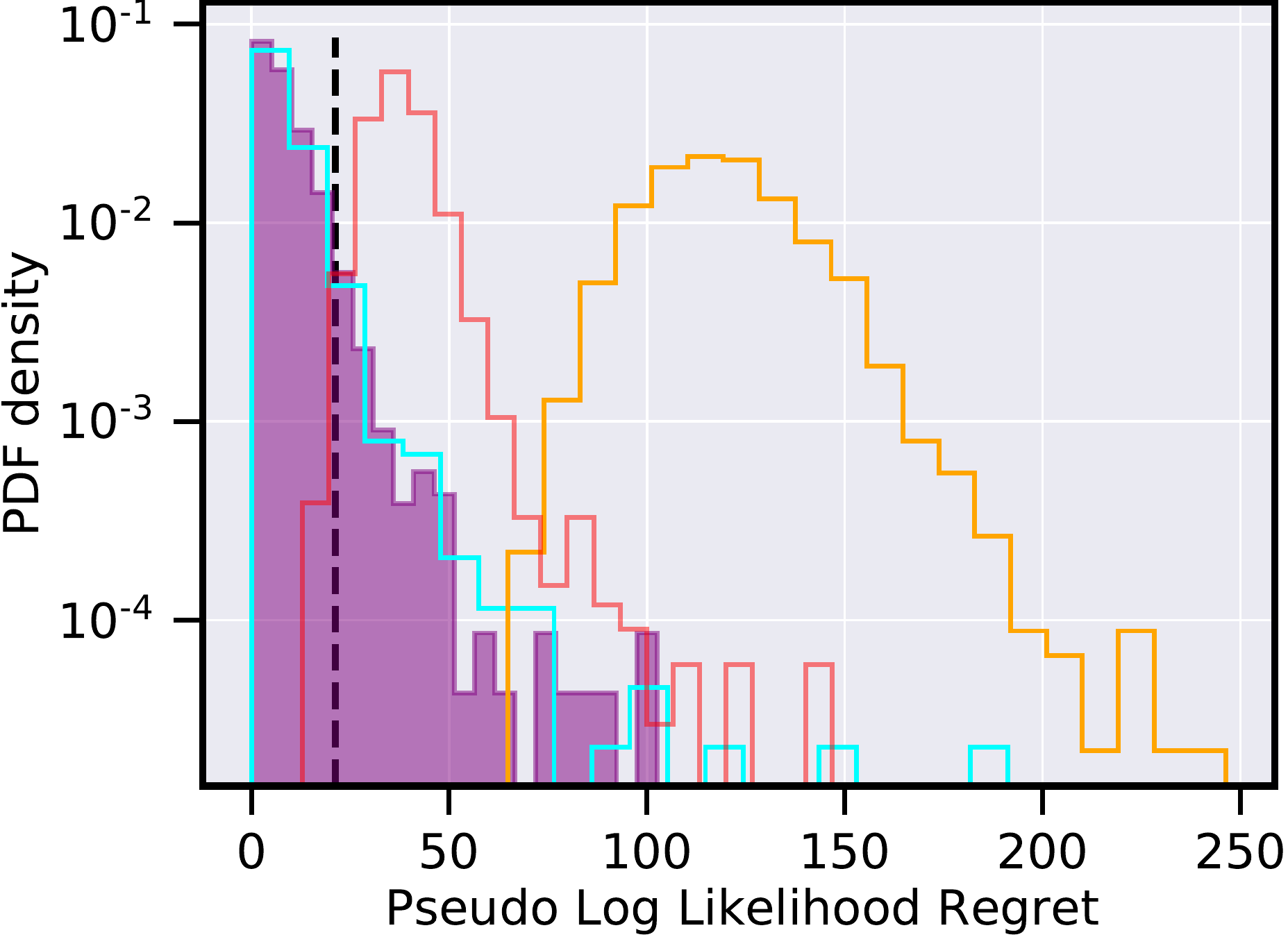}
    \caption{Zoom-in of subfigure~\textbf{(d)}, focusing on the hard OoD data sets.}
    \label{fig:vae_hist_regret_zoomed}
\end{subfigure}
\caption{We visualize the discriminative ability of various metrics to separate Out-of-Distribution (OoD) samples from In-Distribution (ID) samples. Perhaps surprisingly, we observe that epistemic uncertainties, whether calibrated or uncalibrated, are poor metrics for OoD detection. This is our motivation for the RVAE, which directly captures the likelihood of the training data-generating distribution. \textbf{(c)} and \textbf{(e)} show that  log-likelihood is an excellent, if not perfect, discriminator that can separate ID training and test data from even slightly OoD synthetic data created by adding Gaussian noise of the indicated standard deviation ($\sigma$) to normalized training data. In \textbf{(d)} and \textbf{(f)} we calculate the log-likelihood regret \citep{likelihood_regret}, as explained in Section \ref{sec:rvae}. Comparing \textbf{(e)} and \textbf{(f)}, we see that regret is a slightly better OoD detector.  In these plots the training data histogram is more concentrated, and the modes of the histograms of the two noisy data sets are farther away from the mode of the histogram of the training data set.}
\label{fig:common_ood_detections}
\end{figure*}

Figures \ref{fig:mdn_CI_vs_mpiq1} and \ref{fig:mdn_CI_vs_mpiq2} show smoothed averages of the aleatoric, epistemic, and total uncertainties, for both calibrated and uncalibrated models. We highlight a few important aspects. First, as expected, $\sigma_{\rm al}$ is unchanged by calibration since we do not calibrate aleatoric uncertainty. (The light-blue and dark-blue plots coincide so we don't see both.)  Second, as a function of increasing MPIQ, $\sigma_{\rm al}$ (and the identical $\sigma_{\rm al, cal}$ curve) start from a low value, decrease slightly, and then increases almost $3\times$. The initial dip can be attributed to the high density of data points near the mode of the MPIQ distribution at $\sim0.7''$. The  increase at higher MPIQ is likely due to the decreasing density of data points (see the red curve in the left sub-figure of Figure \ref{fig:hist_preliminaries}).  As the model has access to fewer and fewer points it becomes challenging to learn latent representations discriminative enough to be able make good predictions.  Hence, the aleatoric error increases. Third, comparing $\sigma_{\rm epis}$ to $\sigma_{\rm epis, cal}$ we observe that calibration increases epistemic uncertainty.  This justifies our suspicion that the probabilistic MPIQ predictions are over-confident, and that our decision to calibrate them {\it post-hoc} was sensible. Fourth, $\sigma_{\rm epis}$ and $\sigma_{\rm epis, cal}$ follow the same pattern as the aleatoric uncertainty;  they initially dip to a minimum and then rise  with increasing true MPIQ. That said, relative to their starting values, they dip down to lower levels, and rise asymptotically to about $1.25\times$ their respective starting levels. 
Since epistemic uncertainty quantifies the degree to which a sample is out-of-distribution (OoD), these curves imply that, compared to the samples near the median MPIQ of $0.7''$, samples at both the low and high ends of the measured MPIQ distribution are slightly OoD. (We do note that using predicted epistemic uncertainties is not an reliable way to filter out OoD samples, as expounded upon in Section \ref{sec:resultsImprovingIQ} and Figure \ref{fig:common_ood_detections}). We believe that both $\sigma_{\rm al}$ and $\sigma_{\rm epis}$ can be reduced by weighing the loss function for the MDN so that samples with poorer predictions are given more attention by the network. Another strategy would be to over- and under-sample data points near the ends and the mode of the MPIQ distribution, respectively.  This will make the curve be less peaked. By attacking the class-imbalance problem at both the algorithm- and data-level, we expect to de-bias our predictions.

Finally in Figures \ref{fig:mdn_calibration_curve_al}, \ref{fig:mdn_calibration_curve_epis} and \ref{fig:mdn_calibration_curve_total} we demonstrate the effect of probability calibration on the three uncertainties. The $x$- and $y$-axes respectively quantify the expected and observed confidence levels.  If we sample the 50\% CI spread around the median of the predicted MPIQ PDF from the MDN, 50\% of samples should have their measured MPIQ values be covered by the predicted intervals. Hence the black dashed 1:1 line in all three plots is the ideal calibration plot. In the inserts, we also quantify the difference that calibration makes via the ACE and IS metrics, defined in Section \ref{sec:metrics_prob}.  The values to the left of the vertical bar (`$\vert$') in the wheat-colored inserts are for calibrated results, while those to the right for uncalibrated results. Since we only calibrate $\sigma_{\rm epis}$, only Figure \ref{fig:mdn_calibration_curve_epis} shows an improvement. This comes at the cost of poorer post-calibration results for $\sigma_{\rm al}$ and $\sigma_{\rm total}$.

\begin{figure*}
\begin{subfigure}{0.49\textwidth}
    \centering
    \includegraphics[width=\linewidth]{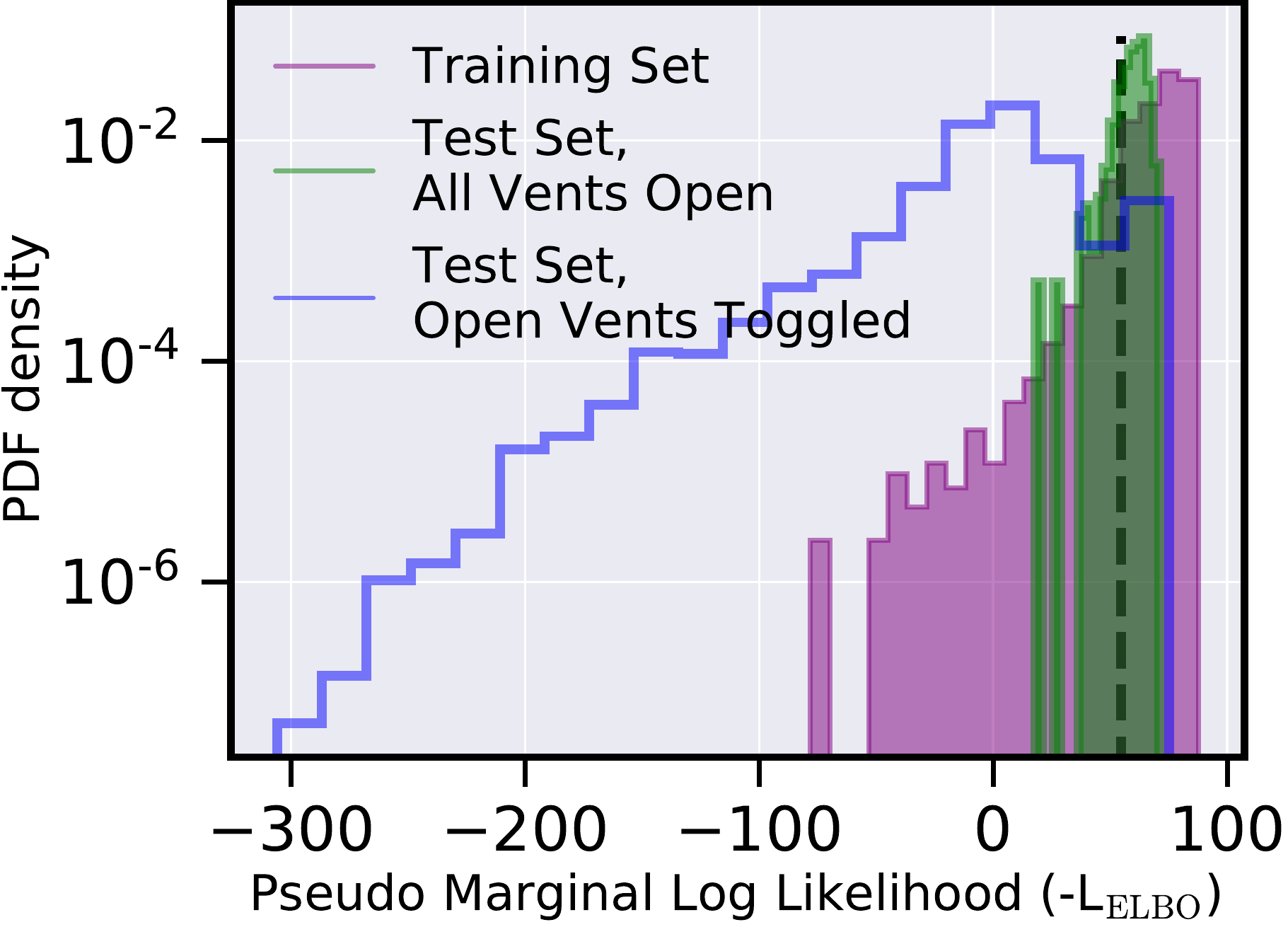}
    \caption{Histograms of the pseudo marginal log likelihood (-L$_{\rm ELBO}$) for three data sets based on one of ten training folds: the training set, the subset of the test set where all vents are open, and the permutated data set constructed by toggling each of the twelve vents for each sample in the test subset. The dashed line indicates the $95^{\rm th}$ percentile  for -L$_{\rm ELBO}$ of the training data set.}
    \label{fig:vae_hist_mll_ood}
\end{subfigure}
\hfill
\begin{subfigure}{0.49\textwidth}
    \centering
    \includegraphics[width=\linewidth]{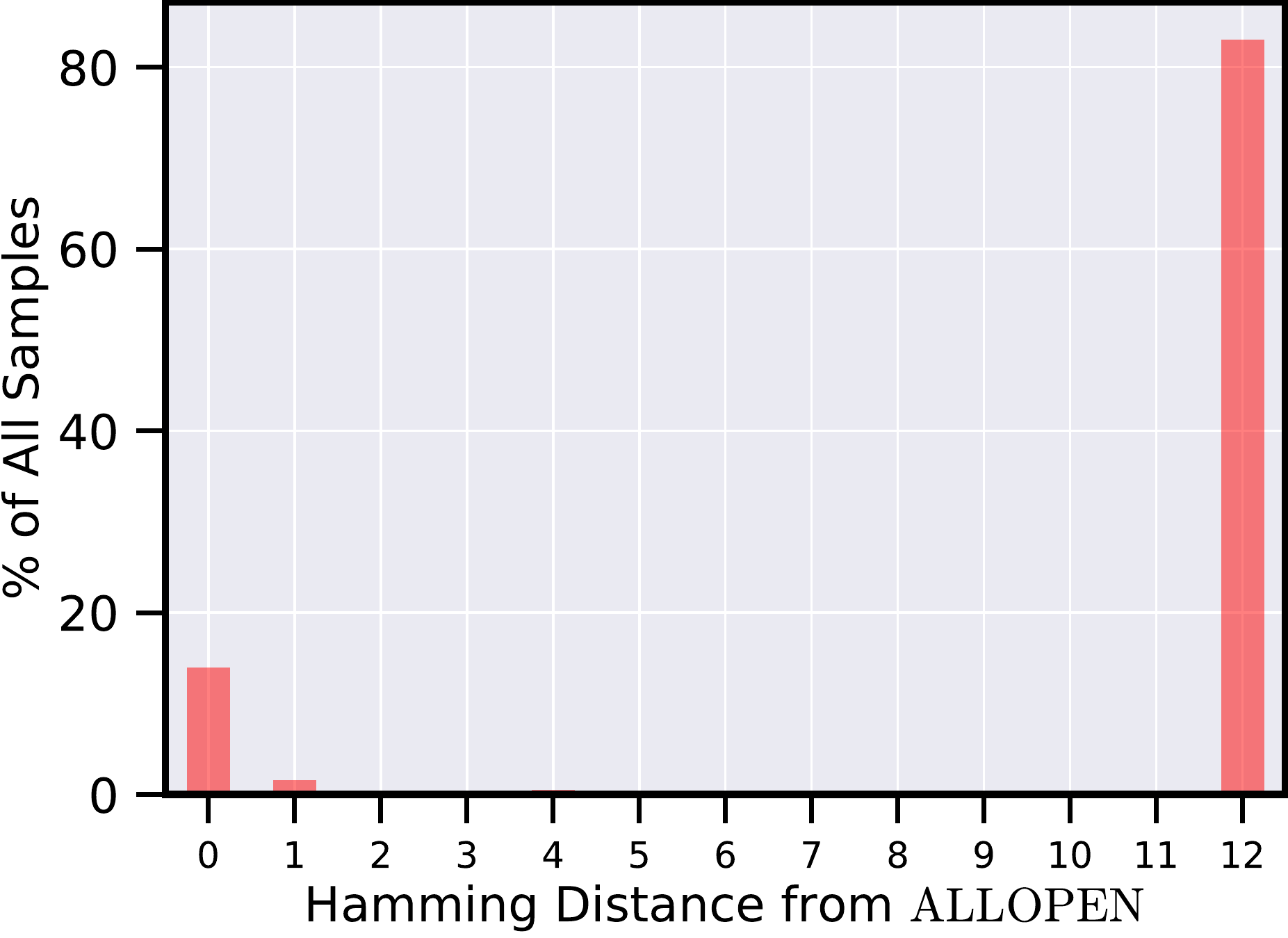}
    \caption{For the training set (purple curve in \textbf{(a)}), we visualize the percentage of observed vent configurations by plotting their Hamming distance from the all-open configuration. With twelve vents, each of which can be either open or closed, we can only have a total of thirteen Hamming distances, with 0 denoting all-open and 1 denoting all-closed.}
    \label{fig:vae_hds_train}
\end{subfigure}
\caption{Identifying valid, ID samples from the corpus of all toggled vent configurations. From samples in the test set with all twelve vents open, we select only those about which we can make confident predictions of MPIQ using our MDN. Only a very small subset of the blue curve is ID.  This makes sense, since the training set consists mostly of samples where the vents are almost all-open or all-closed.  Hence, most samples in the toggled data set with other vent configurations are classified as OoD by our RVAE.}
\label{fig:vae_moneyplot}
\end{figure*}

\begin{figure*}
\begin{subfigure}{0.99\textwidth}
    \centering
    \includegraphics[width=\linewidth]{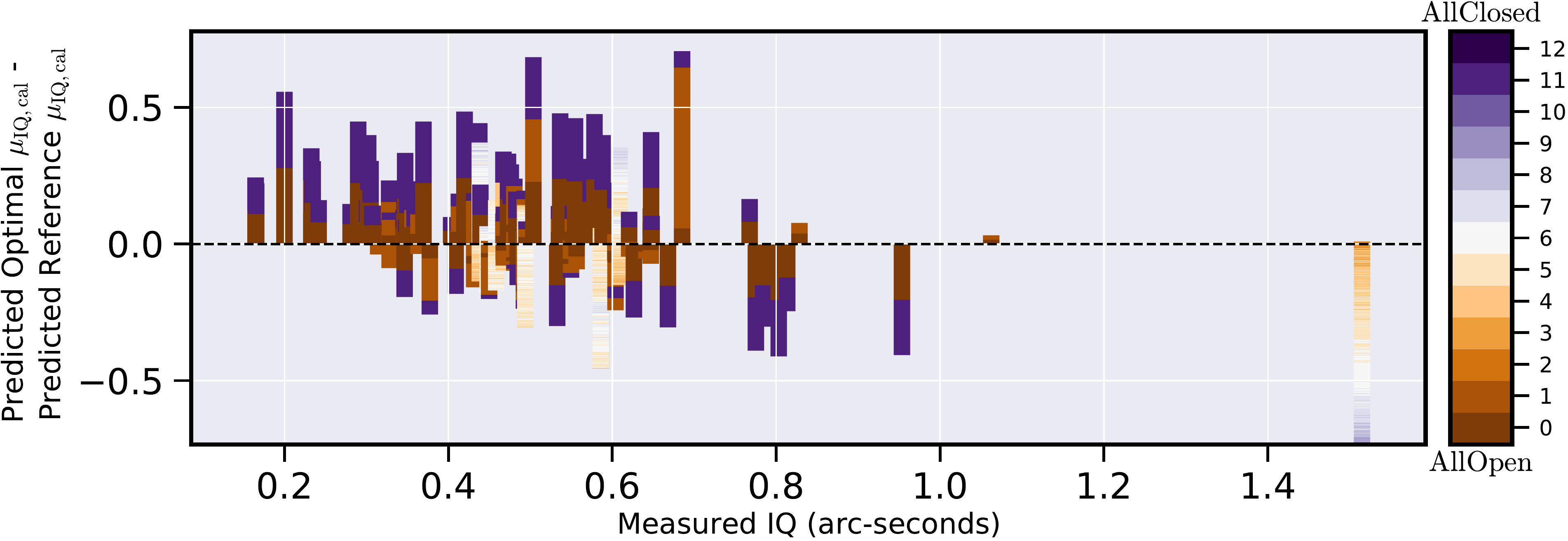}
    \caption{Predicted change in MPIQ corresponding to each possible vent configuration for $100$ randomly selected samples, actuating vent configurations across all in-distribution settings for each. Lower is better. The general trend is that the higher the measured IQ the more room there is for improvement.}
    \label{fig:common_barplot}
\end{subfigure}
\begin{subfigure}{0.99\textwidth}
    \centering
    \includegraphics[width=\linewidth]{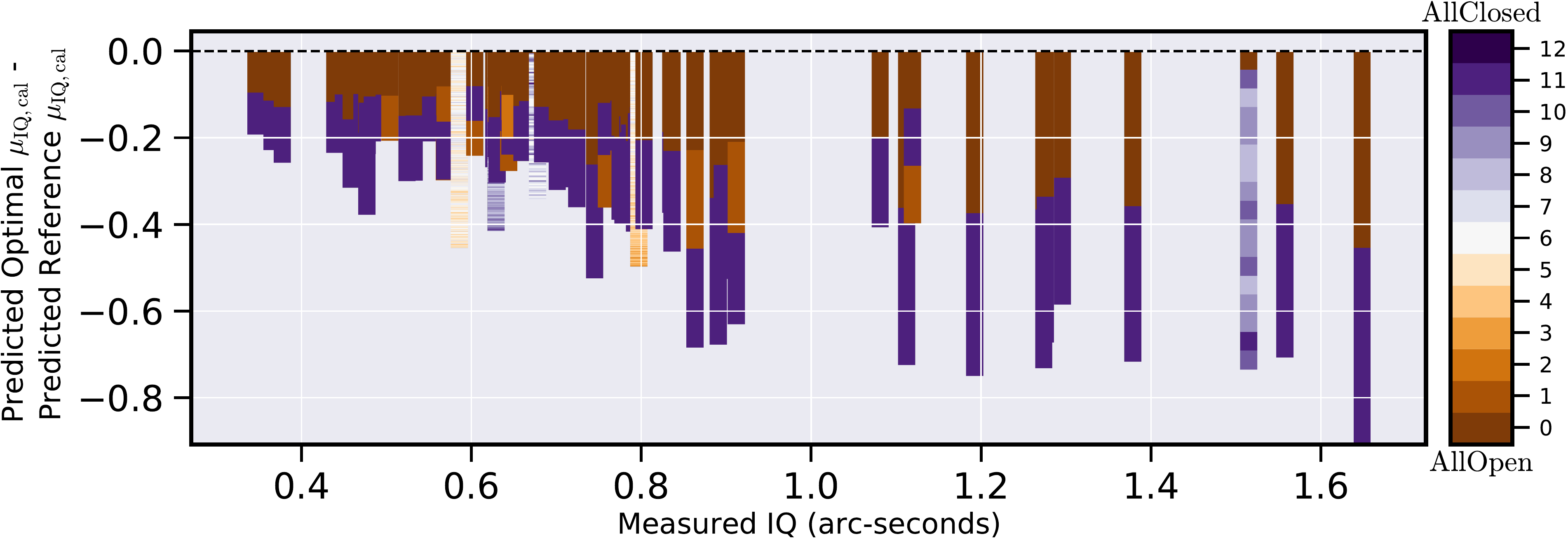}
    \caption{A similar plot to \textbf{(a)} except that now we restrict ourselves to a robust sample, as predicted by total uncertainty.  Of the roughly $600$ samples, according to this measure only $62$ are robust.  For these we see that the predicted change is an improvement (a decrease) in IQ.  Note that this subset of $62$ samples need not overlap significantly or at all with those samples in \textbf{(a)}.}
    \label{fig:common_barplot_robust}
\end{subfigure}
\caption{Visualizing the predicted effect on IQ of optimizing vent configurations.}
\label{fig:common_barplots}
\end{figure*}

In Figure~\ref{fig:MDN_moneyplot_catboost} we show comparative plots for predictions made using our gradient-boosted tree models.  This model is described in Section~\ref{sec:gbdt}. Comparing Figure~\ref{fig:mdn_one_to_one_cat} to Figure~\ref{fig:mdn_one_to_one}, we see that GBDT models significantly underestimate $\sigma_{\rm epis}$. Calibrating the GBDT model using CRUDE does not result in substantial improvement. This is why we use the MDN as our workhorse for MPIQ predictions. For sake of completeness, we compare predictions from {\sc catboost} with those from MDN, and hypothesize reasons for deficient performance of {\sc catboost}, in Appendix \ref{sec:app_mdnvscatboost}.

In Table~\ref{tab:ml_vs_dl_results} we collate the results on the five metrics, for both calibrated and uncalibrated predictions from the MDN. We compare these predictions from those from the boosted-tree GBDT model.  These results demonstrate that the MDN outperforms the GBDT, again supporting the choice to use it as the workhorse model for MPIQ prediction.

\subsection{Actuating dome parameters to improve IQ}\label{sec:resultsImprovingIQ}
\subsubsection{Separating in-distribution (ID) from out-of-distribution (OoD) actuations}\label{sec:results_idvsood}
In addition to predicting MPIQ, one of our driving motivations is to learn how to actuate observatory operating parameters to improve MPIQ.  One set of easily actuatable parameters is the dome vents. Indeed, as mentioned in the discussion of related work in Section~\ref{sec:relatedWork}, fluid flow models were developed in the vent design process to predict the effect on MPIQ of various vent configurations. These models predicted that the optimal MPIQ is achievable with intermediate vent configurations, where the 12 vents are neither all-open nor all-closed \citep{wind_tunnel_test}.
In contrast, in most usage to date vents have been configured either to the all-open or to the all-closed setting. We therefore explore what our MDN model predicts -- how much improvement a modified vent configuration might have on MPIQ reduction. We note that we must be cautious when pursuing this exercise as some vent configurations are not within the training sample. As we describe in Section~\ref{sec:method} and Figure~\ref{fig:vae_hist_mll}, we use the pseudo marginal log likelihood, -L$_{\rm ELBO}$, from the RVAE model as a filter to discriminate in-distribution samples from out-of-distribution ones. In Figure~\ref{fig:common_ood_detections} we justify our choice to use this metric to detect distribution shift. 

In Figure \ref{fig:vae_hist_epis} the pink and cyan curves are the histograms for $\sigma_{\rm epis}$ for the training and test sets for one of ten folds. To simulate out-of-distribution data, we synthesize four data sets. The uniform noise data set, depicted in green, is generated by drawing $50,000 \times 119$ samples, independently, from the uniform distribution between 0 and 1. The constant noise data set, depicted in blue, is generated by drawing $50,000 \times 1$ samples, independently, from the uniform distribution between 0 and 1, and copying this over $119$ times. The orange and red curves are noisy versions of the training data set, where we add Gaussian noise with $\mu = 0$ and $\sigma = 0.10$ and $0.05$, respectively. Since we do not train the MDN with noisy versions of the training data (we use the MoEx data augmentation method only, as described in Section \ref{sec:method}), the uncertainty in predicting MPIQ as a result of noisy versions of training data is classified as epistemic and not aleatoric. The dashed vertical black line marks the $95^{\rm th}$ percentile value for $\sigma_{\rm epis, train}$ -- we classify all values to its right as out-of-distribution. We plot the density in log scale to better capture different ranges. Figure \ref{fig:vae_hist_calepis} is the same as Figure \ref{fig:vae_hist_epis}, except it plots histograms for calibrated epistemic uncertainty. In both figures, it is apparent that epistemic uncertainty, whether calibrated or uncalibrated, is a poor detector of a distribution shift. Distribution shift identification using discriminative models such as the MDN is an area of active research, and we relegate further exploration of this limitation to future work. In this paper, we instead use the RVAE as a proxy for our data distribution, and justify our decision in Figures \ref{fig:vae_hist_mll}, \ref{fig:vae_hist_regret}, \ref{fig:vae_hist_mll_zoomed}, and \ref{fig:vae_hist_regret_zoomed}.

\begin{figure*}
\begin{subfigure}{0.49\textwidth}
    \centering
    \includegraphics[width=\linewidth]{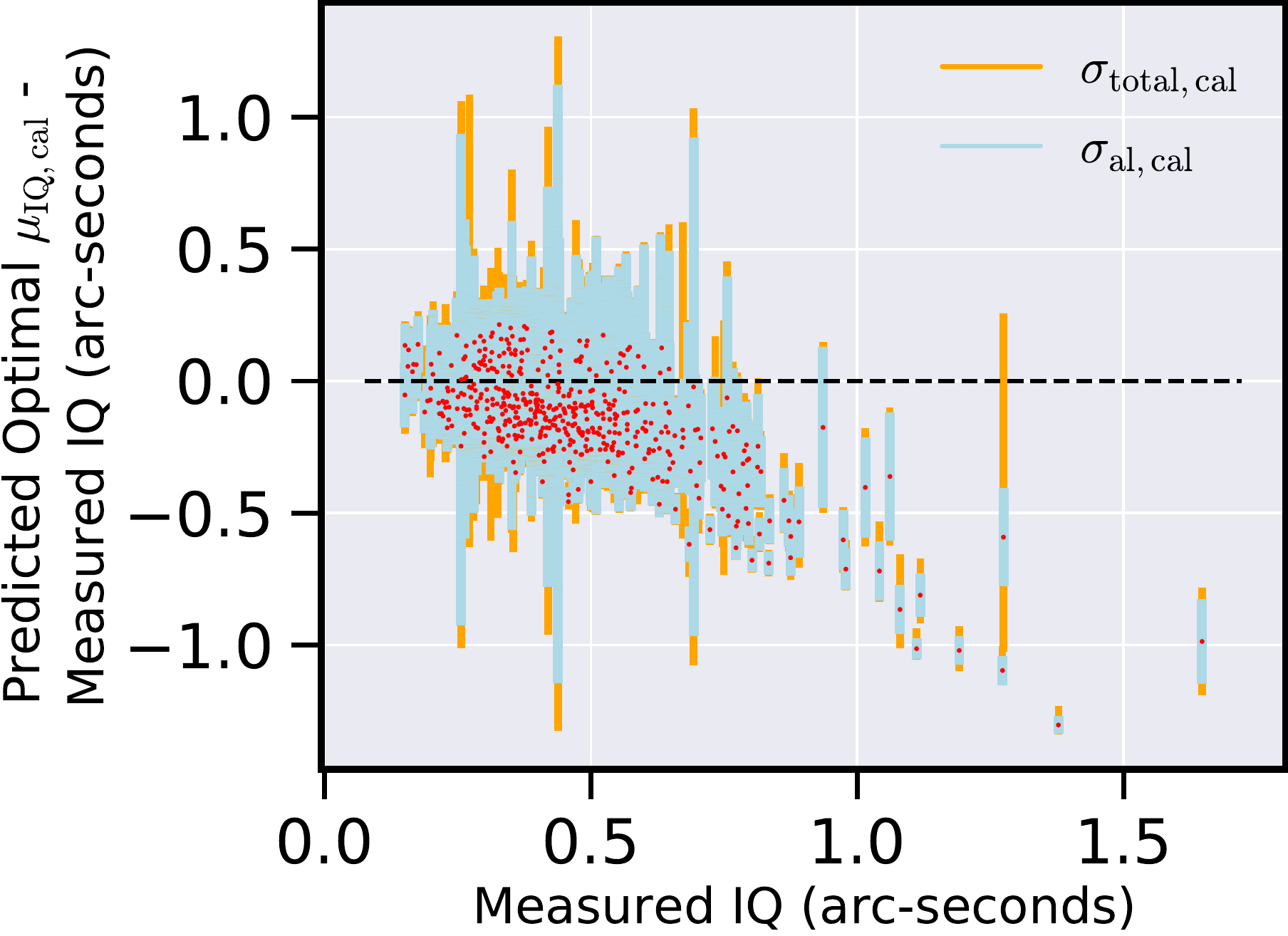}
    \caption{Improvement in MPIQ predicted by the MDN for the optimal vent configurations.  We plot with respect to the measured MPIQ. Negative is better.}
    \label{fig:common_deltaoptimalnominal_vs_nominal_iqs}
\end{subfigure}
\hfill
\begin{subfigure}{0.49\textwidth}
    \centering
    \includegraphics[width=\linewidth]{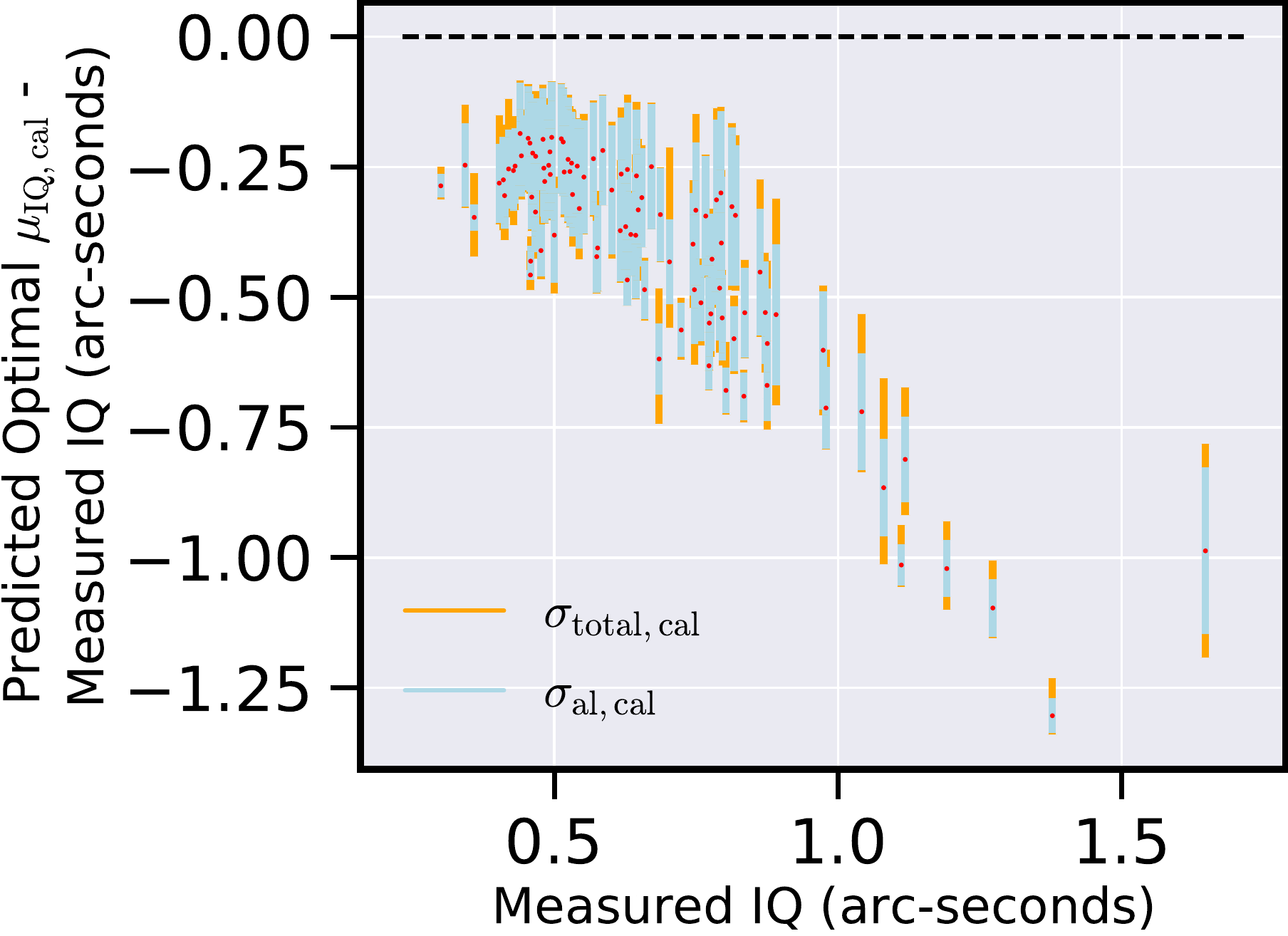}
    \caption{Same as \textbf{(a)}, but  only for those samples where the $84^{\rm th}$ percentile of predicted optimal MPIQ $\leq$ measured MPIQ.}
    \label{fig:common_deltaoptimalnominal_vs_nominal_iqs_robust}
\end{subfigure}
\newline
\begin{subfigure}{0.49\textwidth}
    \centering
    \includegraphics[width=\linewidth]{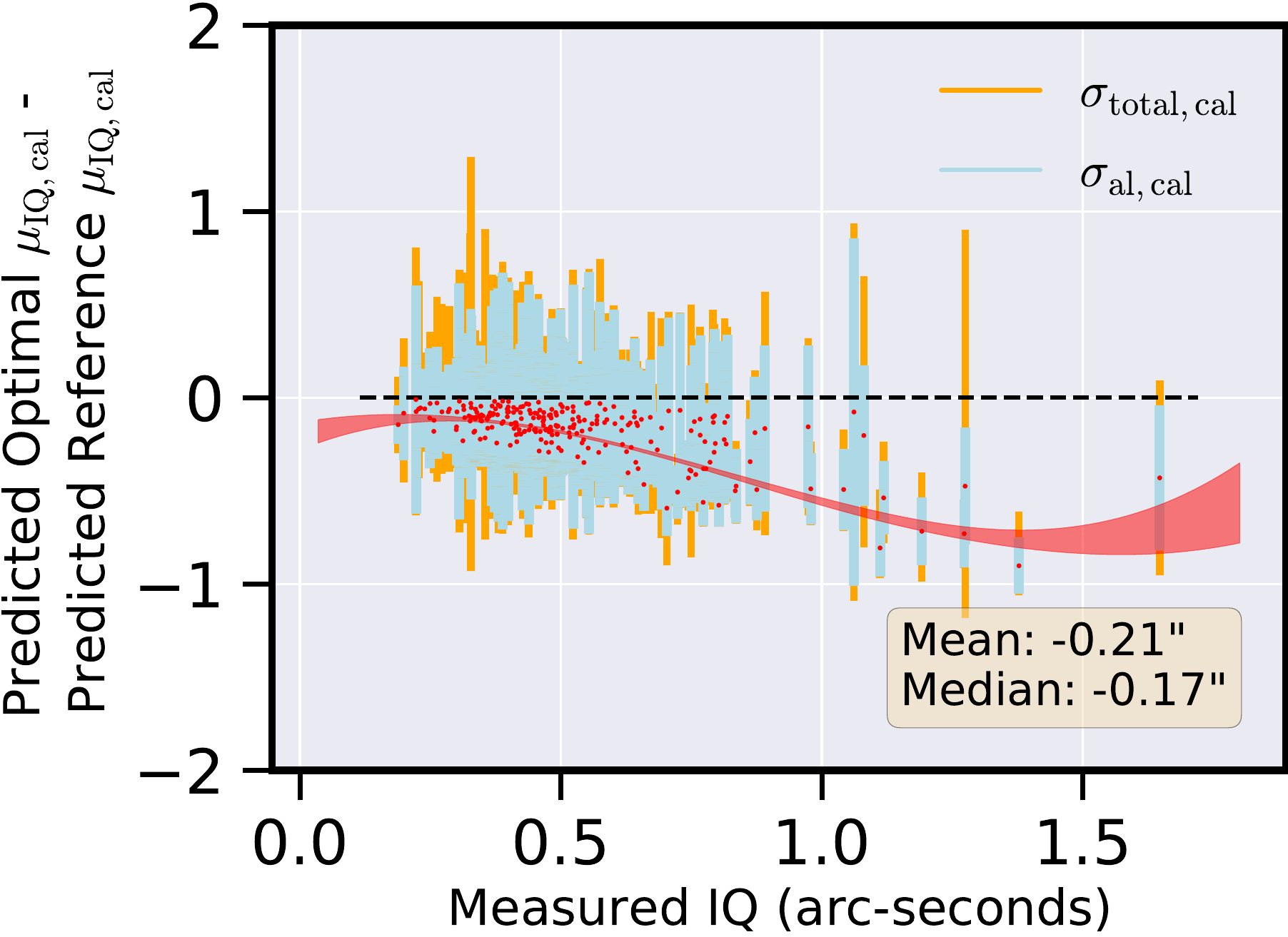}
    \caption{Same as \textbf{(a)}, with the change that here the improvement is calculated with respect to the predicted MPIQ for the \textit{all-open} samples.}
    \label{fig:common_deltaoptimalreference_vs_nominal_iqs}
\end{subfigure}
\hfill
\begin{subfigure}{0.49\textwidth}
    \centering
    \includegraphics[width=\linewidth]{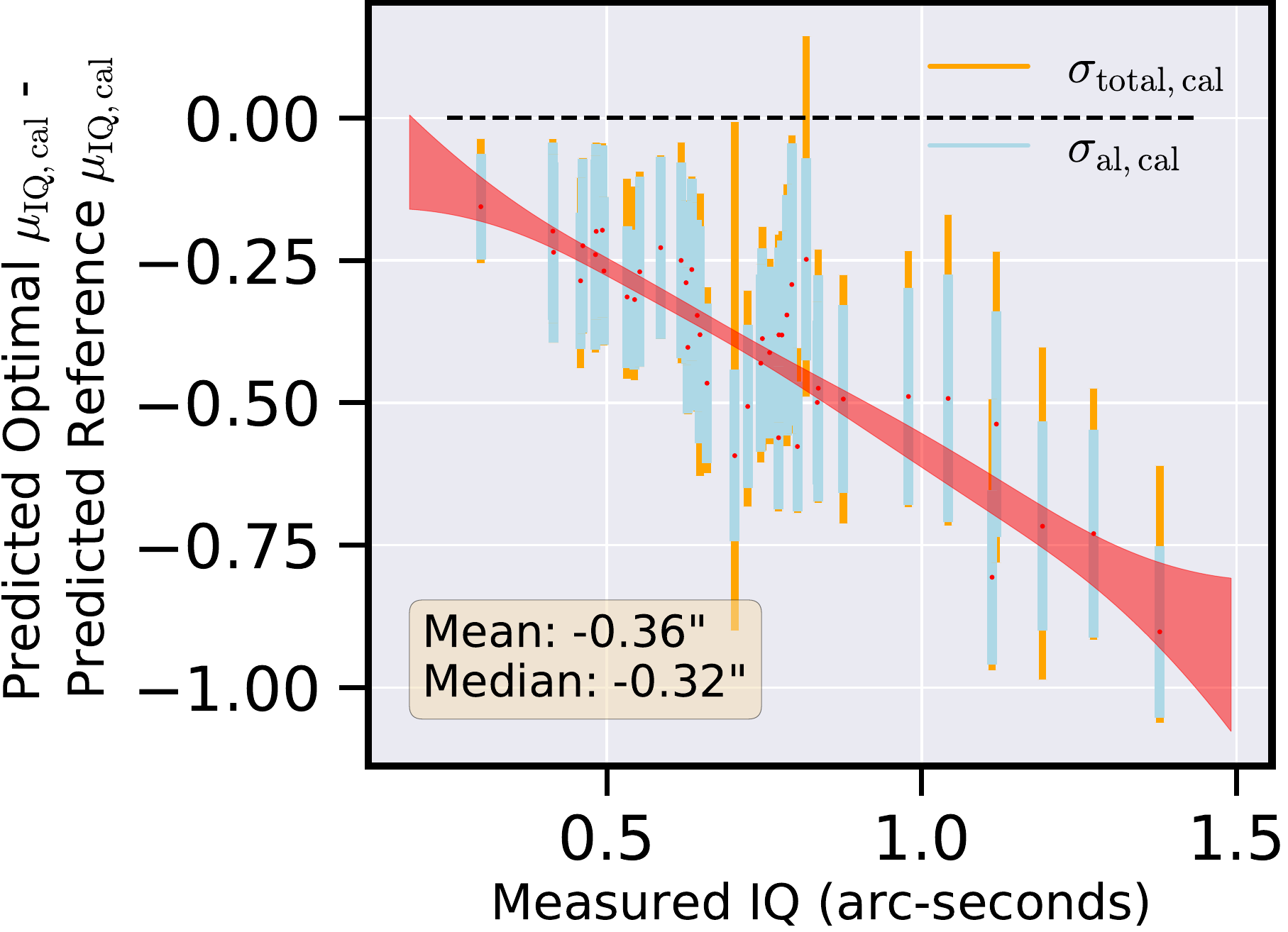}
    \caption{Same as \textbf{(c)}, but for only those samples where the $84^{\rm th}$ percentile of predicted optimal MPIQ $\leq$ $50^{\rm th}$ percentile of predicted reference MPIQ.}
    \label{fig:common_deltaoptimalreference_vs_nominal_iqs_robust}
\end{subfigure}
\caption{We visualize the gains in terms of improved MPIQ prediction that can be achieved by our proposed methodology of toggling the twelve vents open and close individually as a function of environmental and observatory characteristics. The baseline configuration is all-open. After restricting ourselves to a subset of in-distribution ``togglings'', in \textbf{(a)} and \textbf{(b)} we plot the improvement over the measured MPIQ values, whereas in \textbf{(c)} and \textbf{(d)} we plot the improvement over MPIQ values predicted for the same samples in \textit{all-open} configuration. In \textbf{(b)} and \textbf{(d)}, we sub-sample the data points from \textbf{(a)} and \textbf{(c)} respectively, only presenting those samples for which we are quite confident in our estimates. In \textbf{(c)}, the several $y=0$ red dots in the left-half of the plot signify that for those samples, the \textit{all-open} vent configuration is in fact the optimal vent configuration. Finally, we present third-order polynomial fits in \textbf{(c)} and \textbf{(d)}, and estimate total gains achievable using our predicted, optimal vent configurations, using weighted mean and weighted median metrics. These fits are used to calculate average expected reduction in observing times to achieve a fixed signal-to-noise ratio (SNR).}
\label{fig:common_moneyplot1}
\end{figure*}

In Figures \ref{fig:vae_hist_mll} and \ref{fig:vae_hist_mll_zoomed}, we plot the L$_{\rm ELBO}$ for the same data sets described above; the latter figures focuses on the ``hard'' cases of noisy training data. The samples to the left are out-of-distribution. It is immediately clear that the marginal likelihood is a much better distribution shift detector than is the epistemic uncertainty.  This follows because the uniform and constant noise data sets are very easily separable, and the supports of the two Gaussian datasets are both also almost completely to the left of the vertical line. In Figures \ref{fig:vae_hist_regret} and \ref{fig:vae_hist_regret_zoomed}, instead of using L$_{\rm ELBO}$ as the discriminative metric, we instead use the {\rm pseudo log likelihood regret}. Proposed in \cite{likelihood_regret}, likelihood regret for a sample is derived by passing it through the trained RVAE and caching L$_{\rm ELBO}$. We then fix the weights of the decoder and train the encoder to minimize L$_{\rm ELBO}$ for  that single sample. The difference between L$_{\rm ELBO, sample}$ and L$_{\rm ELBO}$ is the pseudo log likelihood regret.  In~\cite{likelihood_regret} this is shown to be a better detector of out-of-distribution samples than L$_{\rm ELBO}$. By definition, regret is always non-negative. Comparing Figures \ref{fig:vae_hist_mll_zoomed} and \ref{fig:vae_hist_regret_zoomed}, we verify that regret is indeed a better separator -- the pink curve is less spread out, and the modes of both the red and orange curves are farther away from the black line. Given these results, it is natural to question our design choice of using L$_{\rm ELBO}$ as the metric we used to identify out-of-distribution samples. We use L$_{\rm ELBO}$ rather than regret because while pseudo log likelihood regret is more robust, it also takes about $50$ times longer to calculate than L$_{\rm ELBO}$. This is because the calculation of regret  requires retraining of the encoder, once per input sample.  Even if we divide our GPU\footnote{NVIDIA Titan RTX 24Gb} into multiple virtual cores for parallel processing, it took about two hours to calculate regret for $4096$ samples ($4096 = 2^{12}$ is the number of permuted vent configurations possible for each input sample). For these computational reasons we use L$_{\rm ELBO}$.  Depending on computing resources, in the future we may move to distributed computing framework to make the use of regret practical.

\subsubsection{Predicted reduction in MPIQ using only on in-distribution (ID) vent configurations}\label{results_idventconfigs}
We now use L$_{\rm ELBO}$ to identify identify the vent settings that are not too ``out-of-distribution'' for which our model will be able to make reliable MPIQ predictions.  As we will develop in the following, these robust predictions indicate that substantial MPIQ improvement is possible by optimizing the vent configuration.  In future work we plan to extend our dataset to reduce the set of out-of-distribution vent configurations, thereby enabling a wider range of reliable predictions, and helping us to realize even greater MPIQ improvements.

\begin{figure*}
\begin{subfigure}{0.49\textwidth}
    \centering
    \includegraphics[width=\linewidth]{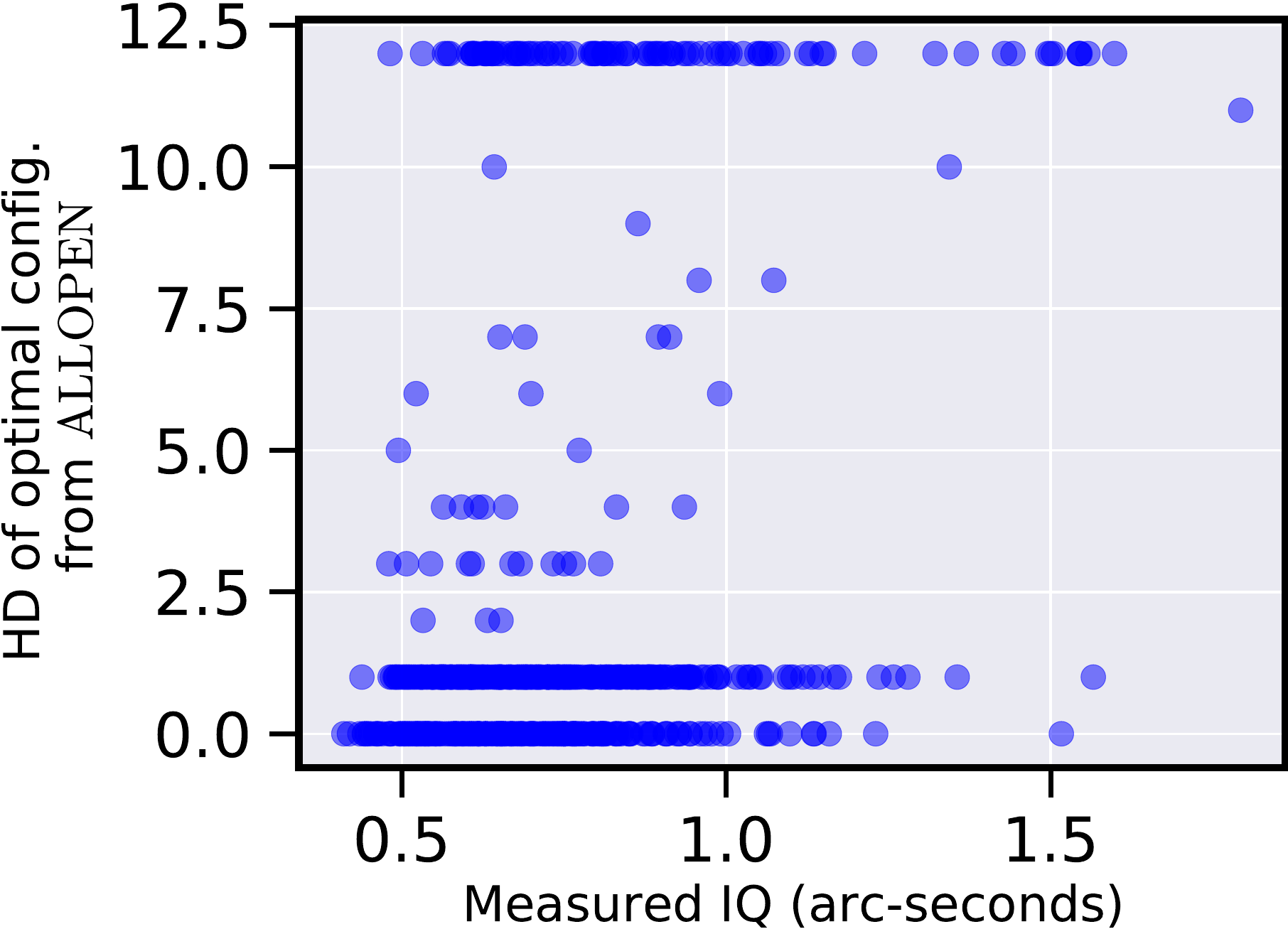}
    \caption{Hamming distance of optimal vent configurations from the all-open configuration.  We plot versus measured MPIQ. Low values on the y-axis (close to 0) indicate that majority of the vents are open.  High values (close to 1) indicate that most of vents are closed.}
    \label{fig:common_hds_vs_iqs}
\end{subfigure}
\hfill
\begin{subfigure}{0.49\textwidth}
    \centering
    \includegraphics[width=\linewidth]{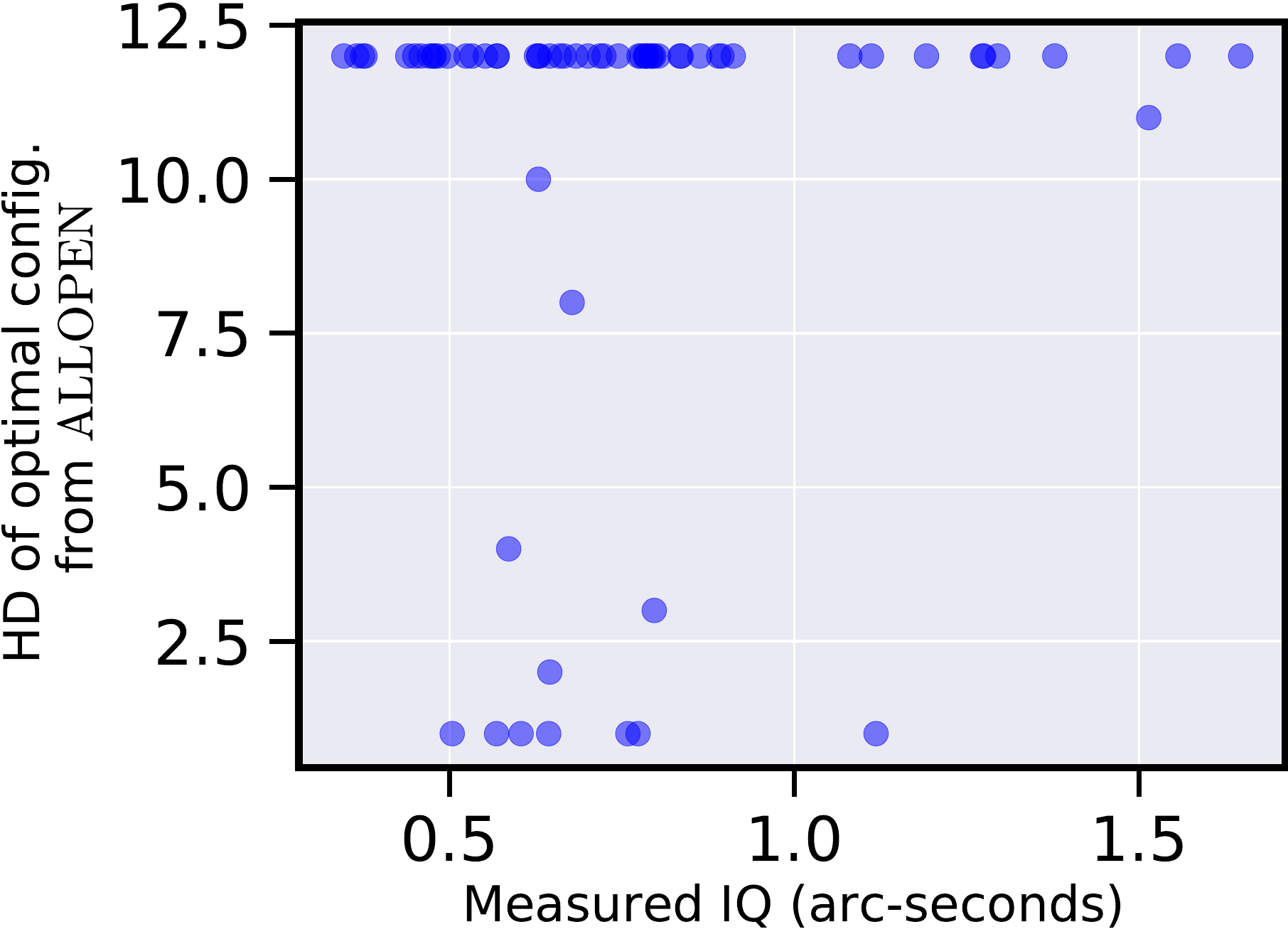}
    \caption{Same as \textbf{(a)}, but only for those samples where the $84^{\rm th}$ percentile of the predicted optimal MPIQ probability distribution function $\leq 50^{\rm th}$ percentile of the predicted reference MPIQ PDF. These are the same samples used to plot Figure \ref{fig:common_deltaoptimalreference_vs_nominal_iqs_robust}. We call these {\it robust} samples.}
    \label{fig:common_hds_vs_iqs_robust}
\end{subfigure}
\newline
\begin{subfigure}{0.49\textwidth}
    \centering
    \includegraphics[width=\linewidth]{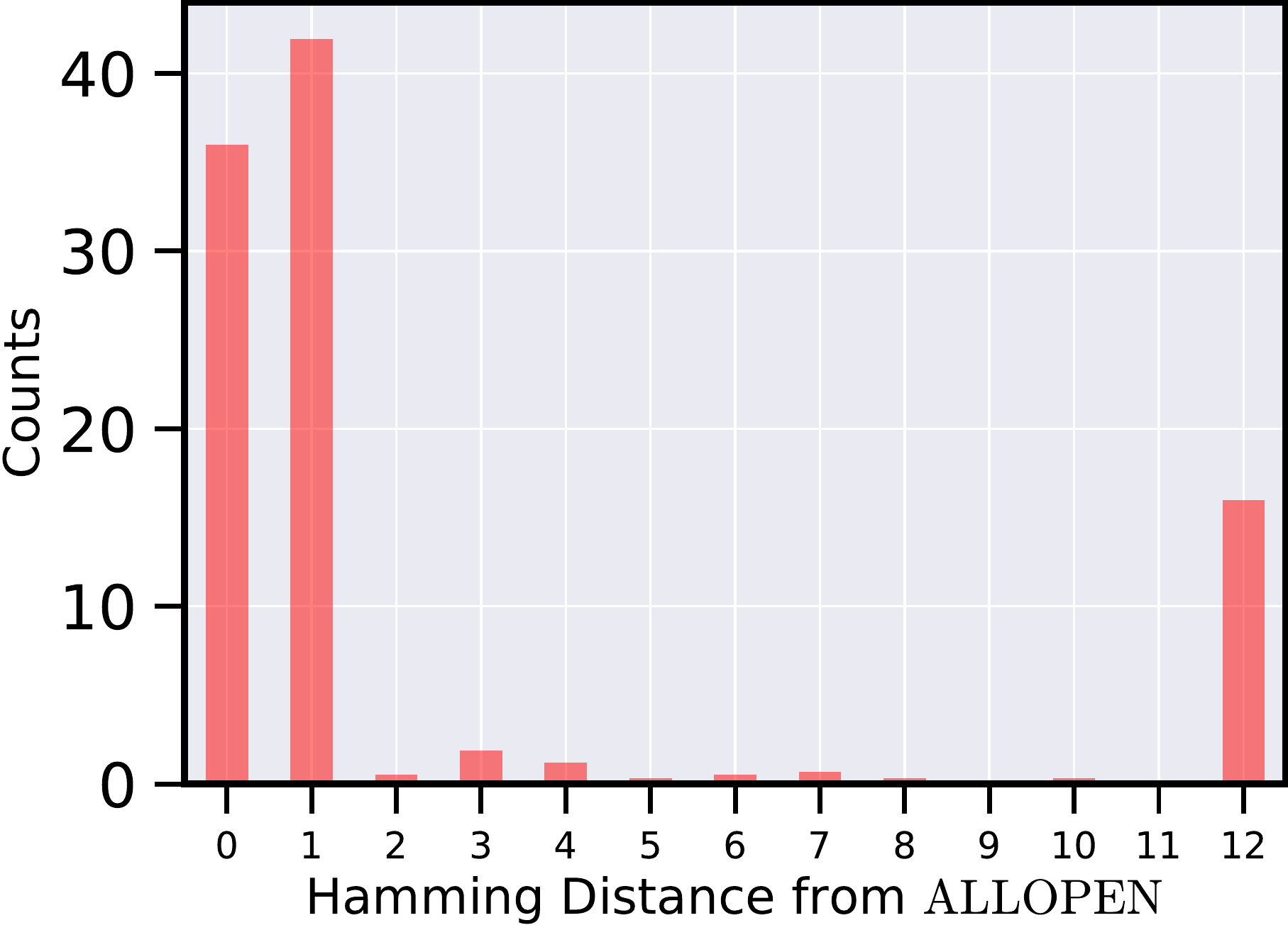}
    \caption{Histogram of Hamming distances of predicted optimal vent configurations. This is the histogram of the y-axis values in \textbf{(a)}.}
    \label{fig:vae_hist_hds}
\end{subfigure}
\hfill
\begin{subfigure}{0.49\textwidth}
    \centering
    \includegraphics[width=\linewidth]{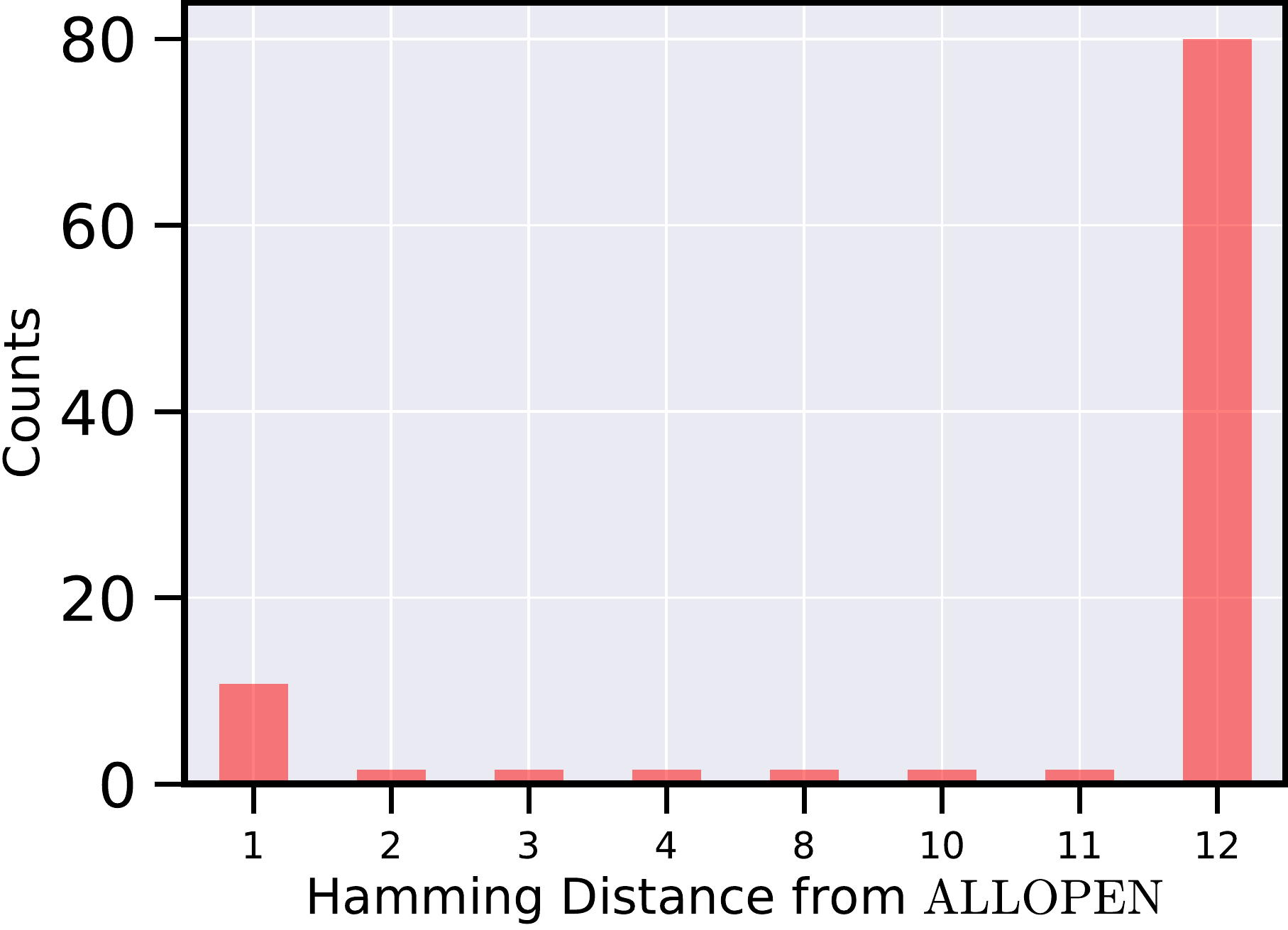}
    \caption{Histogram of Hamming distances of predicted optimal vent configurations for the {\it robust} subset. This is the histogram of the y-axis values in \textbf{(b)}.}
    \label{fig:vae_hist_hds_robust}
\end{subfigure}
\caption{Distribution of predicted optimal vent configuration.  Distribution is measured as Hamming distance from the all-open baseline.}
\label{fig:common_moneyplot2}
\end{figure*}

Figure \ref{fig:vae_moneyplot} demonstrates  results from the process we use to identify, among all possible vent configurations, those for which we can make reliable MPIQ predictions.  By this process we filter out those data records that are OoD. (The workflow that led to these results was described towards the end of Section \ref{sec:putting_it_all_together} and is illustrated in Figure \ref{fig:cfht_rvae_overview}.)
Is is this restricted, or ``filtered'', set of vent configurations that we use to assess the possible improvement.  In Figure \ref{fig:vae_hist_mll_ood} we show  results for one of the ten splits of $\mathcal{D_{F_S,S_S}}$ and predict the MPIQ that would results for all possible vent settings. On average, each test split of $\sim6000$ data records results in $\sim600$ samples.  For each of these $6000$ samples on average only about $4$  other vent configurations (out of a possible $2^{12} -1 = 4095$) are not OoD given the training distribution. For each of these vent configurations we use the MDN to predict the MPIQ three-tuple ($\mu$, $\sigma_{\rm al}$, and $\sigma_{\rm epis}$).  We compare these predictions to the MPIQ three-tuple predictions for the respective base samples (with all vents open).
The results of this exercise, which we will discuss next, are presented in Figures \ref{fig:common_barplots}, \ref{fig:common_moneyplot1} and \ref{fig:common_moneyplot2}. 

In Figure~\ref{fig:common_barplot} we plot bar-charts for the change in MPIQ with optimal vent configurations, with respect to the predicted (calibrated) mean MPIQs for the respective reference test samples with all vents open. For each of the 10 train-test splits of $\mathcal{D_{F_S,S_S}}$ we recall that there are $\sim600$ viable all-open data samples.  Of these viable samples we randomly sub-sample $100$ and, for each of these, predict their mean MPIQ for the all-open configuration using our MDN. We then calibrate these predictions using the CRUDE method, as described in Section \ref{sec:DL_probability_calibration}. We note this prediction will be somewhat different from the measured MPIQ associated with these data records. Finally, we make predictions for each of the ID vent configurations for all 100 samples (roughly $100 \times 4 = 400$) and subtract each of these MPIQ predictions from the predicated calibrated median MPIQs for the all-open configuration.  In Figure~\ref{fig:common_barplot} we plot these difference against the measured MPIQ values. Values above dashed zero-level imply a worsened (predicted) MPIQ in comparison to the baseline of keeping all twelve vents open.  Values below the dashed zero-level suggest that another ID vent configuration will likely result in reduced seeing. Note that we predict the difference in {\it predicted} seeing levels as these are the levels the model would predict were it {\it not} to have a measurement of MPIQ for the baseline all-open configuration.  While in our data set we {\it do} have the baseline MPIQ, in real-time operation that baseline MPIQ value would not be available prior to the observation when the observer would be using the model to decide how to actuate the vent configurations.

To better understand how to read the vent configurations that lead to an improvement in predicted MPIQ, we have color-coded Figure~\ref{fig:common_barplot}. Bars that are dark purple correspond to the all-close configuration; dark brown to all-open. The color gradient corresponds to Hamming distances of the configuration vectors from all-open.  As one would suppose, the model predicts all-closed to be a worse setting more often than not.  This is in keeping with the original motivation for installing the vents, discussed in Section \ref{sec:relatedWork}.  By and large, opening vents improves MPIQ by allowing air currents built up inside the dome to flush. All-close is the same as having no vents, thus represents the  scenario that was meant to improve upon by installing the vents. As we consider higher measured MPIQs (moving from left to right on the x-axis), we see that the optimal configurations tend to be closer to all-close. This is also in line with intuition previously developed at CFHT.  Higher measured MPIQs are typically obtained in high wind speed scenarios, where it makes sense to close the vents.

Figure~\ref{fig:common_barplot_robust} consists of \textit{only} those test samples from the same $\sim600$ that were used to draw from in Figure~\ref{fig:common_barplot}, where the predicted, calibrated upper $84^{\rm th}$ quantiles of the total uncertainties ($\mu_{\rm cal} + \sigma_{\rm total, cal}$) are $\leq$ the predicted calibrated medians for the respective test samples with all vents open. Our idea here is to create a {\it robust} subset. On average, for all 10 train-test splits, this results in $62$ \textit{robust} samples. We emphasize that that this is a second, distinct sampling, with no guarantees of overlap with the $100$ randomly selected samples used to plot Figure \ref{fig:common_barplot}. If we were to mis-actuate the vents and decrease the MPIQ, there would be a negative effect on the downstream science applications.  To mitigate this risk here we chart only those instances where, if our models were to be put into production, we would be very confident in directing the telescope operator to move the vents according to our predictions. From Figure~\ref{fig:common_barplot_robust} we observe that all-open is not the optimal configuration in most situations.  This is true even at lower measured MPIQ values. In fact, significant gains in MPIQ can be realized  by switching each observation from the all-open configuration to the best configuration (from out limited choice of in-distribution and therefore ``viable'' configurations) is significant. When we consider higher MPIQs ($\sim \geq 1''$), the optimal configuration is likely to be all-close. 

\subsubsection{Best achievable MPIQ: A new regime}\label{results_bestmpiq}
In Figure \ref{fig:common_moneyplot1} we present results on the improvement in MPIQ predicted by our MDN given the (hypothetically) optimal vent configurations selected from the restricited set of in-distribution configurations selected by the RVAE.  Figures~\ref{fig:common_deltaoptimalnominal_vs_nominal_iqs} and~\ref{fig:common_deltaoptimalnominal_vs_nominal_iqs_robust} plot the improvement versus measured IQ, while Figures~\ref{fig:common_deltaoptimalreference_vs_nominal_iqs} and~\ref{fig:common_deltaoptimalreference_vs_nominal_iqs_robust} plot the improvement versus predicted and calibrated median MPIQ. Further, the two right-hand plots, Figures~\ref{fig:common_deltaoptimalnominal_vs_nominal_iqs_robust} and~\ref{fig:common_deltaoptimalreference_vs_nominal_iqs_robust} plots the improvement for {\it robust} sub-samples (the samples from the $84^{\rm th}$ quantiles) discussed in the last paragraph.
We note two differences here, compared to Figure \ref{fig:common_barplots}. First, we use all $\sim600$ viable test samples (for Figures \ref{fig:common_deltaoptimalnominal_vs_nominal_iqs} and \ref{fig:common_deltaoptimalreference_vs_nominal_iqs}) in a given train-test split, and not sub-sample of 100; in Figure \ref{fig:common_barplots} we were forced to sub-sample due to space constraints. Second, here we show the difference in predicted optimal MPIQs with respect to both the predicted all-open MPIQs, and the measured MPIQs. For Figures \ref{fig:common_deltaoptimalnominal_vs_nominal_iqs} and \ref{fig:common_deltaoptimalreference_vs_nominal_iqs}, we pick the optimal vent configuration for each of the $\sim600$ viable samples, use the MDN to derive calibrated predictions of MPIQ, and plot the difference between predicted calibrated optimal median and predicted calibrated median for the corresponding test sample with all twelve vents open. For Figures \ref{fig:common_deltaoptimalnominal_vs_nominal_iqs_robust} and \ref{fig:common_deltaoptimalreference_vs_nominal_iqs_robust}, we do the same, but only for the \textit{robust} samples from the $\sim600$ samples. We observe that as we consider larger measured MPIQ values there is increased advantage to optimizing the vent configuration.

In the four sub-figures in Figure~\ref{fig:common_moneyplot2} we visualize the distribution of Hamming distances for the optimal vent configurations chosen for the samples in Figures \ref{fig:common_moneyplot1}. Either from the density of points plotted in Figure~\ref{fig:common_hds_vs_iqs}, or from the distribution of Hamming distances from the all-open configuration plotted in Figure~\ref{fig:vae_hist_hds}, we observe that most optimal vents configurations are close to either all-open or all-closed.  This should not come as a surprise.  As discussed earlier, most vent configurations that are {\it not} close to either the all-open or all-closed configurations were OoD and so were filtered out of the ``viable'' set of configurations for which we consider the MPIQ predictions.  At the risk of repeating some of the broader context provided earlier, the fact that the intermediate vent configurations are OoD is a result of the way the observatory has been operated to date.  Most often the vents have been configured either all-open or all-closed, to a large degree because observers have had no reasoned methodology to follow to choose alternate configurations.  The training data we have access to therefore clusters around the all-open and all-closed configurations.  In a sense then, Figure~\ref{fig:common_moneyplot2} is  another illustration of a main motivation for our work; we want to expand the range of options for the observers so they can better tune observatory performance.

An important observation from Figures~\ref{fig:common_hds_vs_iqs} and~\ref{fig:vae_hist_hds} is that the model predicts about 60\% of samples would have resulted in improved MPIQ had a different setting been chosen.  Figures~\ref{fig:common_hds_vs_iqs_robust} and~\ref{fig:vae_hist_hds_robust}, which are the predictions for the ``robust'' subset discussed earlier are even more definitive -- a ful 85\% of samples would have benefitted from a different vent configuration.  About half of the adjustments would have be to close a {\it single} vent, while the other half would have closed all 12 vents.  Only a smattering of predictions falls between these two choices.  Of course, as just discussed, the intermediate range is mostly OoD.\footnote{At this point is is very helpful to refer back to Figure~\ref{fig:vae_hds_train} and observe that `single-vent-closed' is the third most frequent vent setting in the training set. This explains why our model finds that this type of close-a-single-vent adjustment is in-distribution.
Further, the $\sim 38\%$ prevalence of one-closed-vent configurations in the robustified results of Figures~\ref{fig:common_hds_vs_iqs_robust} and~\ref{fig:vae_hist_hds_robust} tells us that this option is of great use in improving the (robustly predicted) MPIQ resulst plotted in
Figures~\ref{fig:common_deltaoptimalnominal_vs_nominal_iqs_robust} and~\ref{fig:common_deltaoptimalreference_vs_nominal_iqs_robust}.
} That said, the peakiness at one-closed-vent and at all-12-closed-vents is, for us, a strong indication that the true optimal configuration lies somewhere in the middle.  Not till we can collect additional data on this intermediate range to bring it in distribution will we be able to make robust predictions in that range that we can use to advise -- with confidence -- how the observer might more productively operate the telescope. 

\subsection{Quantifying feature importance in prediction of MPIQ}
\label{sec:resultsRelContribIQ}

\begin{figure*}
\begin{subfigure}{0.49\textwidth}
    \centering
    \includegraphics[width=\linewidth]{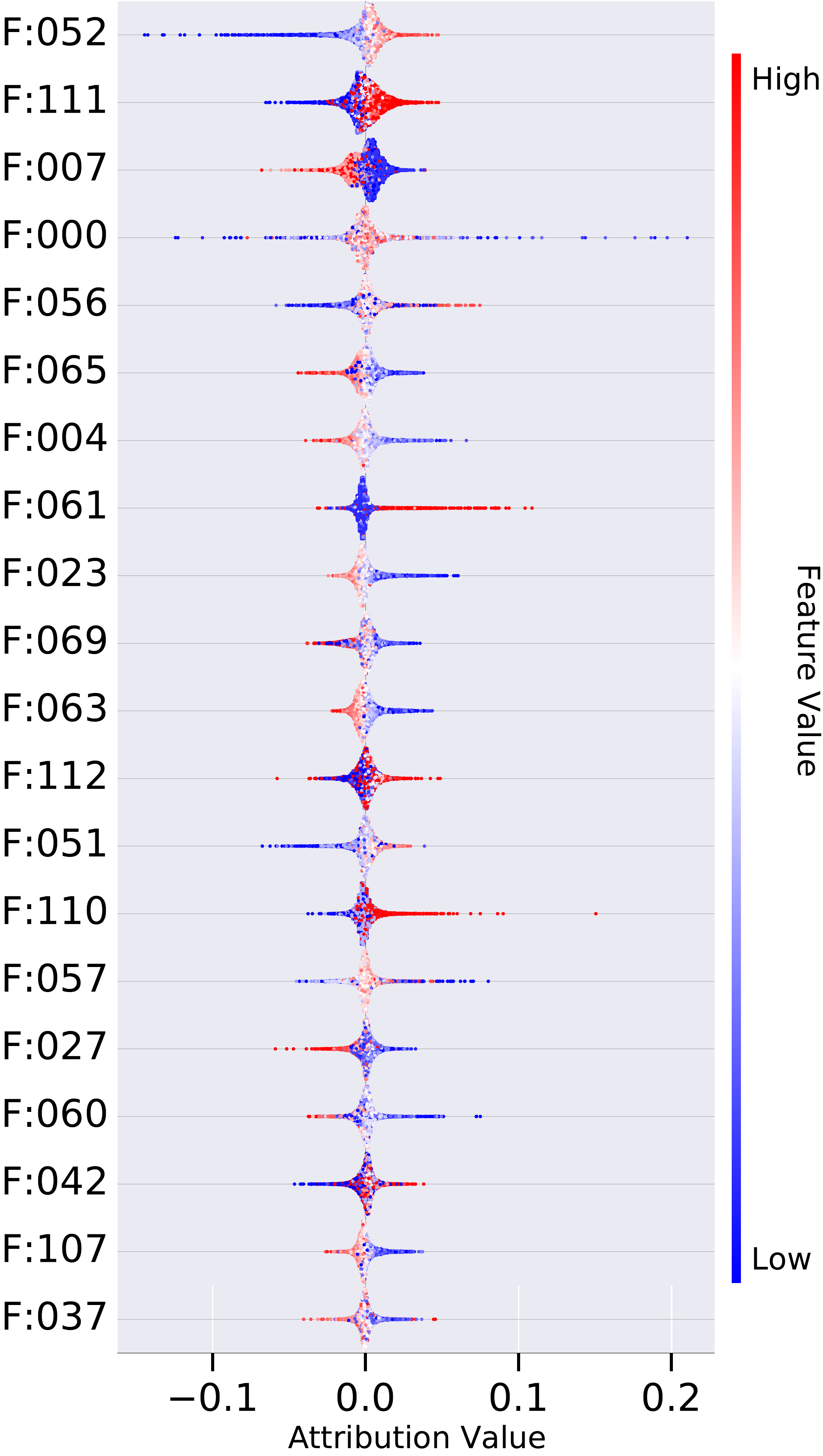}
    \caption{Attribution plot for all samples. This gives a bird's eye view of which features the model sees as most predictive.}
    \label{fig:attplot0}
\end{subfigure}
\hfill
\begin{subfigure}{0.49\textwidth}
    \centering
    \includegraphics[width=\linewidth]{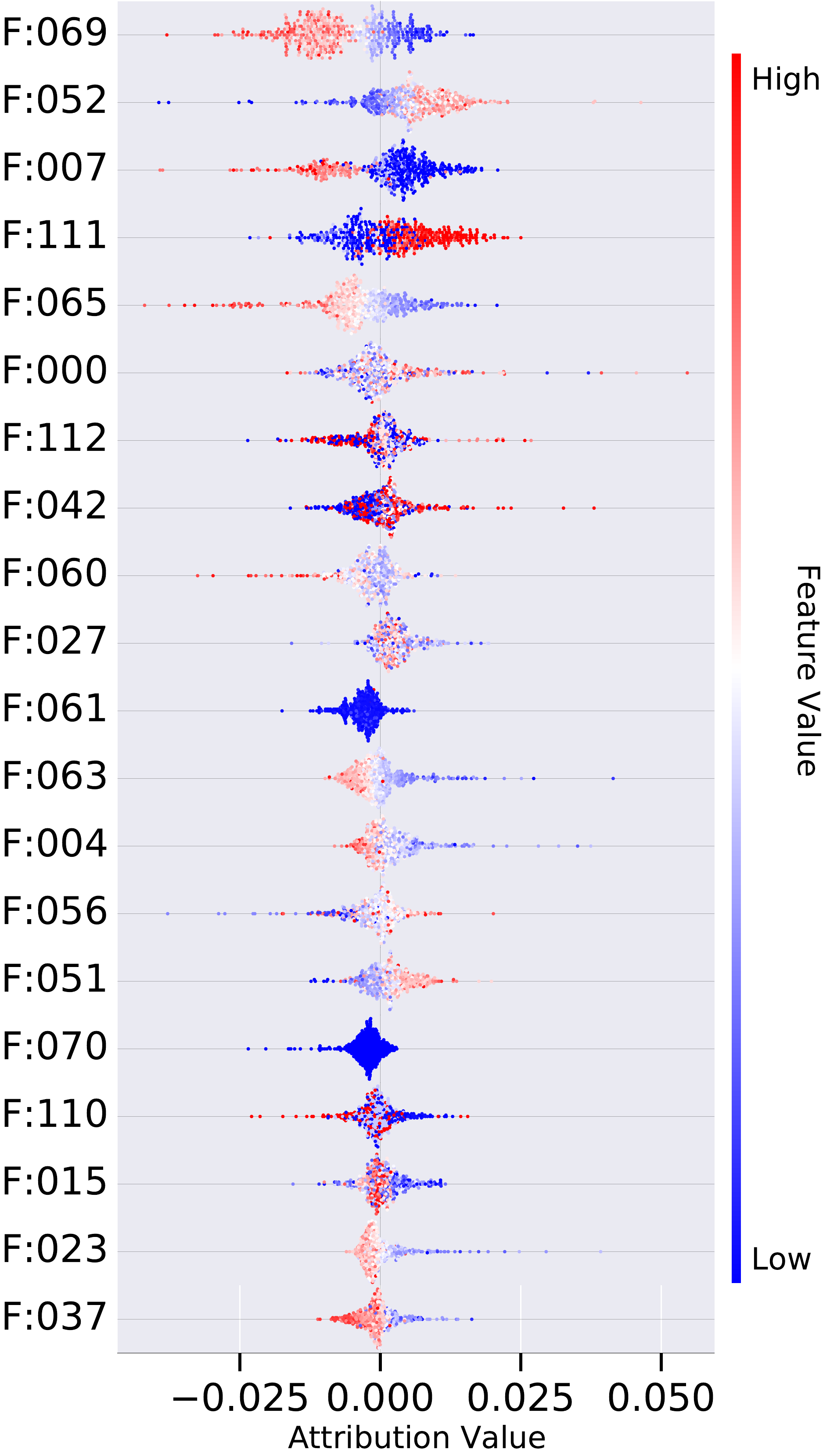}
    \caption{Attribution plot for in-distribution (ID) samples with all vents open. This enables us to isolate those features which make samples ID.}
    \label{fig:attplot1}
\end{subfigure}
\caption{Attribution (a.k.a. summary) plots. By using expected gradients \citep{explaining_explanations_hessians}, we obtain the impact of each feature on the mixture density network's predicted output. These are then collated for all samples in the test set (for a given fold), and collated again for all test sets from all folds. \textbf{(a):} Using \textit{all} $\sim60000$ samples. \textbf{(b):} Using only the $\sim6000$ in-distribution (ID) samples for which there exist vent configurations that result in lower predicted median MPIQ than their counterparts with all 12 vents open.}
\label{fig:attplots}

\captionof{table}{Most predictive features identified in Figure \ref{fig:attplots}.}
\begin{tabular}{cccc}
\toprule
Abbreviation &                                        Feature & Abbreviation &                                      Feature \\
\midrule
       F:000 &                             Barometric pressure &          F:057 &           Rear observing room air temperature \\
       F:004 &                        Catwalk temperature, north &         F:060 &      Thrust bearing surface temperature, south beam \\
       F:007 &                                 Filter central wavelength &         F:061 &                Top ring air temperature, east \\
       F:015 &                                     Current altitude &         F:063 &                     Top ring air temperature, west \\
       F:023 &                             Dome top temperature &         F:065 &         Truss surface temperature, north halfway-up \\
       F:027 &                        Dome wall temperature, west &         F:069 &          Truss surface temperature, west halfway-up \\
       F:037 &           Fourth floor crawlspace air temperature &         F:070 &                                   Vent L1 \\
       F:042 &                                $\sin$(hour of day) &         F:107 &                      Weather tower temperature \\
       F:051 &         Mirror surface temperature, east underside &         F:110 &                    Weather tower wind speed \\
       F:052 &  Mirror surface temperature, south underside spigot &         F:111 &                              $\cos$(week of year) \\
       F:056 &                   Observing room air temperature &         F:112 &                              $\sin$(week of year) \\
\bottomrule
\end{tabular}
\label{tab:attribution_features}
\end{figure*}

As a final contribution, we quantify the relative importance of each of the 119 features in predicted median MPIQ. By leveraging the integrated gradients technique~\citep{explaining_explanations_hessians}, we can attribute a \textit{Shapley} score to each feature for each sample. This score measures the linear change in the predicted output (with respect to the average of the MPIQs across all samples in the training set, which is called the `offset') that is induced by a small change in any given feature (i.e., the gradient). A positive score for a feature \textit{f} in sample \textit{x} implies that in \textit{x}, \textit{f} acts to increase the predicted MPIQ, while a negative score points to \textit{f}'s role in decreasing the predicted MPIQ. The larger the magnitude is of this Shapley'' score, the bigger the role of \textit{f} in determining MPIQ for \textit{x}. We average such scores for all 119 features across all samples of interest, and plot {\it attribution plots} in Figure \ref{fig:attplots}. In Figure \ref{fig:attplot0}, we carry out this exercise for all test samples for each train-test split of  $\mathcal{D_{F_S,S_S}}$, and collate the results from all 10 cycles (assuming the number of splits is 10). In Figure \ref{fig:attplot1}, we repeat this exercise, but only for the significantly smaller test set samples ($\sim600$ per train-test split) that are in-distribution (ID). 

Concentrating on in-distribution samples (\ref{fig:attplot1}), we can see that the most important features in the prediction of MPIQ can be grouped roughly into three groups, related respectively to dome convection, seasonal variations and filter central wavelengths. In the first group, additional dome turbulence can be sourced by air convection, which is sourced by (positive) internal temperature gradients within the dome, with respect to altitude in the dome. Therefore, MPIQ is expected to increase with the mirror temperature that acts as a source of convection, and decrease with the temperature of the upper structures of the telescope (truss), that tends to reduce the temperature gradient. This is precisely what we see in the attribution plots of the two most important features. The third feature in importance shows a correction of the predicted MPIQ with respect to the $\lambda^{1/5}$ law used in Equation~\ref{eqn:iq_filter_plus_zenith_correction}, with a predicted MPIQ larger at smaller wavelength compared to the theoretical scaling. Finally, the fourth ranked feature shows the seasonal variation of image quality, with better average seeing during the summer months due to more clement weather. Further features in the list can most of the time be attributed to one of the groups described above.

\section{Conclusions and Future Work} \label{sec:conclusion}

In this paper we present what we envisage to be the first in a series of studies that will ultimately lead to dynamically optimized scheduling at the Canada-France-Hawaii Telescope. We have initiated that program herein by developing machine-learning based data-driven methods of image quality (IQ) prediction.  
We present results for two models.  The first is a feed-forward mixture density network (MDN) used in conjunction with a robust variational autoencoder. We trained both on a new dataset that comprises eight years of data collected at CFHT since the installation of the dome vents.  The MDN produces probabilistic predictions of image quality, while the autoencoder estimates the marginal distribution of the data. On average, IQ can be predicted to within 0.07'' accuracy based on environmental conditions and telescope operating parameters. By varying the configuration of the dome vents (in an in-distribution way) in response to environmental conditions, our model predicts that IQ can be improved by about 10\% over historical patterns, with the gains increasing when the nominal IQ value is large. For SNR-based observations, this represents gains of up to 10-15\%.  These gains, in turn, can be equated to  approximately 1M USD in operating costs per year of SNR-based observing.  Such gains would be realized in the form of additional observations made and experiments conducted; additional science accomplished.

We see several important avenues for further inquiry. Perhaps most immediate, the improvements in IQ that we present are predicted by extrapolating over hypothetical vent configurations.  While the uncertainties predicted by our model suggests that these out-of-distribution predictions are robust, we need to verify our predictions by collecting additional observation data in these operating regimes. By collecting such data we will be able to extend our model and robustly predict IQ for the more intermediate vent configurations ``half-way'' between all-open and all-closed.  In doing so we aim, finally, to realize the full utility of the dome vents.

Second, in this study we have treated each data record as an independent sample.  In reality of course, exposures are temporally related.  Numerous exposures are collected each night. By treating data records as independent, we do not leverage what we anticipate are quite important temporal relationships extant in the data. By incorporating temporal models, our aim is be to be able to produce real-time robust forecasts of IQ some five-to-twenty minutes into the future.  The realization of such capabilities will enable the adaptive reorganization of a nominal observation schedule, the real-time scheduling protocols we mention in the introduction. 

Connected to the second point, in this paper we work exclusively with data records from MegaCam.  Going forward we plan to augment our data set with records from other CFHT instruments. While MegaCam's IQ measurements are the most accurate, other instruments also measure IQ with acceptable accuracy, and equally importantly, operate when MegaCam is offline. CFHT schedules instruments in blocks or ``runs'' of several consecutive nights, e.g., swapping out instruments twice a month according to their sensitivity to moonlight. The consequence of this is that training data from MegaCam is not temporally contiguous. This makes it more difficult to use in training a scheduler.

Fourth, we do not currently take into account the physical locations of the different sensors. While discarding this spatial information made modeling and data analysis easier for this initial study, it leaves out useful side  information that can connect the placement of sensors and their relative values.  Going forward we will incorporate such information into our models.

Fifth, the approach we take to {\it post-hoc} calibration of epistemic uncertainties -- CRUDE \citep{crude_probability_calibration} -- has its own set of limitations. While state-of-the-art in terms of improving sharpness and calibration, CRUDE implicitly assumes a symmetric distribution of uncertainty.  In our context this means we do not fully leverage the {\it asymmetric} uncertainties output by our $\beta$ posteriors. CRUDE also assumes that, once normalized (by their standard deviations), all  errors are drawn from the same distribution.  Therefore CRUDE weights all data points equally. In our context one implication is that while calibrating the probability distribution function (PDF) for a test sample with nominal (true) IQ of $2''$, which is near the right tail of the distribution (see the red curve in the left sub-figure of Figure \ref{fig:hist_preliminaries}), we are strongly impacted by samples with nominal IQ values near 0.6 arc-seconds, which is the mode of the distribution. This uniformity of treatment is not ideal. We plan to address this shortcoming in the future.

Sixth, we note that we are cautious about the thresholding method we apply to detect out-of-distribution samples (see~Figure \ref{fig:vae_hist_mll_ood}). In this work we take the $95^{\rm th}$ percentile of the pseudo marginal log-likelihoods for the training set samples as the out-of-distribution threshold. However, this is an {\it ad-hoc} choice based on intuition.  We do not claim that it is the optimal method to filter out out-of-distribution samples. We also show that log-likelihood regret is a more accurate metric than is log-likelihood when aiming to separate the two types of distributions.  We refrain from using regret in this work due to practical concerns about run-time. In future work, we will leverage distributed computing to  integrate this superior metric into our pipeline, and we will explore more principled ways to set the threshold.

Finally, in a slightly different direction, we note that a subset of the authors are collaborating with a concurrent and complimentary study of dome seeing at CFHT. In that study direct measurements of local in-dome optical turbulence are being collected using AIRFLOW instruments~\cite{lai2019}. AIRFLOW sensors are always-on optical turbulence sensors and, as discussed in the introduction, turbulence is  highly correlated with instrument IQ, the metric of interest herein. The current work informs the AIRFLOW study in that it can provide insight into which locations sensors should be placed.  Conversely,  data from the AIRFLOW study can provide a new data stream for the current study.   Taken together, these two studies will offer unique insights into the nature of dome seeing and ways that effects which degrade seeing can be mitigated.

\section*{Acknowledgements}
This research was based on observations obtained at CFHT, which is operated from the summit of Maunakea by the National Research Council of Canada, the Institut National des Sciences de l’Univers of the Centre National de la Recherche Scientifique of France, and the University of Hawaii. The authors wish to recognize and acknowledge the very significant cultural role that the summit of Maunakea has always had within the indigenous Hawaiian community. S.D. and S.G. thank CFHT for hosting them at the beginning of the project. S.G. thanks Eric Zelikman (Stanford University), Haleh Akrami (University of Southern California), Joseph Jazinek (Department of Computer Science, University of Washington), and Dr. Yusuke Nomura (Department of Radiation Oncology, Stanford University) for productive e-mail correspondence. S.P. thanks Olivier Lai for fruitful discussions. The authors would like to thank the CFHT staff, and Stephen Gwyn at the Canadian Astronomy Data Centre. Y.S.T. is grateful to be supported by the NASA Hubble Fellowship grant HST-HF2-51425.001 awarded by the Space Telescope Science Institute.

\section*{Data Availability}
The data underlying this article are available in Zenodo \citep{cfht_dataset_zenodo}, at \url{https://dx.doi.org/10.5281/zenodo.5295743}.





\bibliographystyle{mnras}
\bibliography{main.bib} 


\appendix

\section{Workflow and Training Details} \label{sec.workflowFigs}

\begin{figure*}
    \centering
    \includegraphics[width=0.89\textwidth]{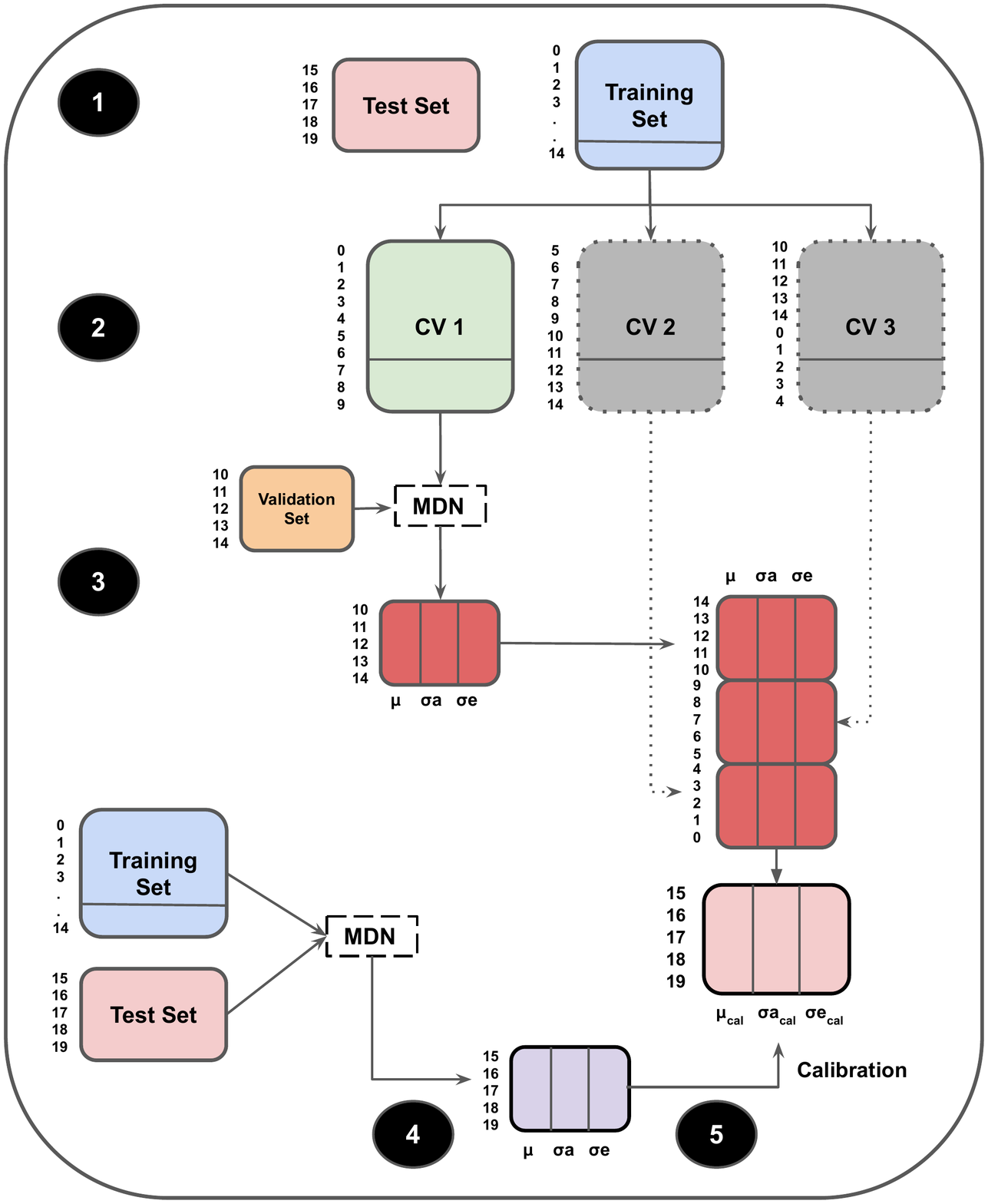}
    \caption{Workflow for generating predictions for MPIQ using our mixture density network (MDN, Figure \ref{fig:rvae_plus_mdn}). \textbf{First}, we divide the input data set $\mathbf{X}$ into $N_{\rm CV,1} = 10$ cross-validation folds, at any point referring to the collation of $N_{\rm CV,1}-1$ of them as $\mathbf{X_{TRAIN}}$, and the remaining fold as $\mathbf{X_{TEST}}$. We repeat this $N_{\rm CV,1}$ times to cover all samples in $\mathbf{X}$, but depict only one such iteration here for illustration. \textbf{Second}, we sub-divide $\mathbf{X_{TRAIN}}$ into $N_{\rm CV,2} = 3$ CV folds. Same as before, $N_{\rm CV,2}-1$ folds are collated while the remaining fold, referred to as the validation set $\mathbf{V}$, is set aside. \textbf{Third}, each of the $N_{\rm CV,2}$ CV folds is used to create $N_{\rm bags} = 3$ `bags' by randomly shuffling its data and picking the same number of samples with replacement. The MDN is trained on one of such folds, and predictions on the validation set give us the $16^{\rm th}$, $50^{\rm th}$, and $84^{\rm th}$ quantile predictions, plus the epistemic uncertainty per sample in $\mathbf{V}$. This process is repeated $N_{\rm CV,2} - 1$ more times to get predictions for all samples in $\mathbf{X_{TRAIN}}$.  \textbf{Fourth}, the MDN is now trained on the entire training set, and predictions collected for samples in $\mathbf{X_{TEST}}$. \textbf{Fifth}, we use {\sc CRUDE} \citep{crude_probability_calibration}, and the predictions from the third step to calibrate the predicted values from the fourth.}
    \label{fig:cfht_mdn_overview}
\end{figure*}

\begin{figure*}
    \centering
    \includegraphics[width=0.89\textwidth]{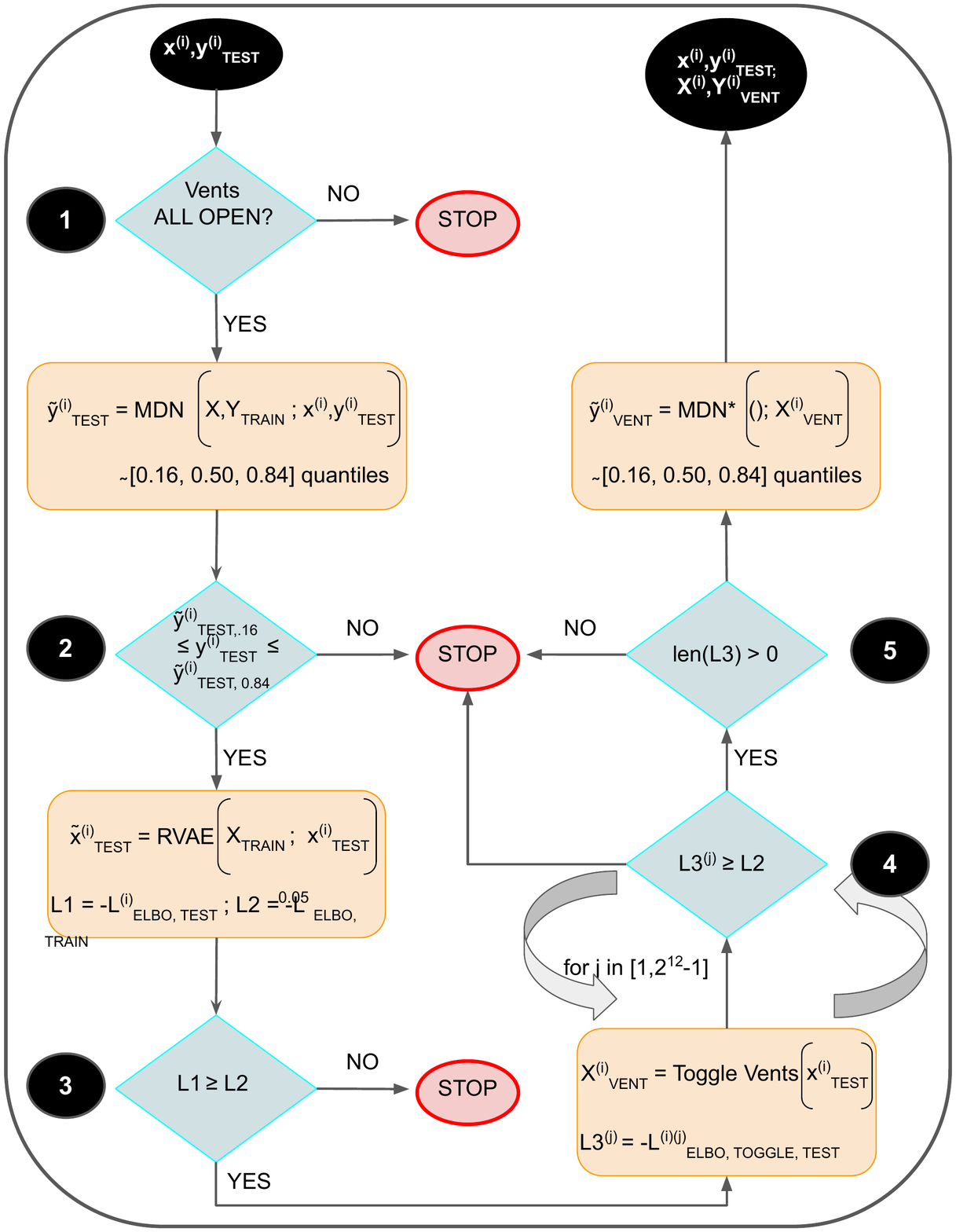}
    \caption{Workflow for identifying acceptable, in-distribution (ID) vent configurations, for each sample in the test set. \textbf{First}, we isolate samples with all twelve vents open. \textbf{Second}, we further pick only those samples for which the predicted $84^{\rm th}$ and $16^{\rm th}$ quantiles, generated by adding the epistemic and aleatoric uncertainties in quadrature, envelope the true MPIQ. `MDN' stands for Mixture Density Network, see Figure \ref{fig:cfht_mdn_overview}. \textbf{Third}, we use the training set with our robust VAE, calculate the $5^{\rm th}$ percentile of the pseudo-marginal log likelihood loss, and use that as a lower cut-off to separate in-distribution test samples from out-of-distribution (OoD) ones. \textbf{Fourth}, for the test samples thus filtered, we generate $2^{12} - 1$ samples by toggling the 12 vents into open and close positions, but skip the all-open configuration since that is the  base case. From these hypothetical samples, we find the ID ones by repeating the procedure of step 3. \textbf{Finally}, we throw out those test samples for which none of the hypothetical cases passed the cut-off test.}
    \label{fig:cfht_rvae_overview}
\end{figure*}

\begin{figure*}
    \centering
    \begin{minipage}{.97\textwidth}
        \begin{subfigure}{0.49\textwidth}
            \centering
            \includegraphics[width=.98\linewidth]{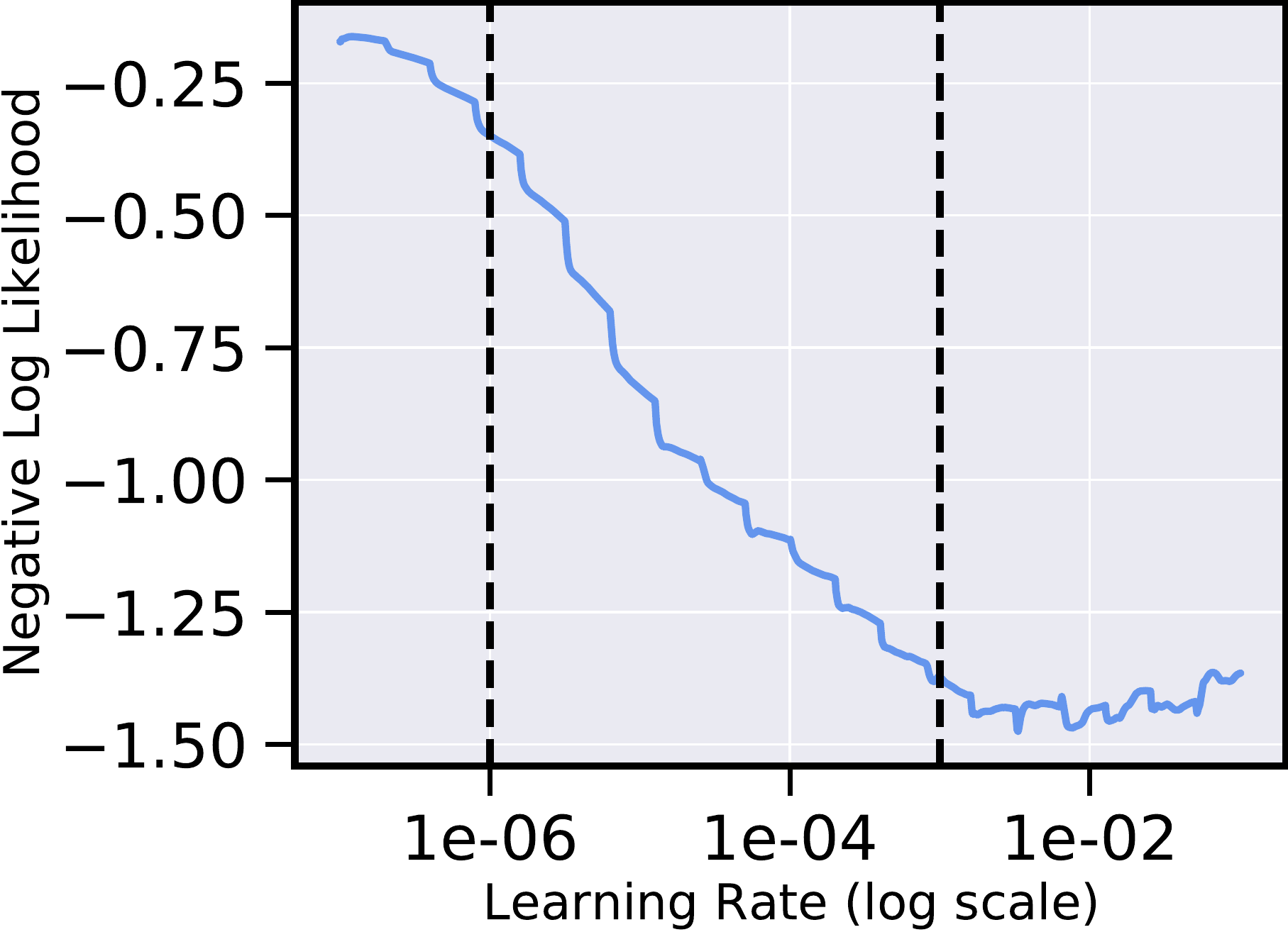}
            \caption{Choosing lower and upper learning rates for the cyclic learning rate scheduler, as described in Section \ref{sec:putting_it_all_together}. The vertical lines indicate the chosen limits of $10^{-6}$ and $10^{-3}$.}
            \label{fig:mdn_find_lr}
        \end{subfigure}
        \hfill
        \begin{subfigure}{0.49\textwidth}
            \centering
            \includegraphics[width=.98\linewidth]{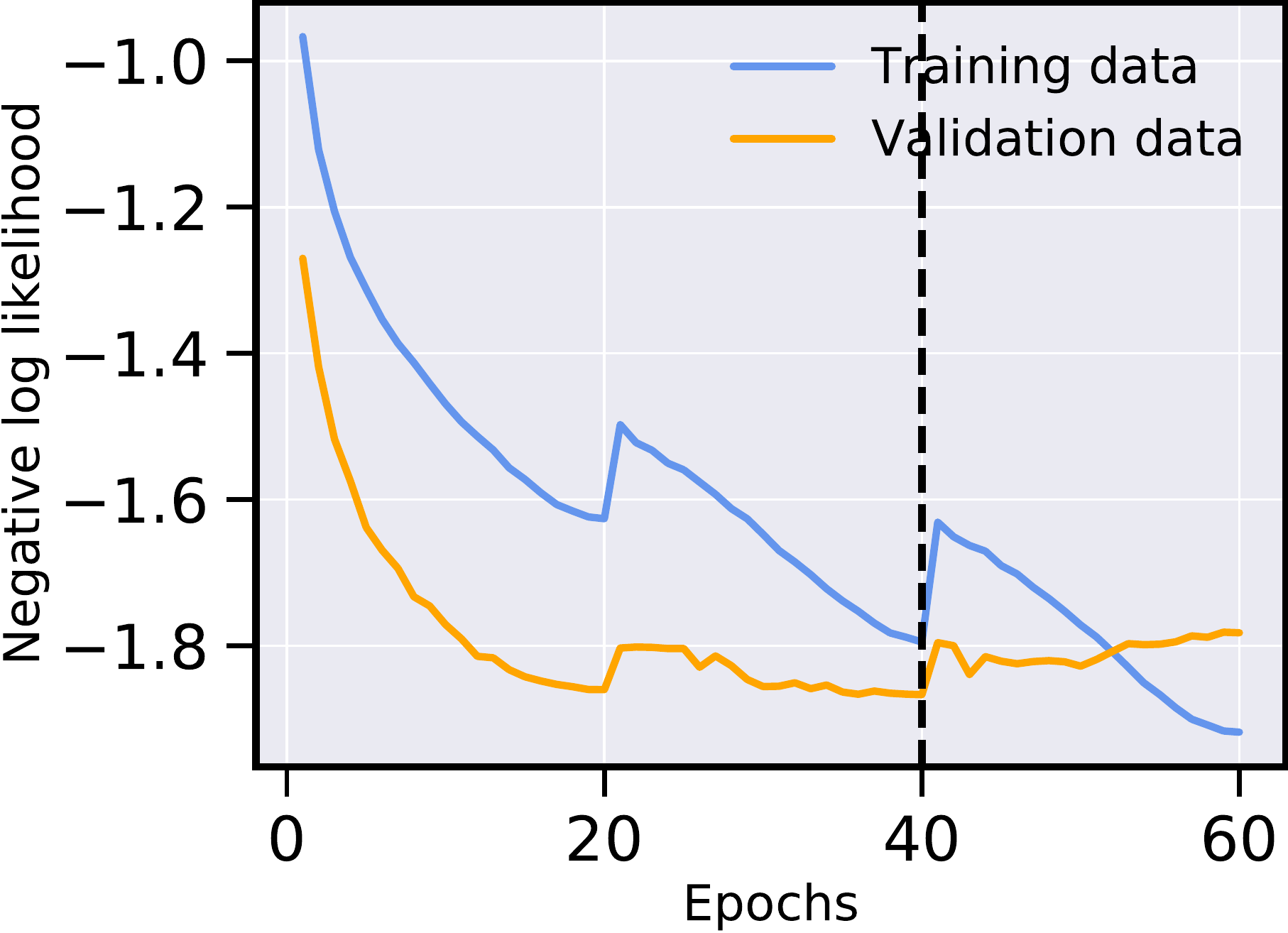}
            \caption{Training and validation curves for one of ten folds when using the Mixture Density Network, as described in Section \ref{sec:putting_it_all_together}. The vertical line indicates the epoch of minimum validation loss.}
            \label{fig:mdn_training_curve}
        \end{subfigure}
        \caption{Loss as a function of learning rate and epochs for one of ten folds when training using the MDN. In $\textbf{(b)}$, the training loss is higher than the validation loss owing to the MoEx augmentation \citep{moex}, as we explain in Section \ref{sec:mdn}.}
        \label{fig:mdn_ancillary}
    \end{minipage}

    \begin{minipage}{.97\textwidth}
        \begin{subfigure}{0.49\textwidth}
            \centering
            \includegraphics[width=.98\linewidth]{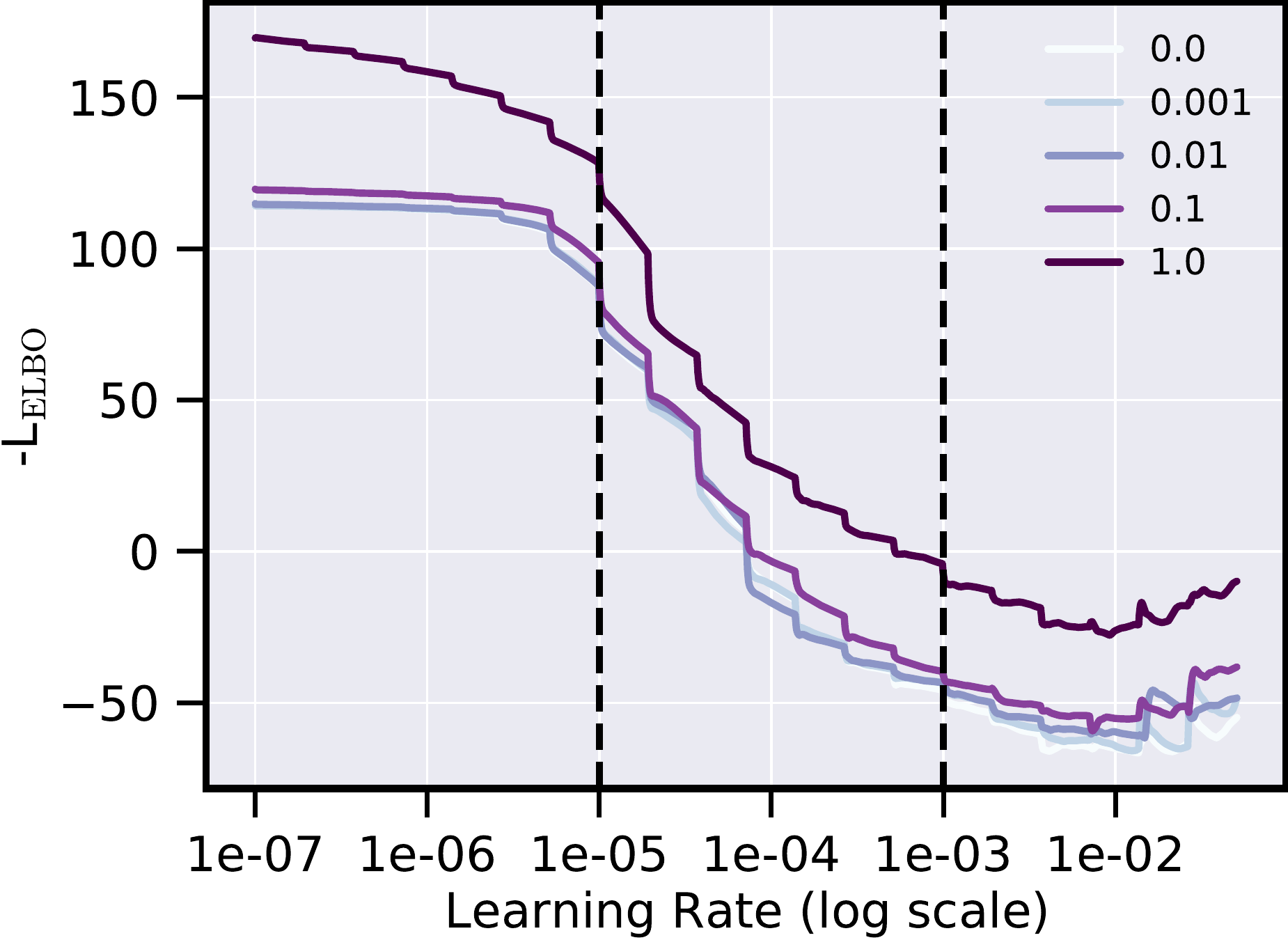}
            \caption{Choosing lower and upper learning rates for the cyclic learning rate scheduler. Since we also anneal W$_{\rm KL}$, it is important to find LR limits that encourage quick convergence over the entire domain of W$_{\rm KL}$.}
            \label{fig:vae_find_lr}
        \end{subfigure}
        \hfill
        \begin{subfigure}{0.49\textwidth}
            \centering
            \includegraphics[width=.98\linewidth]{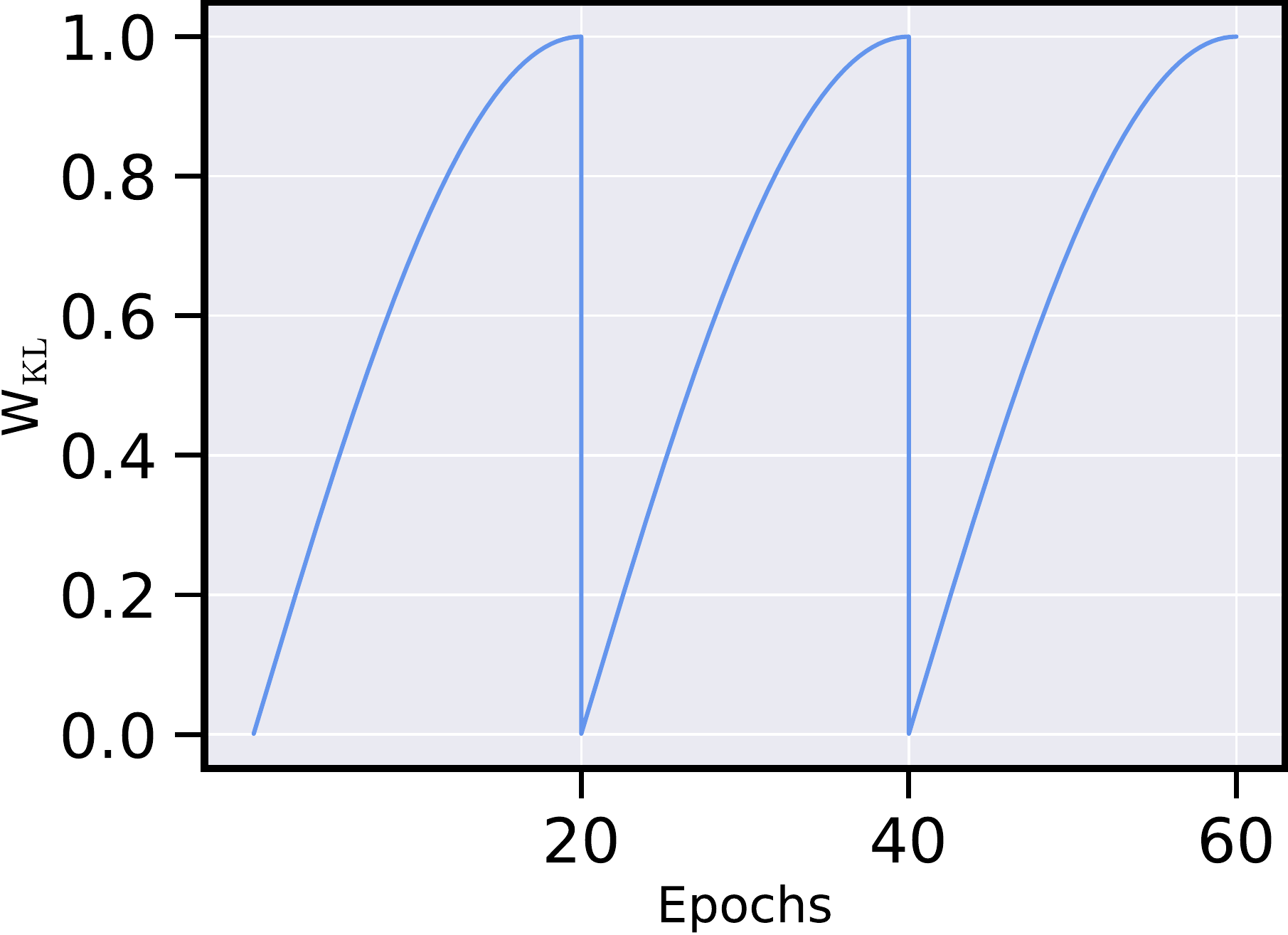}
            \caption{Annealing W$_{\rm KL}$ from 0 to 1 in each cycle of 20 epochs.Finding optimal $\beta$ for the robust variational autoencoder. This prevents L$_{\rm KL}$ from collapsing to 0, and is based upon findings of \cite{cyclical_wkl_annealing}.}
            \label{fig:vae_weightkl_vs_epoch}
        \end{subfigure}
        \newline
        \begin{subfigure}{0.49\textwidth}
            \centering
            \includegraphics[width=.98\linewidth]{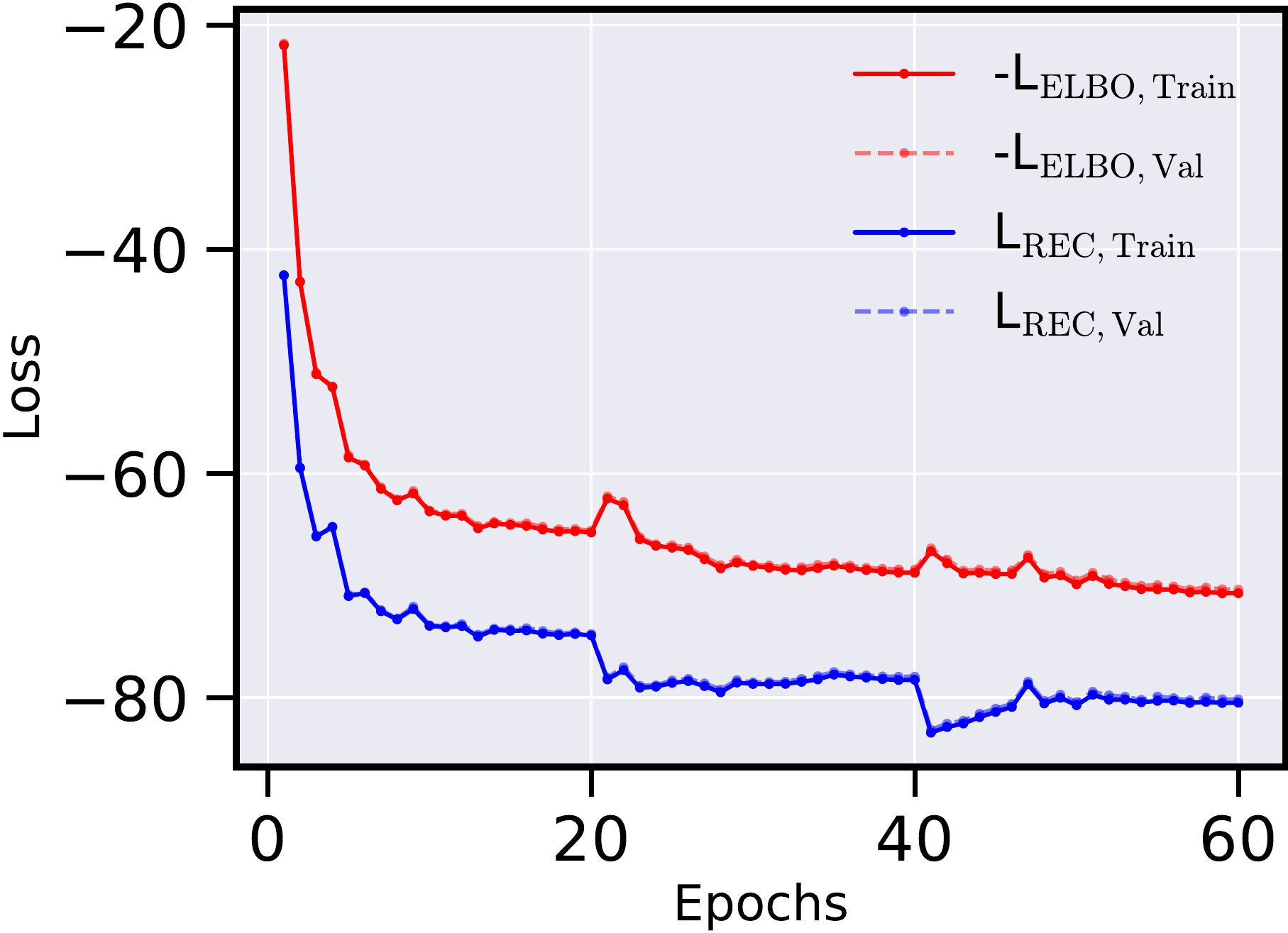}
            \caption{Training and validation curves for the total loss and the reconstruction loss for one of ten folds.}
            \label{fig:vae_trainingcurve1}
        \end{subfigure}
        \hfill
        \begin{subfigure}{0.49\textwidth}
            \centering
            \includegraphics[width=.98\linewidth]{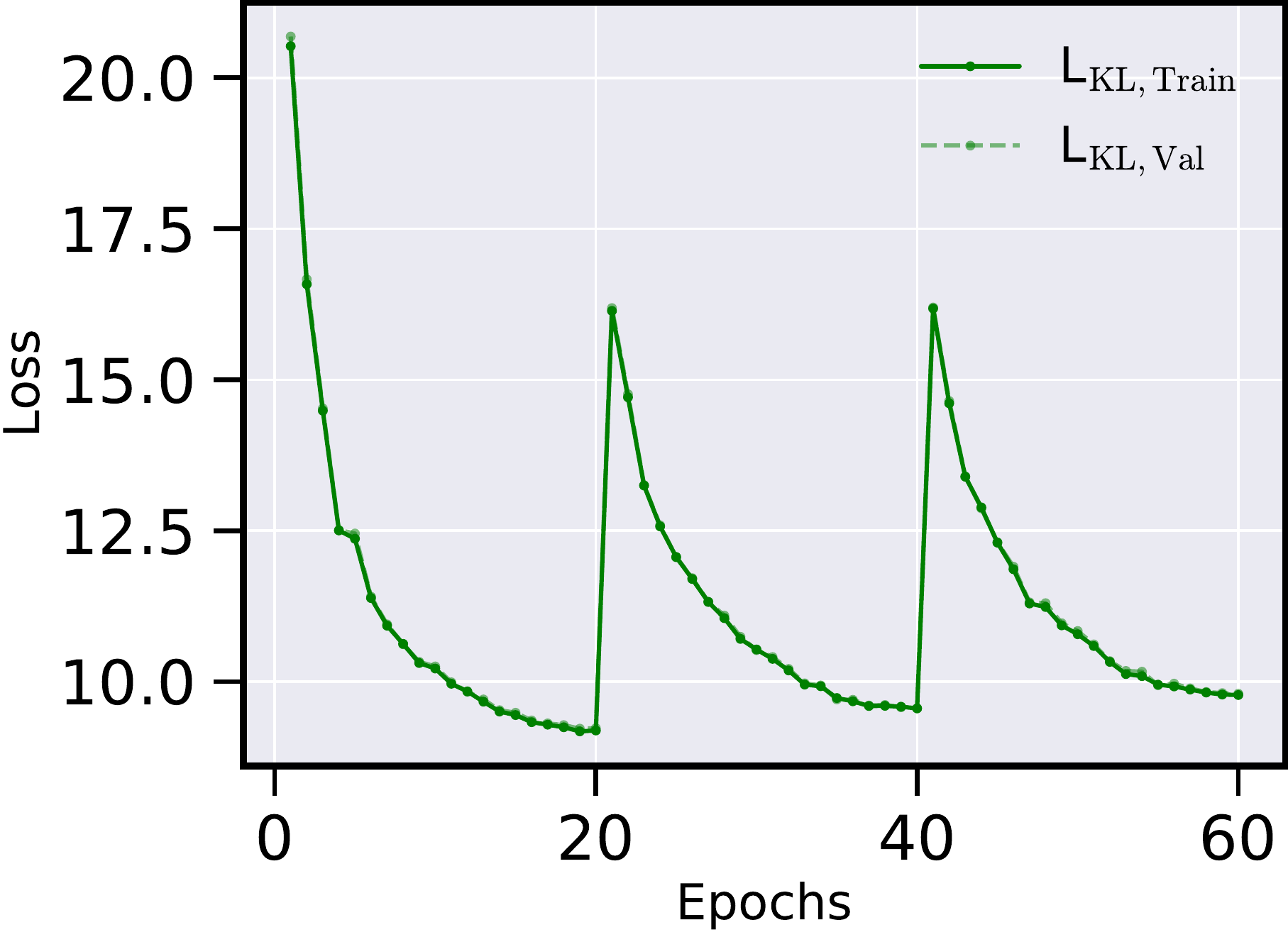}
            \caption{Training and validation curves for the KL-divergence loss for one of ten folds.}
            \label{fig:vae_trainingcurve2}
        \end{subfigure}
        \caption{Losses as a function of learning rate, W$_{\rm KL}$, and epochs, for one of ten folds when training using the RVAE. While the MDN has a single loss (see Figure \ref{fig:mdn_ancillary}), the RVAE has two individual losses.  In conjunction these form the final loss that is minimized by mini-batch gradient descent (-L$_{\rm ELBO}$ = L$_{\rm REC}$ + L$_{\rm KL}$, see Section \ref{sec:rvae}). Different from Figure \ref{fig:mdn_training_curve}, \textbf{(c)} and \textbf{(d)} here are not `live' plots, but constructed once the RVAE has been fully trained.}
        \label{fig:vae_ancillary}
    \end{minipage}
\end{figure*}

In this appendix we present a few figures that we anticipate will help the reader better understand our implementation of the machine learning techniques we use.  In particular, in Figure~\ref{fig:cfht_mdn_overview} we illustrate our MDN training process.  In Figure~\ref{fig:cfht_rvae_overview} we illustrate the approach taken to identify in-distribution vent configurations amongst all possible $2^{12}$ configurations.  Our predictions are restricted to  only in-distribution vent configurations.  Finally, in Figures~\ref{fig:mdn_ancillary} and~\ref{fig:vae_ancillary} we present some details on how we select our learning rate.

\section{MDN vs. Catboost} \label{sec:app_mdnvscatboost}

\begin{figure*}
\begin{subfigure}{0.99\textwidth}
    \centering
    \includegraphics[width=.98\linewidth]{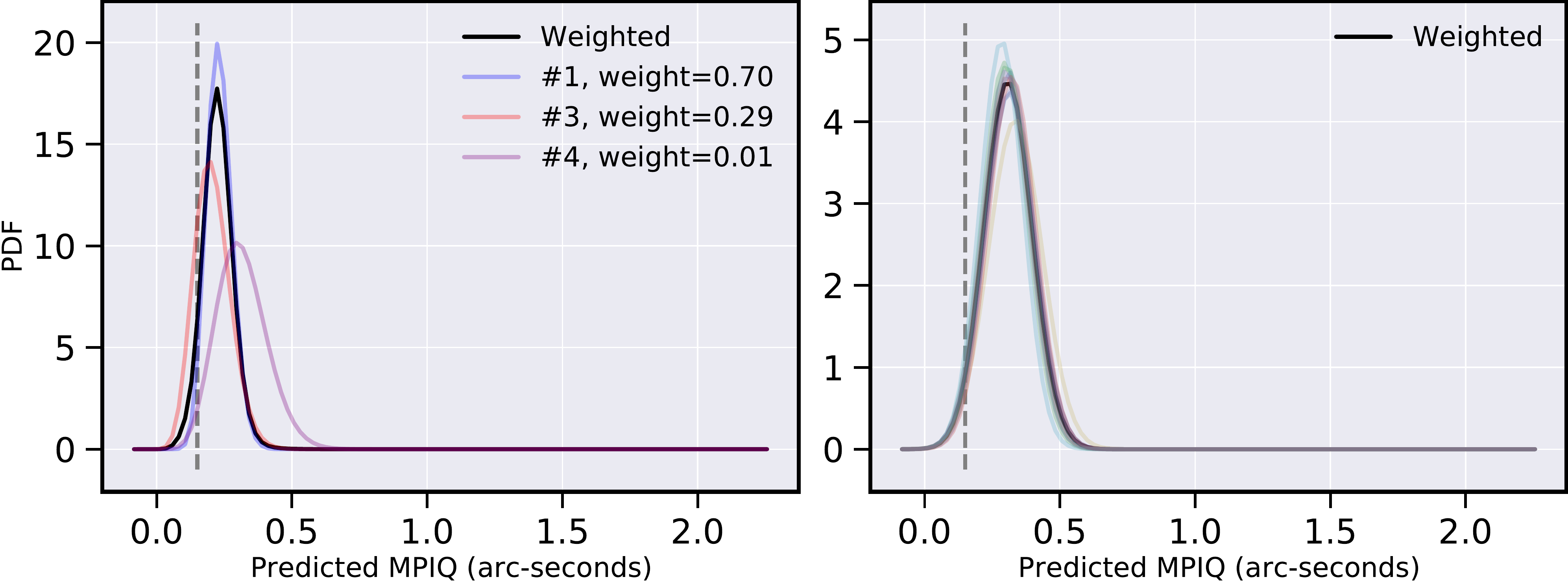}
    \caption{Predictions for training sample with true MPIQ $\sim0.15''$. \textbf{Left:} PDFs for components of the MDN with non-zero weights, along with the final weighted PDF in black. The first, third, and fourth MDN component have non-trivial weights of 0.70, 0.29 and 0.10, respectively, and hence are the only three weights plotted. \textbf{Right:} PDFs for the 10 equally-weighted components of the {\sc catboost} ensemble, along with the final PDF in black.}
    \label{fig:mdn_vs_cb_train_low}
\end{subfigure}
\newline
\begin{subfigure}{0.99\textwidth}
    \centering
    \includegraphics[width=.98\linewidth]{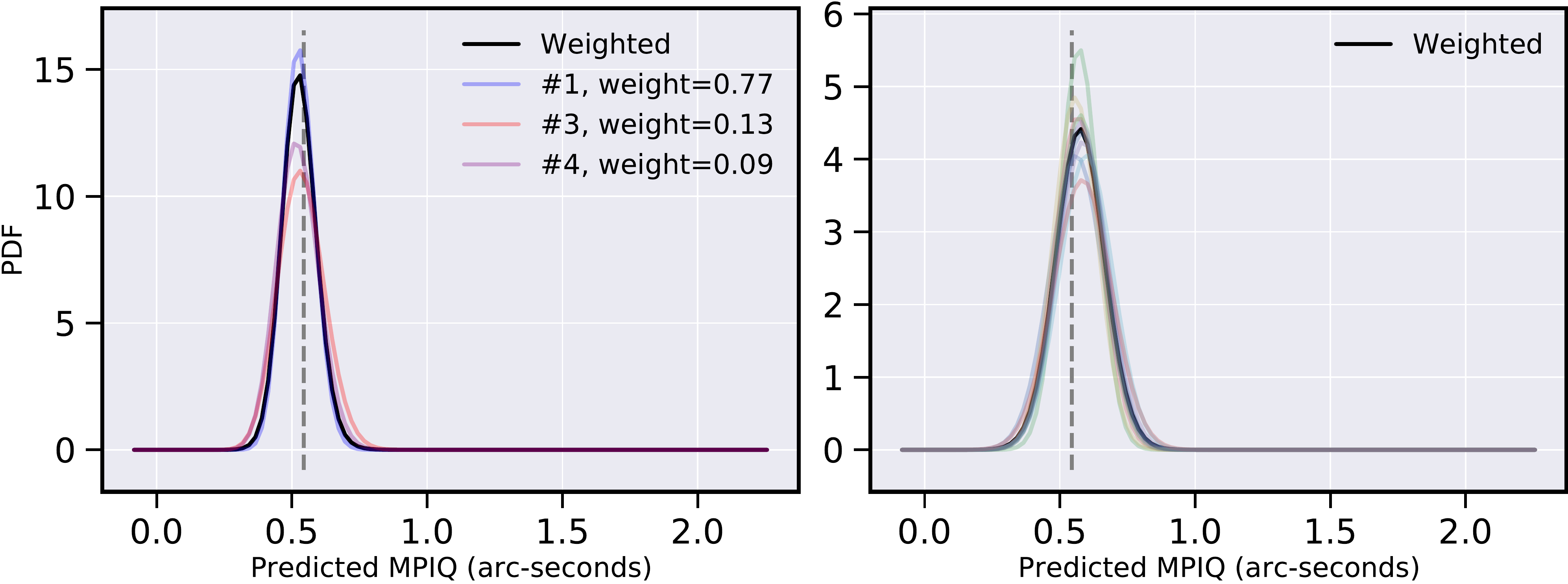}
    \caption{Similar to Figure \ref{fig:mdn_vs_cb_train_low}, except for training sample with true MPIQ $\sim0.52''$. \textbf{Left:} The first, third, and fourth MDN component have non-trivial weights of 0.77, 0.13 and 0.09, respectively.   \textbf{Right:} PDFs for the 10 equally-weighted components of the {\sc catboost} ensemble, along with the final PDF in black.}
    \label{fig:mdn_vs_cb_train_med}
\end{subfigure}
\newline
\begin{subfigure}{0.99\textwidth}
    \centering
    \includegraphics[width=.98\linewidth]{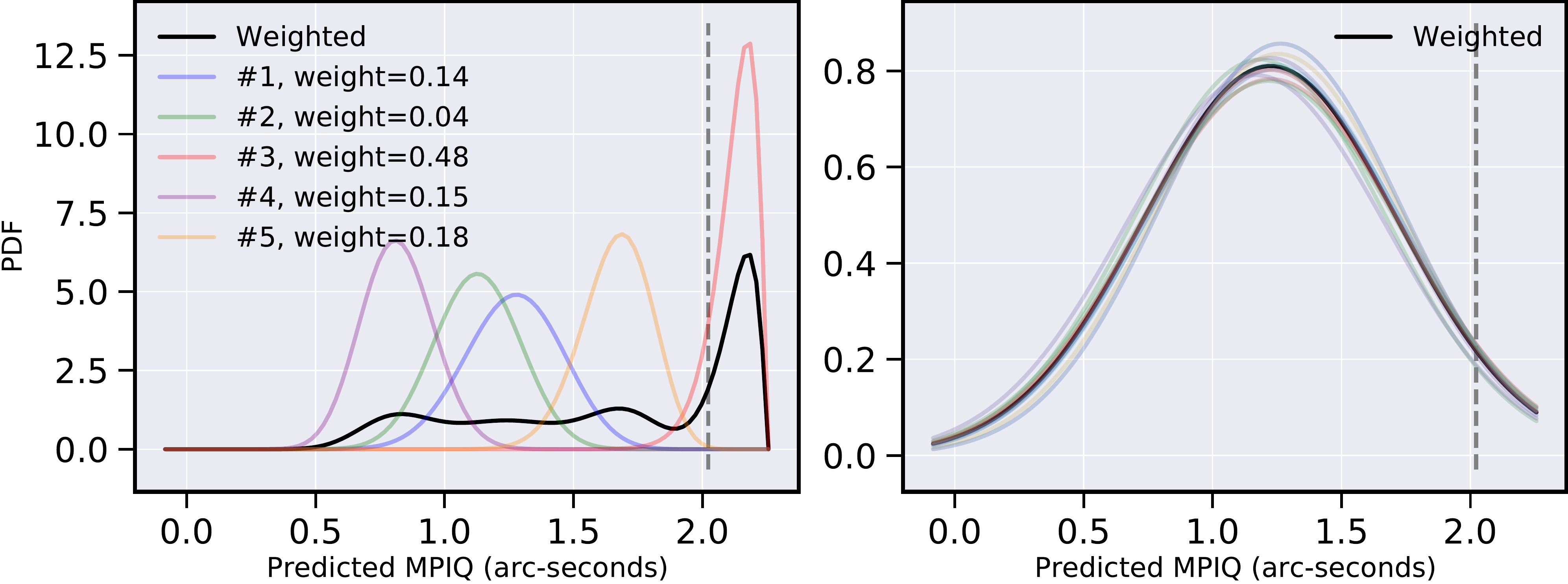}
    \caption{Similar to Figures \ref{fig:mdn_vs_cb_train_low} and \ref{fig:mdn_vs_cb_train_med}, except for training sample with true MPIQ $\sim2.0''$. \textbf{Left:} All five components of the MDN have non-zero weights, per the legend. \textbf{Right:} PDFs for the 10 components of the {\sc catboost} ensemble, each with an equal weight of 0.10, along with the final PDF in black.}
    \label{fig:mdn_vs_cb_train_high}
\end{subfigure}
\caption{Probability distribution functions (PDFs) of model components for both  MDN and {\sc catboost} ensembles, for three samples from the training set (at low, medium, and high MPIQ). In each subfigure, the colored curves represent model components PDFs.  The solid black curve in each subfigure is the weighted, final PDF.  The dashed, gray vertical line indicates the true or measured MPIQ.}
\label{fig:mdn_vs_cb_train}
\end{figure*}

\begin{figure*}
\begin{subfigure}{0.99\textwidth}
    \centering
    \includegraphics[width=.98\linewidth]{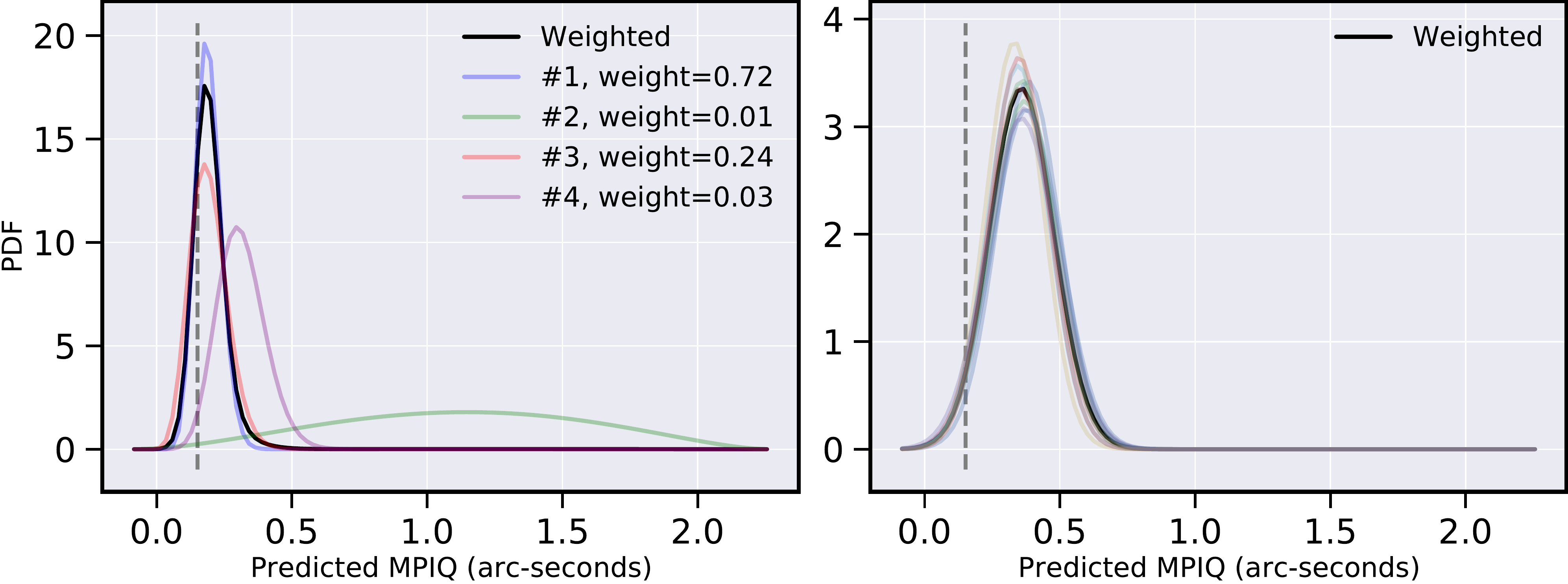}
    \caption{Predictions for test sample with true MPIQ $\sim0.15''$. \textbf{Left:} PDFs for components of the MDN with non-zero weights, along with the final  PDF in black. \textbf{Right:} PDFs for the 10 components of the {\sc catboost} ensemble each with equal weight, along with the final  PDF in black.}
    \label{fig:mdn_vs_cb_test_low}
\end{subfigure}
\newline
\begin{subfigure}{0.99\textwidth}
    \centering
    \includegraphics[width=.98\linewidth]{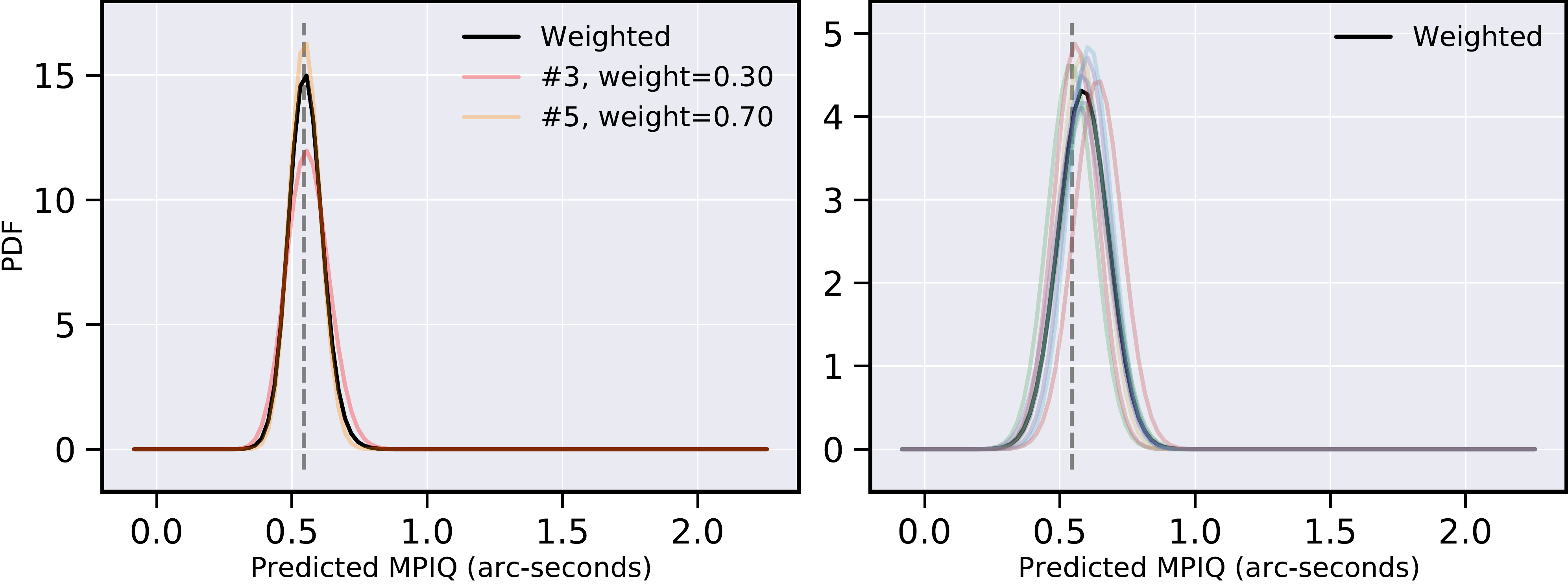}
    \caption{Similar to Figure \ref{fig:mdn_vs_cb_test_low}, except for training sample with true MPIQ $\sim0.52''$.}
    \label{fig:mdn_vs_cb_test_med}
\end{subfigure}
\newline
\begin{subfigure}{0.99\textwidth}
    \centering
    \includegraphics[width=.98\linewidth]{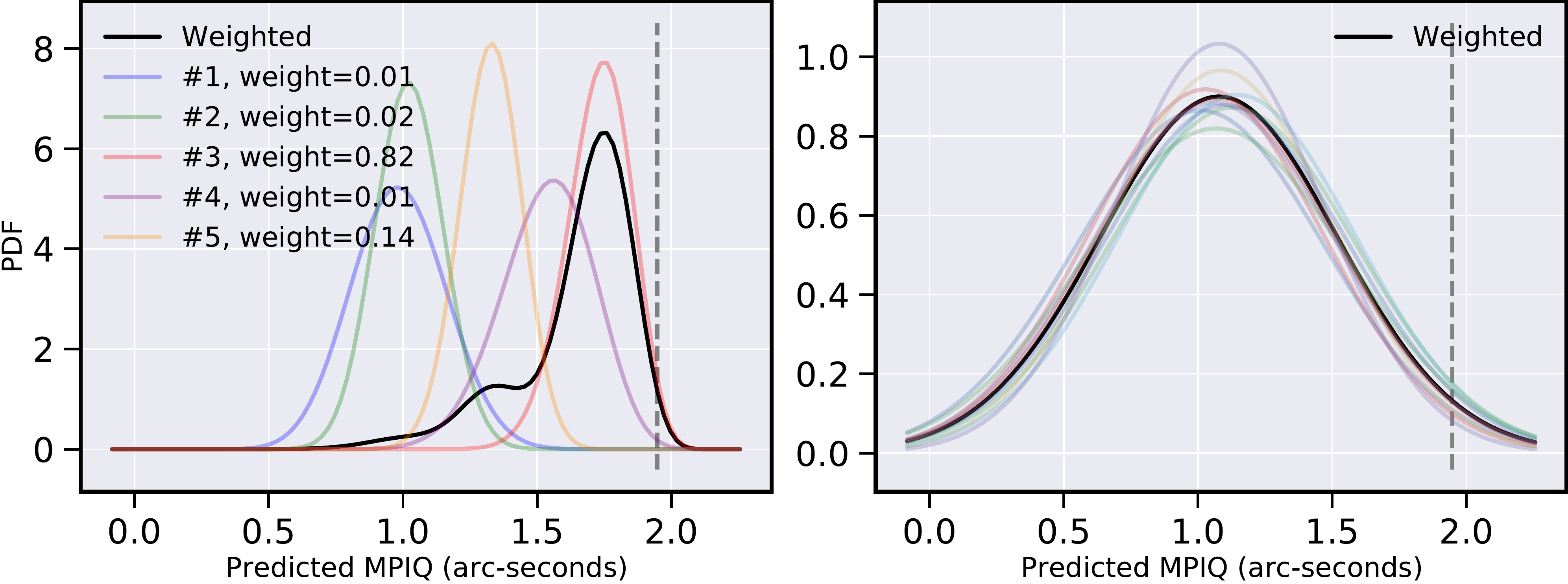}
    \caption{Similar to Figures \ref{fig:mdn_vs_cb_test_low} and \ref{fig:mdn_vs_cb_test_med}, except for training sample with true MPIQ $\sim1.98''$.}
    \label{fig:mdn_vs_cb_test_high}
\end{subfigure}
\caption{This figure is similar to Figure \ref{fig:mdn_vs_cb_train}, but now for three samples (at low, medium, and high MPIQ) selected from the test set of $\sim6600$ samples (as opposed to the training set in Figure \ref{fig:mdn_vs_cb_train}).}
\label{fig:mdn_vs_cb_test}
\end{figure*}

In Figures \ref{fig:mdn_vs_cb_train} and \ref{fig:mdn_vs_cb_test}, we plot the predicted probability distribution functions (PDFs) for three samples each from the training and the test sets, using both the mixture density network (MDN) and the {\sc catboost} ensemble. Note that while GBDT (gradient boosted decision tree) is the specific algorithm, and the term we have repeatedly used in the main text of this paper, the specific implementation of it used here is {\sc catboost}\footnote{\url{https://catboost.ai/}}.These samples are those with those with the minimum, median, and maximum measured MPIQ, in both the training and test sets, respectively. We aim to visualize both the difference in performance between the two models at various levels of measured MPIQ, as well as the full PDF for all model components for both models.

We make several observations. First, all 10 components of the {\sc catboost} ensemble have equal weights, whereas each of the 5 components of the MDN can have variable weights depending on the sample. Second, the diversity in model components, both in \textit{x-} and \textit{y-} values, is much larger for MDN than for {\sc catboost}. Third, uncertainties as quantified by the full-width-at-half-maximum are significantly lower for PDFs predicted by MDN than they are for PDFs predicted by {\sc catboost}. Fourth, for both low and high MPIQ values, the $\beta$ likelihoods in MDN components enable a significantly more flexible representation, and hence a more realistic, asymmetric PDF. Specifically, the low-MPIQ ($\sim0.15''$) and high-MPIQ ($\sim2''$) both have quite distinct MDN components (in terms of the predictions) whereas the mid-MPIQ ($\sim0.5''$) MDN predictions are all quite similar, more akin to the {\sc catboost} ensemble. As $\sim0.5''$ is near the mode of the MPIQ distribution (see the top plots in Figure \ref{fig:hist_preliminaries}),  both models are able to capture the relationship between the input features and the output MPIQ values quite well.  In contrast, the behavior of the two models is quite different at the tails of the distribution. The added flexibility if the MDN results in significantly better predictions than the {\sc catboost} ensemble can accomplish.  This is especially the case at low and high MPIQ values, which are close to OoD.

\begin{figure*}
    \centering
    \begin{minipage}{.97\textwidth}
        \begin{subfigure}{0.49\textwidth}
            \centering
            \includegraphics[width=.98\linewidth]{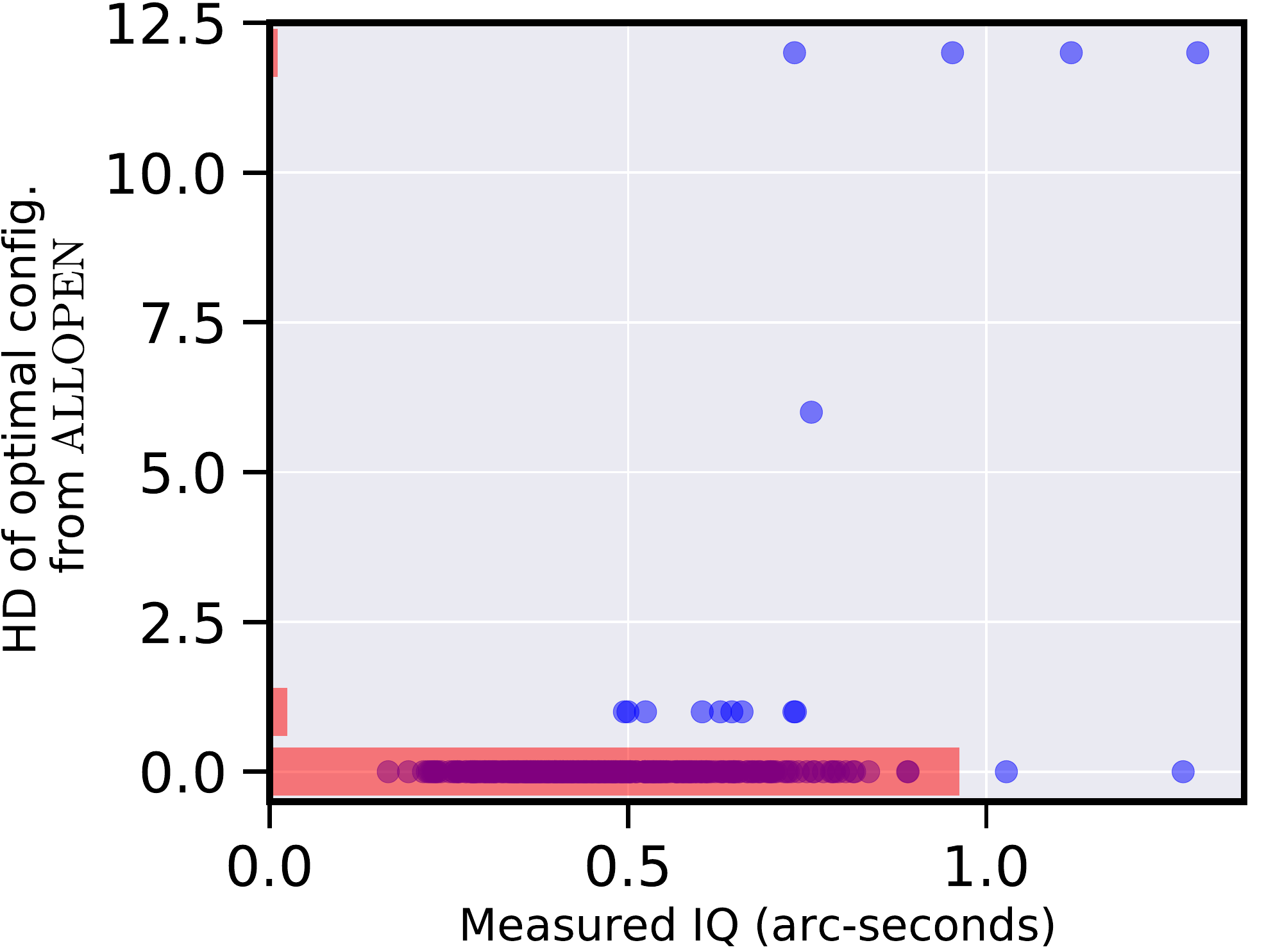}
            \caption{Hamming distance (HD) of optimal vent configurations with reference to the all-open baselines, plotted against measured MPIQ values. We overlay this against a bar-chart showing the percentage of these configurations.}
            \label{fig:hdsvsiqs_and_histhds_cb}
        \end{subfigure}
        \hfill
        \begin{subfigure}{0.49\textwidth}
            \centering
            \includegraphics[width=.98\linewidth]{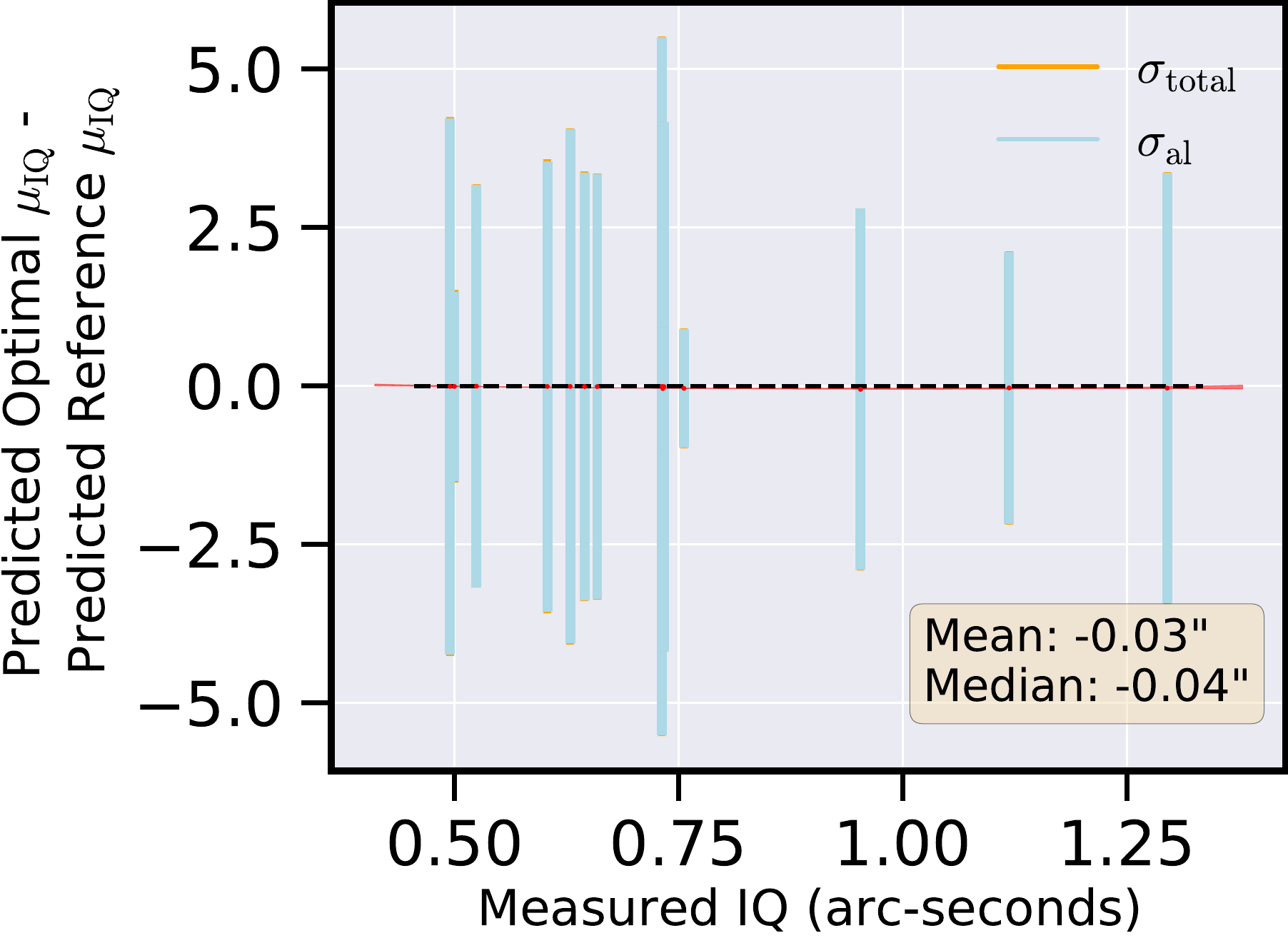}
            \caption{Improvement in MPIQ for the optimal vent configurations over the all-open baselines. The {\sc catboost} ensemble predicts only 14 test samples to have alternative vent configurations that lower mean predicted MPIQ.}
            \label{fig:deltaoptimalreference_vs_nominal_iqs_all_cb}
        \end{subfigure}
        \newline
        \begin{subfigure}{0.99\textwidth}
            \centering
            \includegraphics[width=.98\linewidth]{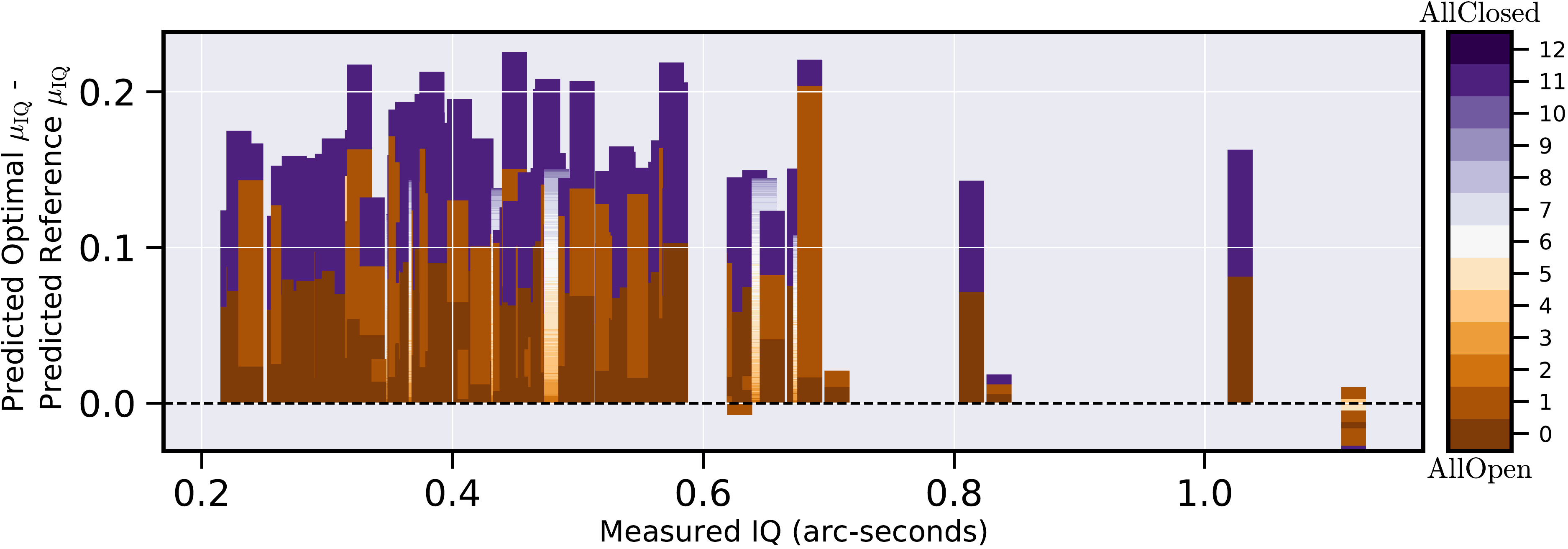}
            \caption{Change in MPIQ predicted by the {\sc catboost} ensemble (as opposed to the predictions of the MDN model presented in Figure \ref{fig:common_barplot}) for each possible vent configuration for $100$ randomly selected samples, actuating vent configurations across all ID settings for each. Lower is better.}
            \label{fig:barplot_cb}
        \end{subfigure}
        \caption{We use the {\sc catboost} ensemble to find the optimal vent configurations that would result in the lowest MPIQ values for each of the $\sim660$ test samples with all vents open. Our results indicate that by and large, the {\sc catboost} ensemble in unable to find any superior alternative configurations.}
        \label{fig:cb_appendix}
    \end{minipage}

    \begin{minipage}{.97\textwidth}
        \begin{subfigure}{0.49\textwidth}
            \centering
            \includegraphics[width=.98\linewidth]{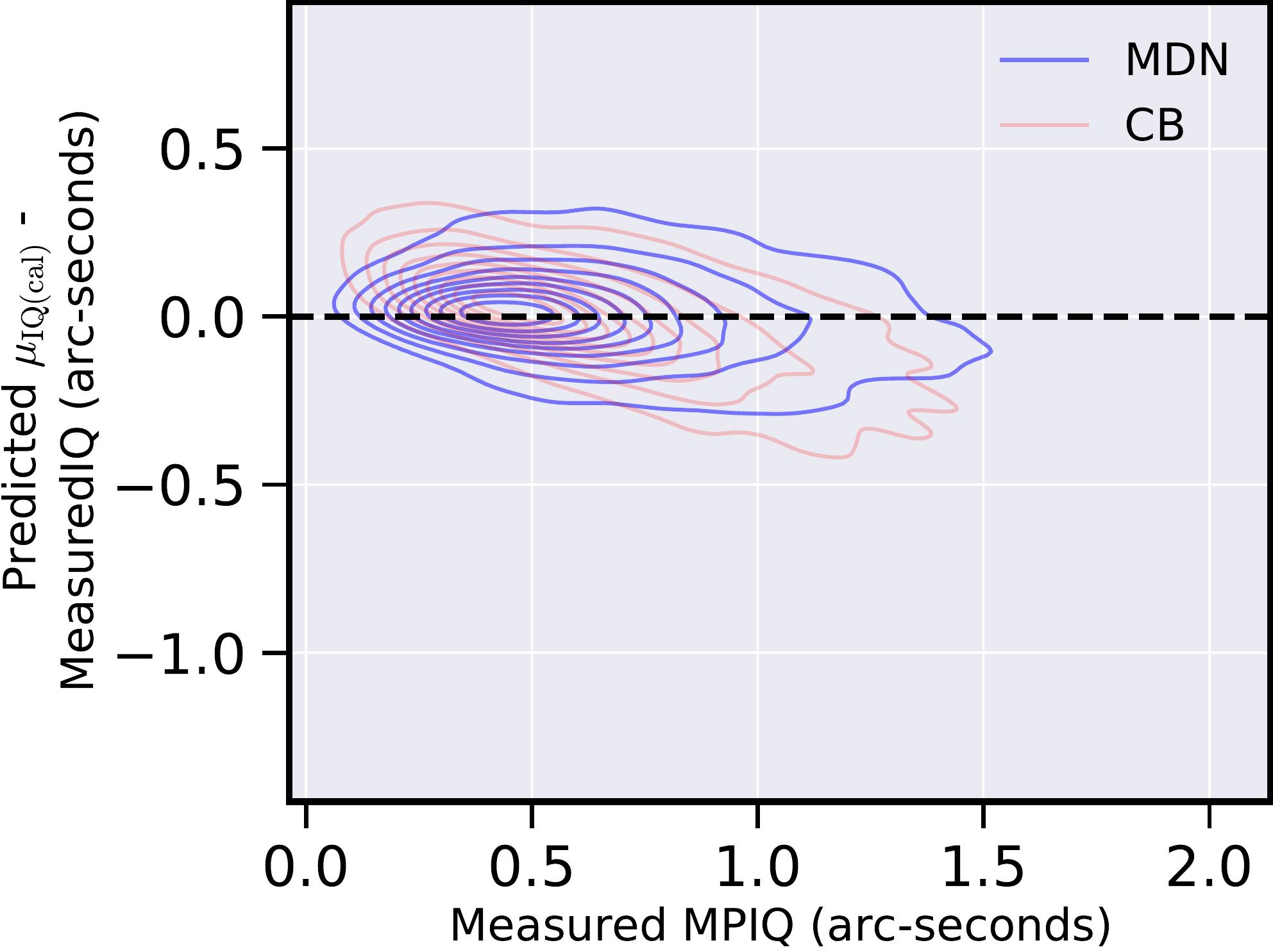}
            \caption{Kernel density estimate plots for the errors in predictions for the $\sim6600$ test samples, from the MDN in blue and the {\sc catboost} ensemble in red.}
            \label{fig:mdncbminusmeasured_vs_measured_density}
        \end{subfigure}
        \hfill
        \begin{subfigure}{0.49\textwidth}
            \centering
            \includegraphics[width=.98\linewidth]{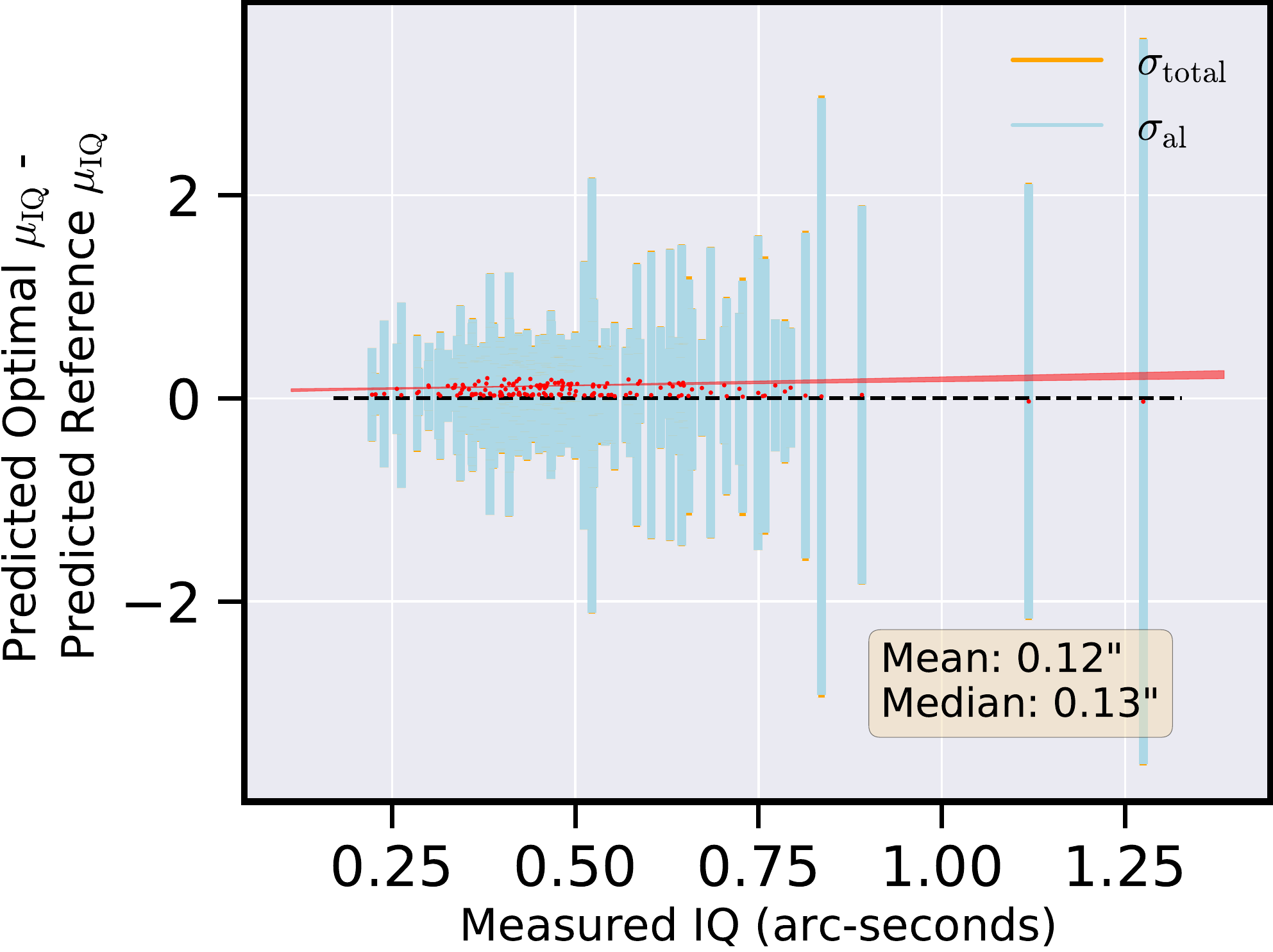}
            \caption{Improvement in MPIQ predicted by the {\sc catboost} ensemble for the optimal vent configurations found by the MDN.}
            \label{fig:deltaoptimalreference_vs_nominal_iqs_all_cbpredonmdn}
        \end{subfigure}
        \caption{{\sc catboost} vs. MDN. In \textbf{(a)}, 
        we highlight that {\sc catboost}'s predictions are highly biased at both low and high MPIQ, whereas the MDN is only slightly biased fpr large MPIQ values. 
        In \textbf{(b)}, we use {\sc catboost} to predict MPIQ PDFs for the hypothetical samples with optimal vent configurations identified by the MDN. {\sc catboost} predicts an \textit{increase} in MPIQ, further verifying that it does not extrapolate well beyond densely sampled data regions.}
        \label{fig:cbonmdn}
    \end{minipage}
\end{figure*}

Finally, in Figures \ref{fig:cb_appendix} and \ref{fig:cbonmdn} we highlight the {\sc catboost} ensemble's deficiencies as a predictor, justifying our decision to use MDN for all  tasks in this paper. In Figure \ref{fig:cb_appendix} we see that for all except 14 test samples, the {\sc catboost} ensemble is unable to hypothesize vent configurations that can improve over the all-open baseline.  And, even in cases where improvement is hypothesized, the improvements are minuscule. In Figure \ref{fig:mdncbminusmeasured_vs_measured_density} we explicitly highlight the large difference in performance between the MDN and the {\sc catboost} models, especially when it comes to the tails of the MPIQ distributions.  This is something  we also showed in Figures  \ref{fig:mdn_vs_cb_train} and \ref{fig:mdn_vs_cb_test}. In Figure \ref{fig:deltaoptimalreference_vs_nominal_iqs_all_cb} we check how the {\sc catboost} model performs on test samples with vent configurations designated as optimal by the MDN, and observe, unsurprisingly, poor performance. Recall that from Figure \ref{fig:deltaoptimalreference_vs_nominal_iqs_all_cb} we already know that only for a very small fraction of samples (14 out of $\sim660$) does the {\sc catboost} ensemble predict any reduction in MPIQ at all.

\bsp	
\label{lastpage}
\end{document}
